\documentclass[nofootinbib,longbibliography,reprint,aps,prd,showpacs,superscriptaddress,groupedaddress]{revtex4-1}
\usepackage{amssymb,mathrsfs,amsmath}

\usepackage{braket}
\usepackage{graphicx}
\usepackage{dcolumn}   
\usepackage{bm} 
\usepackage{multirow}
\usepackage{footnote}
\usepackage{epstopdf}
\usepackage{rotating}
\usepackage{hyperref}
\usepackage{url}
\usepackage{xcolor}
\usepackage{pifont}

\graphicspath{ {./images/} }

\newcommand{\cmark}{\ding{51}}%
\newcommand{\xmark}{\ding{55}}%

\definecolor{darkblue}{cmyk}{1, 1, 0, 0}
\hypersetup{colorlinks=true,urlcolor=darkblue,citecolor=darkblue,linkcolor=darkblue}

\newcolumntype{C}[1]{>{\centering\let\newline\\\arraybackslash\hspace{0pt}}m{#1}}
\newcolumntype{L}[1]{>{\raggedright\let\newline\\\arraybackslash\hspace{0pt}}m{#1}}
\newcolumntype{R}[1]{>{\raggedleft\let\newline\\\arraybackslash\hspace{0pt}}m{#1}}

\newcommand{\mnras}{Mon.~Not.~R.~Astron.~Soc.}
\newcommand{\aap}{Astron.~Astrophys.}

\newcommand{\physrep}{Phys.~Rep.}

\newcommand{\apjs}{Astrophys.~J.~Supp.}
\newcommand{\ssr}{Space Science Reviews}
\newcommand{\jcap}{J.~Cosmol.~Astropart.~Phys.}

\newcommand{\lambdaB}{\bm{\lambda}}

\newcommand{\muB}{\bm{\mu}}
\newcommand{\mubar}{\bm{\bar{\mu}}}
\newcommand{\deltaB}{\bm{\delta}}
\newcommand{\deltaG}{\bm{\delta}^T\bm{G}\bm{\delta}}

\newcommand{\deltaGhalf}{\tfrac{1}{2}\bm{\delta}^T\bm{G}\bm{\delta}}
\newcommand{\deltaGhalfbar}{\tfrac{1}{2}\bar{\bm{\delta}}^T\bm{\bar{G}}\bar{\bm{\delta}}}
\newcommand{\CBone}{\bm{C^{(1)}}}
\newcommand{\CBtwo}{\bm{C^{(2)}}}
\newcommand{\CB}{\bm{C}}

\newcommand{\IOInumggw}{5.41}
\newcommand{\RNnumggw}{12.96}
\newcommand{\sigmaeqggw}{59.3\%}
\newcommand{\Snumggw}{-432.20}
\newcommand{\SigSnumggw}{-0.63}
\newcommand{\SnumggwP}{4.05}
\newcommand{\SigSnumggwP}{1.51}
\newcommand{\CERnumggw}{2.41}
\newcommand{\SigCERnumggw}{1.39}
\newcommand{\IOInumggl}{4.34}
\newcommand{\RNnumggl}{11.52}
\newcommand{\sigmaeqggl}{56.9\%}
\newcommand{\Snumggl}{-340.20}
\newcommand{\SigSnumggl}{-0.60}
\newcommand{\SnumgglP}{3.06}
\newcommand{\SigSnumgglP}{1.46}
\newcommand{\CERnumggl}{1.84}
\newcommand{\SigCERnumggl}{1.17}

\begin{document}
\title{Cosmological discordances: A new measure, marginalization effects, and application to geometry versus growth current data sets}
\date{\today}
\author{Weikang Lin}
\email{wxl123830@utdallas.edu}
\author{Mustapha Ishak}
\email{mishak@utdallas.edu}
\affiliation{Department of Physics, The University of Texas at Dallas, Richardson, Texas 75080, USA}

\begin{abstract}
The continuous progress toward more precise cosmological surveys and experiments has galvanized recent interest into consistency tests on cosmological parameters and models. At the heart of this effort is quantifying the degree of inconsistency between two or more cosmological data sets. We introduce an intuitive moment-based measure we call the \textit{index of inconsistency} (IOI) and show that it is sensitive to the separation of the means, the size of the constraint ellipsoids, and their orientations in the parameter space. We find that it tracks accurately the inconsistencies when present. Next, we show that parameter marginalization can cause a loss of information on the inconsistency between two experiments and we quantify such a loss using the drop in IOI. In order to zoom on a given parameter, we define the relative residual IOI and the relative drop in IOI. While these two quantities can provide insights on the parameters that are most responsible for inconsistencies, we find that the full IOI applied to the whole parameter spaces is what must be used to correctly reflect the degree of inconsistency between two experiments. We discuss various properties of IOI, provide its eigenmode decomposition, and compare it to other measures of discordance. Finally, we apply IOI to current geometry data sets (i.e. an improved Supernovae Type Ia compilation, baryon acoustic oscillations from 6dF, SDSS MGS and Lyman-$\alpha$ forest, and high-$\ell$ cosmic microwave background (CMB) temperature data from Planck-2015) versus growth data sets (i.e. Redshift Space Distortions from WiggleZ and SDSS, Weak Lensing from CFHTLenS, CMB Lensing, Sunyav-Zeldovich effect, and low-$\ell$ CMB temperature and polarization data from Planck-2015). We find that a persistent inconsistency is present between the two data sets. This could reflect the presence of systematics in the data or inconsistencies in the underlying model.
\end{abstract}

\maketitle


\section[Introduction]{Introduction}
The $\Lambda$CDM standard model of cosmology enjoys a number of observational confirmations and successes. However, it does come with two intriguing conundrums. The first one is that it requires a dark matter component counting for about $26\%$ of matter-energy content in the Universe. The second one is that the expansion of the Universe is accelerating and we do not know what is driving this acceleration. Associated with this cosmic acceleration is a dark energy component that could count for about $69\%$ of the energy budget in the Universe.
These and other questions have motivated studies to consider if there are any problems with the underlying model, theory or assumptions. See, for example, Refs.\,\cite{2012-Clifton-MG,2017-Lopez-Corredoira-test-of-SMcosmo,2008-cosmology-review-Tsagas,Cosmo.test.of.GR,2015-rev-Joyce-Jain,2007-Ishak-Remarks-DE,2016-Joyce-Lombriser-Schmidt-DEvsMG,2015-review-LambdaCDM,2016-Debono-Smoot-GRandCosmology} and references therein.

One route to test this that is attracting more attention lately is to compare results and parameters from different  experiments and then to look for any inconsistencies. Over a decade ago, Ref.\,\cite{2006-Ishak-splitting} used simulated data sets to demonstrate how inconsistencies between the dark energy parameter spaces as constrained by the expansion versus the growth can signal a failure of the underlying gravity theory. A related method called ``parameter splitting'' was applied to real data in Refs.\,\cite{2007-Wang-etal-consistency,2015Ruiz-etal-param-splitting,2016Bernal-etal-param-splitting} and looked for inconsistencies between dark energy parameters as constrained by expansion versus the growth. References.\,\cite{2012-Shafieloo-crossing-stat,2016-Shafieloo-Hazra-Consistency-Planck} introduced a technique called the crossing function, which enables us to see if the best fit from one experiment is consistent with the constraints from another experiment.

Most recently, a number of papers have focused on possible discrepancies between cosmological parameters as constrained by different experiments. While some have faded away, some others seem to persist; see, for example Refs.\,\cite{2016-Bernal-Verde-Riess-H0,2015-Valentino-Melchorri-Silk-Beyon-LCDM,2015bao-sdssIII,2013CFHTlens,2017-Charnock-Battye-Moss}. We take the point of view that whether due to systematic effects in the data or due to underlying physics, these discrepancies need a careful study. Otherwise, a joint analysis will be of questionable outcome. In the next section, we review briefly some of the inconsistencies or tensions that have appeared in the literature.

An important question is to quantify the degree of inconsistency or tension. Very often, the inconsistency between two experiments is qualitatively shown in marginalized one-dimensional (1D) or two-dimensional (2D) likelihood contour plots. However, as we demonstrate in this paper, this method cannot accurately represent the inconsistency. First of all, it is not a quantitative method, rendering it difficult to be interpreted in an accurate way. Secondly, if the underlying model has three or more parameters this method fails to account for the full inconsistency. Moreover, as shown later in this work, listing all the marginalized 1D or 2D plots does not properly represent the full inconsistency either. Therefore some measures have been introduced in the literature in order to quantify inconsistencies between two experiments, and we review most of them in Sec.\,\ref{section-comparison}. An important criterion is that a measure must properly describe experimental discordances.

In this paper, we define and apply a measure of inconsistency that is found to track accurately inconsistencies. It is a moment-based quantity that we call the \textit{index of inconsistency} (IOI). This can be applied to two or more experiments or data sets. We find that in the Gaussian and weak prior limit, most other measures in the literature reduce to or contain a term of IOI. This measure properly describes the factors causing the inconsistency: i.e mean difference, constraint volumes and their orientations.

Importantly, we also show here that parameter marginalization can hide inconsistency. We argue that a  good measure of inconsistency should reflect this. When zooming on specific parameters, we define the relative drop in IOI and the relative residual IOI. This allows one to track down the inconsistencies and find parameters that are most subject to these inconsistencies.

Finally, we apply the new measures and study the inconsistency between the geometry data sets and the growth data sets. For the geometry, we use the type Ia supernovae compilation in Ref.\,\cite{2014supernova740}, the baryon acoustic oscillation (BAO) from the Six Degree Field Galactic Survey (6dF) \cite{2011BAO-6df}, the main galaxy sample from the Sloan Digital Sky Survey (SDSS-DR7) and the SDSS quasar-Lyman-$\alpha$ forest \cite{2015bao-sdssIII}, and the high-$\ell$ CMB temperature data from Planck 2015 \cite{Planck2015XIII-Cos.Param.}. For the growth, we use the low-$\ell$ CMB temperature and polarization data form Planck 2015 \cite{Planck2015XV-lensing}, CMB lensing \cite{Planck2015XV-lensing}, thermal Sunyaev--Zel'dovich effect \cite{2015Planck-SZ-cluster-count}, cosmic shear from the Canada France Hawaii Lensing Survey (CFHTlens) \cite{2013CFHTlens}, and the redshift space distortion (RSD) from the WiggleZ Dark Energy Survey (WiggleZ MPK) \cite{2010WiggleZ-MPK,2012WiggleZ-MPK} and the SDSS DR12 CMASS and LOWZ catalogs \cite{2015sdss-dr12}. This application is motivated by the fact that modified gravity theory can lead to inconsistency on the dark energy properties when fitting $\Lambda$CDM or $w$CDM model separately into the geometry and the growth sets of experiments, and also in order to quantify the degree of inconsistency between these two data sets as reported in some previous works.

We organize our paper as follows: In Sec.\,\ref{section-experiments-inconsistency} we review some inconsistencies among different cosmological experiments. In Sec.\,\ref{section-IOI-definition-properties} we define IOI, demonstrate that it can properly describe the inconsistency between two experiments, discuss its properties, and show that it correctly reflects the effects of marginalization. In Sec.\,\ref{section-comparison} we view other measures of experimental inconsistency in the literature, and point out some limitations. Then in Sec.\,\ref{section-application} we apply IOI along with some other measures in Gaussian and weak prior limit to the geometry and the growth experiments based on the $\Lambda$CDM and the $w$CDM models. Finally we summarize our work in Sec.\,\ref{section-summary}. In this work, ``inconsistency'', ``tension'' and ``discordance'' share the same meaning, but the quantities IOI, the \textit{tension} ($\mathcal{T}$), and the \textit{discordance} are different measures of inconsistency.

\section[Experimental inconsistencies]{Brief survey of reported tensions or inconsistencies}\label{section-experiments-inconsistency}
In this section we list some inconsistencies or tensions between cosmological experiments as reported in the literature. Again, regardless if these are caused by different systematic effects in various data sets or caused by shortcomings in the cosmological model, they require careful examination.

  \textit{Hubble constant tension:} A precise locally measured Hubble constant by ladder distance observation is given as $H_0=73.24\pm1.74~\rm{km~s^{-1}/Mpc^{-1}}$ by Refs.\,\cite{2016Riess-etal-Hubble,2011Riess-etal-hubble}, which is higher than the one derived from the Planck 2015 data $H_0=66.93\pm0.62~\rm{km~s^{-1}/Mpc^{-1}}$ assuming the $\Lambda$CDM model \cite{Planck2016-intermediate-results-XLVI}. The tension is reported as at a 3.4-$\sigma$ confidence level in Ref.\,\cite{2011Riess-etal-hubble}. Reference \cite{2016-Bernal-Verde-Riess-H0} used their measure called the \textit{tension} and reported a high inconsistency between the local measurement of $H_0$ and the Planck result. They found that the odds for the two experiments to be consistent one with another are only $1:116$.

  In order to locally measure the Hubble constant, one needs to hierarchically calibrate the distances to different celestial objects, so papers have pointed out the systematic effects involved. For example, Ref.\,\cite{2014Efstathiou-Hubble} adopted a different outlier rejection criteria in the Cepheid samples and find that the high value of the locally measured Hubble constant in Ref.\,\cite{2011Riess-etal-hubble} could be due to a systematic error in the distance calibration. But recently authors in Ref.\,\cite{2016Riess-etal-Hubble}, using more Cepheid variables calibrated type Ia supernova, confirmed their earlier high value found in Ref.\,\cite{2011Riess-etal-hubble}. Also Ref.\,\cite{2017Casertano-Riess-Lattanzi-H0-Gaia} using Gaia Data Release 1 reported a similarly high value of the Hubble constant ($H_0=73.0~\rm{km~s^{-1}/Mpc^{-1}}$), different from the Planck measurement at $2.5$-$3.5$ $\sigma$ level. So the tension of $H_0$ between local measurements and the Planck 2015 result still remains. The authors of Ref\,\cite{2017-Bonvin-et-al-H0LiCOW} independently measured the Hubble constant based on time delay strong lensing, and reported $H_0=71.9^{+2.4}_{-3.0}~\rm{km~s^{-1}/Mpc^{-1}}$ in the $\Lambda$CDM model consistent with the local measurement.

  On the other hand, this tension might indicate a possible problem with the underlying model. For example, if the $\Lambda$CDM model is extended to the $w$CDM model, it was claimed in Ref.\,\cite{2016-Hubble-reconcile} that a piecewise function of $w$ can reconcile the derived Hubble constant from the Planck data with the one locally measured. Reference \cite{2016-Valentino-Melchiorri-Silk-reconciling-Planck-local} showed that varying 12 cosmological parameters can solve the current tension on $H_0$ between Planck and local measurement. But they also pointed out that tension remains if BAO and distances to supernova are included in the joint analysis.

\textit{Gravitational Lensing and Planck:}
Another persisting tension is between cosmic shear experiments and Planck for the determination of the amplitude of matter density fluctuations as, for example, parametrized by $\sigma_8$. CFHTlenS \cite{2013CFHTlens} found a lower amplitude than Planck 2015 \cite{Planck2015XIII-Cos.Param.}.  In the marginalized $\sigma_8$ vs $\Omega_m$ plane, the CFHTlenS confidence contours are shifted to the upper left compared to the Planck 2015 results. Most recently, the KIDS-450 survey finds a similar discrepancy at a 2.5-$\sigma$ level for $S_8=\sigma_8\sqrt{\Omega_m/0.3}$ \cite{2017-KiDS-Weak-lensing} and claims a substantial discordance in the full parameter space compared to Planck 2015. Finally, the authors of Ref.\,\cite{2016-leauthaud-etal-lensing-low} used galaxy-galaxy lensing measurements of the BOSS CMASS sample using 250 square degrees of weak lensing from CFHTLenS and CS82 and found also a lower value of $S_8$ than that of Planck 2015. They pointed out that this can be caused by systematics or new physics.
Indeed, a number of systematic effects for lensing require more work such as intrinsic alignment of galaxies, baryonic effects, and photometric redshifts \cite{2015-Troxel-Ishak-lensing,2015-KirK-etal-Galaxy-aligments,2015-Eifler-etal-baryonic-effects-WL,2015-Dossett-et-al-Planck-CFHTlenS,2016-Krause-etal-IA-impact}. On the other hand, it was pointed out in Refs.\,\cite{2015-Valentino-Melchorri-Silk-Beyon-LCDM,2016-Valentino-Melchiorri-Silk-reconciling-Planck-local} that the tension between CFHTlenS and Planck can be solved by varying $A_{\rm lens}$ (lensing anomaly parameter).

    \textit{BAO at z=2.34:} The baryon acoustic oscillation in the Lyman-$\alpha$ forest measurement gives the ratio of the Hubble radius to the drag epoch sound horizon $D_H/r_d=9.145\pm0.204$ at $z_{eff}=2.34$ \cite{2015bao-sdssIII}, which combines the Ly$\alpha$ forest autocorrelation \cite{2015bao-auto} and the quasar-Ly$\alpha$ cross-correlation \cite{2014bao-cross} methods. If we take $r_d=147.50$\,Mpc$^{-1}$ form Planck 2015 \cite{Planck2015XIII-Cos.Param.}, the inferred Hubble parameter at $z=2.34$ is $H(z=2.34)\approx 222\pm5~\rm{km~s^{-1}/Mpc}$\footnote{Since $D_H(z)=1/H(z)$.}. On the other hand, the Planck 2015 best fit gives $H_{Pl}(z=2.34)=236.6~\rm{km~s^{-1}/Mpc}$, which is higher than the one derived from BAO Lyman-$\alpha$ forest measurement. It is suggested that this discrepancy could be evidence for the interacting dark energy and dark matter model, see, for example \cite{Interacting.DE.DM2}.

    \textit{CMB Lensing anomaly:} The $\Lambda$CDM model predicts a certain strength of the CMB lensing potential. In Ref.\,\cite{2008-lensing-anomaly} a lensing anomaly parameter defined as $A_{\rm lens}\equiv C_\ell^{\psi,obs}/C_\ell^{\psi,\Lambda{\rm{CDM}}}$ was introduced (where $C_\ell^{\psi,obs}$ and $C_\ell^{\psi,\Lambda{\rm{CDM}}}$ are the observed and predicted lensing-potential power spectra) to test such an amplitude. If $\Lambda$CDM is a consistent model on the CMB temperature, polarization and lensing observation, this lensing anomaly parameter must be unity, $A_{\rm lens}=1$. Early the WMAP (along with ACBAR) data give $A_{\rm lens}=3.1^{+1.8}_{-1.5}$ at the $2\sigma$ confidence level \cite{2008-lensing-anomaly}, meaning the observed lensing potential is significantly higher than expected. However, this high value of $A_{\rm lens}$ did not persist in the Planck data, and dropped to $1.22\pm0.1$ at the $1\sigma$ confidence level \cite{Planck2015XIII-Cos.Param.}. So although there is still a small gap between $A_{\rm lens}$ and unity, it became more consistent with the prediction of the $\Lambda$CDM model. Nevertheless, it was shown that modified-gravity equations for the two scalar potentials (in the Newtonian gauge) can solve this small lensing anomaly \cite{Lensing.Anomaly2}. They claimed that this lensing anomaly gives a preference for modified gravity at a $95\%$ confidence level. The authors in Ref.\,\cite{2016-Munoz-etal-compensated-isocurvature} found that adding compensated isocurvature perturbations to the $\Lambda$CDM model can solve such a CMB lensing anomaly.

    \textit{Redshift Space Distortion (RSD) and Cluster Abundance vs Planck:} Compared to what is inferred from Planck 2015 best fit, RSD measurements generally give smaller growth of the large structure parametrized as $f\sigma_8$, where $f=d\ln\delta/d\ln a$ is the growth rate; see for examples Refs.\,\cite{2011WiggleZ-growth-rate,2016Bernal-etal-param-splitting,2014sdssIII-redshift-space}.
The two most precise galaxy cluster abundance measurements also seem to have significant tensions with the Planck 2015 results in the marginalized $\sigma_8$-$\Omega_m$ plane. One is the x-ray as a mass tracerfrom the Chandra cluster cosmology project \cite{2009-Chandra-clustering}, and the other is the thermal Sunyaev-Zel'dovich effect from Planck itself \cite{2015Planck-SZ-cluster-count}; also see Ref.\,\cite{2016Bernal-etal-param-splitting}.

    \textit{Geometry versus Growth:}
When combined, geometry probes and growth probes provide powerful combinations to constrain cosmology. Moreover, when one is contrasted with another they constitute a useful consistency test of the underlying theory \cite{2006-Ishak-splitting,2006-Koyama-growth-in-MG,2006-Bertshinger-growth-test} or a mean to detect different systematics in the data sets. The authors of Refs.\,\cite{2007-Wang-etal-consistency,2015Ruiz-etal-param-splitting,2016Bernal-etal-param-splitting} used a technique called parameter splitting, in which they separate parameters constraining the dark energy properties (e.g., $\Omega_\Lambda$ and $w$) into a geometry set (e.g., $\Omega_\Lambda^{geom}$ and $w^{geom}$) and a growth set (e.g., $\Omega_\Lambda^{grow}$ and $w^{grow}$). Authors of Ref.\,\cite{2015Ruiz-etal-param-splitting} found that RSD data generally favor a higher $w^{grow}$, and Ref.\,\cite{2016Bernal-etal-param-splitting} found that a subset of cluster abundance data (from Refs.\,\cite{2009-Chandra-clustering,2015Planck-SZ-cluster-count}) mostly cause the deviations of dark-energy metaparameters.

\section[Definition of IOI and its properties]{Index of inconsistency: motivation and definition}\label{section-IOI-definition-properties}
Different works on measures of inconsistency in the literature adapt different notations. For a consistent discussion throughout the paper, we use the following notation so Bayes's theorem reads
\begin{equation}\label{eq-bayes-theorem}
\mathscr{P}(\lambdaB;{\rm Q})=\frac{\mathcal{L}({\rm Q};\lambdaB)\mathcal{P}(\lambdaB)}{E(\rm Q)}\,,
\end{equation}
where $\mathscr{P}$ is the posterior probability distribution, $\mathcal{L}$ the likelihood, $\mathcal{P}$  the prior and $E$ the evidence. Most of the experiments discussed here are assumed to give Gaussian likelihoods $\mathcal{L}^{(i)}(\bm{Q};\lambdaB)$ on the parameters. For Gaussian likelihoods,
\begin{equation}\label{eq-likelihood-of-ith-exp-Gauss}
\mathcal{L}^{(i)}=\mathcal{L}^{(i)}_{max}\exp{\big(-\tfrac{1}{2}(\lambdaB-\bm{\mu^{(i)}})^T\bm{L^{(i)}}(\lambdaB-\bm{\mu^{(i)}})\big)}\,,
\end{equation}
where $\lambdaB$ is the parameter vector, $\bm{\mu^{(i)}}$ the mean ($i=1,2$ for two experiments), and $\bm{L^{(i)}}$ the Fisher matrix. For mildly non-Gaussian distributions, the above equation is treated as an approximation to the real distribution, with $\bm{\mu^{(i)}}$ and $\bm{L^{(i)}}$ given by the mean and inverse of the covariance matrix of the real distribution. The mean $\bm{\mu^{(i)}}$ is a function of the data vector $\bm{Q}$. Experiments are assumed to have the same Gaussian prior distribution
\begin{equation}
\mathcal{P}=\tfrac{\sqrt{|\bm{P}|}}{(2\pi)^{n/2}}\exp\big(-\tfrac{1}{2}(\lambdaB-\bm{\mu^{(p)}})^T\bm{P}(\lambdaB-\bm{\mu^{(p)}})\big)\,,
\end{equation}
where $|\bm{P}|\equiv\det(\bm{P})$ and $\bm{P}$ is the inverse of the covariance matrix of the prior. A weak prior limit is taken as $\bm{P}\rightarrow\bm{0}$. The Gaussian and weak prior limit refers to a situation where likelihoods are Gaussian on the parameters and the prior is weak. Posteriors are normalized, and are denoted as $\mathscr{P}^{(i)}$. The inverse of the covariance matrix of a posterior is denoted as $\bm{F^{(i)}}$, and the mean as $\bm{\bar{\mu}^{(i)}}$. Some common notations in this work are summarized in Tables\,\ref{table-notations} and \ref{table-other-notation}.

\begin{table*}[pt]
\caption[Notations for distributions, means and Fisher matrices]{\label{table-notations}Table of notations: Probability distributions, their means and elements of means, Fisher matrices and elements of the Fisher matrices for the likelihood of the $i$th experiment, prior, and the posterior of the $i$th experiment. Likelihoods are \emph{not} normalized in the parameter space, while the prior and posteriors are.}
\begin{ruledtabular}
\begin{tabular}{lccccc}
Distributions & Notations & \multicolumn{1}{b{0.15\textwidth}}{Inverse of\newline covariance matrix } & \multicolumn{1}{b{0.13\textwidth}}{Elements of\newline Fisher matrices} & Means & \multicolumn{1}{b{0.08\textwidth}}{Elements\newline of means} \\
\hline
$i$th Likelihood & $\mathcal{L}^{(i)}$ & $\bm{L^{(i)}}$ & $L^{(i)}_{~jk}$ & $\bm{\mu^{(i)}}$ & $\mu^{(i)}_{~j}$\\
Prior & $\mathcal{P}$ & $\bm{P}$ & $P_{jk}$ & $\bm{\mu^{(p)}}$ & $\mu^{(p)}_{~j}$ \\
$i$th Posterior & $\mathscr{P}^{(i)}$ & $\bm{F^{(i)}}$ & $F^{(i)}_{~jk}$ & $\bm{\bar{\mu}^{(i)}}$ & $\bar{\mu}^{(i)}_{~j}$\\
\end{tabular}
\end{ruledtabular}
\caption{\label{table-other-notation}Other frequently used notations in this work.}
\begin{ruledtabular}
\begin{tabular}{ccccc}
Parameter vector& Observable vector & Mean difference & Covariance matrix & $G$ matrix\\
$\lambdaB$  & $\bm{Q}$ & $\deltaB$ & $\bm{C}$ & $\bm{G}=(\bm{C^{(1)}}+\bm{C^{(2)}})^{-1}$ \\
\end{tabular}
\end{ruledtabular}
\end{table*}

\subsection[Motivation of the definition of IOI]{Motivation for IOI}\label{subsection-motivation-IOI}
\begin{figure}[!tbp]
\includegraphics[width=\linewidth]{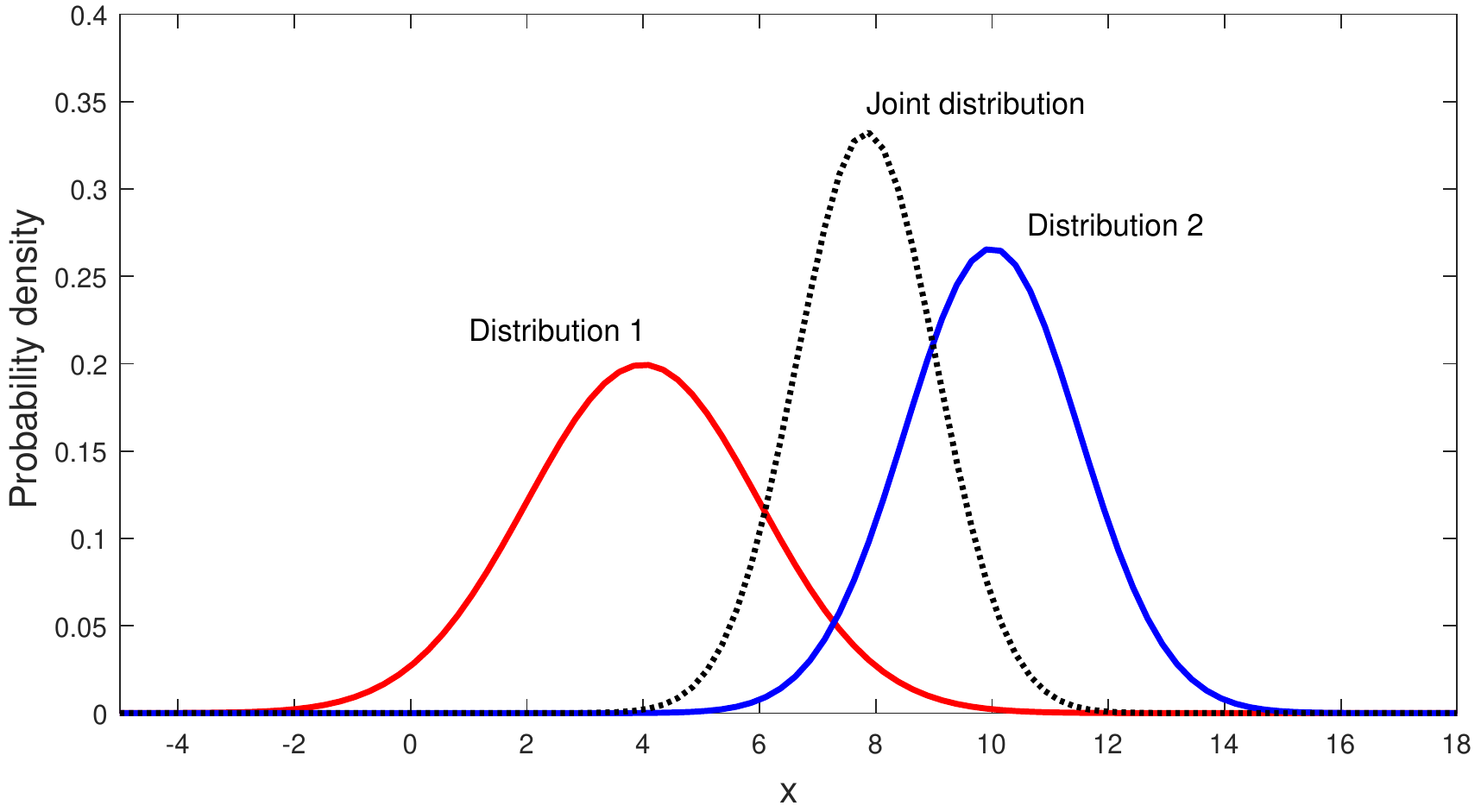}
\caption[Motivation of the definition of IOI]{\label{fig-motivation-IOI}Two toy one-dimensional Gaussian probability density distributions (red and blue) and their joint distribution (black dotted).}
\end{figure}

In the Bayesian parameter estimation, experiments give probability density distributions of the parameters in a given model. Different experiments usually give different distributions. The joint distribution is the one obtained by simultaneously analyzing the data from two or more experiments. Figure\,\ref{fig-motivation-IOI} shows two toy 1D Gaussian distributions (red and blue) given by two different experiments, and their joint distribution (black dotted) on parameter $x$.

We can see in Fig.\,\ref{fig-motivation-IOI} that the red distribution favors the value of $x$ at $4$, while the blue one favors at $10$, so there is an inconsistency between the two experiments. One might want to use the difference of the two means to quantify the inconsistency, i.e., $\delta=x_{blue}-x_{red}=10-4=6$. But this method cannot be right, because the mean difference is not invariant under a parameter scaling. If we let $x\rightarrow x'=2x$, the new means become $20$ and $8$, and the difference becomes $12$. The inconsistency between two experiments for a model should not depend on the scaling.

One then realizes that the mean difference needs to be normalized by the uncertainties of the distributions, since the uncertainties will be scaled inversely as the mean difference. And this normalization makes perfect sense: if the uncertainties are very large compared to the mean difference, the inconsistency will actually be small. But which distribution's uncertainty should be used to normalize the mean difference? Intuition tells us that we should not use just one but somehow both of them. One might want to use the uncertainty of the joint distribution. For example, the figure of bias (Fob) defined in Ref.\,\cite{2009-C-Shapiro-Fob} reduces to this type of form in one dimension. But this choice has problems for the reasons that follow. Suppose the uncertainties of the red and the blue distributions are $\sigma_{(1)}$ and $\sigma_{(2)}$; the uncertainty of the joint distribution is given by $\sigma^{-2}=\sigma_{(1)}^{-2}+\sigma_{(2)}^{-2}$. Normalizing $\delta$ by $\sigma$ will give a quantity such as $\frac{\delta^2}{\sigma^2}=\delta^2(\tfrac{1}{\sigma_{(1)}^2}+\tfrac{1}{\sigma_{(2)}^2})$. If one distribution has a much smaller uncertainty than the other, say $\sigma_{(2)}\ll\sigma_{(1)}$, that quantity can be approximated as $\frac{\delta^2}{\sigma^2}\simeq\frac{\delta^2}{\sigma_{(2)}^2}$. So in this situation the mean difference is normalized only by $\sigma_2$, and it diverges as $\sigma_{(2)}\rightarrow0$. This is very counterintuitive. If one uncertainty is very small but the other is very large, the inconsistency should be small. That is because the distribution with a large uncertainty can extend to the mean of the other distribution that has a small uncertainty. So if $\sigma_{(2)}\ll\sigma_{(1)}$, the mean difference should be normalized by $\sigma_{(1)}$ instead of $\sigma_{(2)}$. Therefore, instead of $\delta^2(\tfrac{1}{\sigma_{(1)}^2}+\tfrac{1}{\sigma_{(2)}^2})$, it is a quantity such as $\tfrac{\delta^2}{\sigma_{(1)}^2+\sigma_{(2)}^2}$ that should be used.

Here, we approach the inconsistency in a slightly different way from above. Indeed, the above discussion did not consider the mean of the joint distribution. Instead of using the difference between the means of the two distributions, we use two differences. One difference $\delta^{(1)}$ is between the first distribution mean and the joint distribution mean, and the other difference $\delta^{(2)}$ is between the second distribution mean and the joint distribution mean. Then we normalize the first difference by the first uncertainty, and the second difference by the second uncertainty. Finally we take the average of the normalized mean differences (squared) as the measure of inconsistency. More precisely, we use the following quantity to measure the inconsistency,
\begin{equation}\label{eq-normalized-differences}
\rm{Inconsistency}=\frac{1}{2}\left(\frac{(\delta^{(1)})^2}{\sigma_{(1)}^2}+\frac{(\delta^{(2)})^2}{\sigma_{(2)}^2}\right)\,.
\end{equation}
The meaning of Eq.\,\eqref{eq-normalized-differences} is as follows: it is the average of two terms, each of which measures the ``difficulty'' for the corresponding distribution to ``support'' or ``favor'' the mean of the joint distribution. The mean of the joint distribution will be closer to the mean of the distribution with a smaller uncertainty. For a metaphor, if two people are separated, one is at location A and the other is at location B. We can use the minimum time taken by them to meet to quantify the degree of how they are separated. Surely both of them need to move, and the one who travels faster or moves easier needs to make a longer distance.

We will show in Sec.\,\ref{subsection-IOI-definition} that the above quantity actually turns out to be
\begin{equation}\label{eq-normalized-diff-turns-out}
  \rm{Inconsistency}=\frac{1}{2}\frac{\delta^2}{\sigma_{(1)}^2+\sigma_{(2)}^2}\,,
\end{equation}
which has the same form as the one we logically obtained in the previous paragraph.

So far we have only considered 1D distributions, but let us ``extrapolate'' the result to multidimensional distributions. For multidimensional Gaussian distributions, the uncertainty (squared) is specified by the covariance matrix $\bm{C}$. So, it is reasonable to guess the general form of Eq.\,\eqref{eq-normalized-diff-turns-out} to be $\tfrac{1}{2}\deltaB^T(\bm{C^{(1)}}+\bm{C^{(2)}})^{-1}\deltaB$, and it turns out to be a proper guess as we demonstrate in Sec.\,\ref{subsection-IOI-definition}.

\subsection[Definition of IOI]{Definition of IOI}\label{subsection-IOI-definition}
We first define the term $\tfrac{1}{2}\Delta\chi^2(\muB)$ as
\begin{equation}\label{eq-Dchi-square-definition}
\tfrac{1}{2}\Delta\chi^2(\muB)\equiv\tfrac{1}{2}\big(\Delta\chi^2_{(1)}(\muB)+\Delta\chi^2_{(2)}(\muB)\big)\,,
\end{equation}
with $\Delta \chi^2_{(i)}(\muB)=\chi^2_{(i)}(\muB)-\chi^2_{(i)}(\bm{\mu^{(i)}})$. We can see that $\tfrac{1}{2}\Delta\chi^2(\muB)$ between two experiments is defined as an average of two terms, namely,
\begin{description}
  \item[~~$\Delta\chi^2_{(1)}(\muB)$]  The difficulty for the first experiment to support the mean of the joint analysis.
  \item[~~$\Delta\chi^2_{(2)}(\muB)$]  The difficulty for the second experiment to support the mean of the joint analysis.
\end{description}
So $\tfrac{1}{2}\Delta\chi^2(\muB)$ is the averaged difficulty for the two experiments to support the joint mean. This definition of $\tfrac{1}{2}\Delta\chi^2$ is equivalent to $\tfrac{1}{2}\Delta\chi^2\equiv\tfrac{1}{2}\chi^2_{min}-(\tfrac{1}{2}\chi^2_{(1),min}+\tfrac{1}{2}\chi^2_{(2),min})$, where $\chi^2=\chi^2_{(1)}+\chi^2_{(2)}$ and ``min'' means minimum of $\chi^2$ (and maximum of likelihood).

\begin{figure*}[!htbp]
\centering
\includegraphics[width=0.32\textwidth]{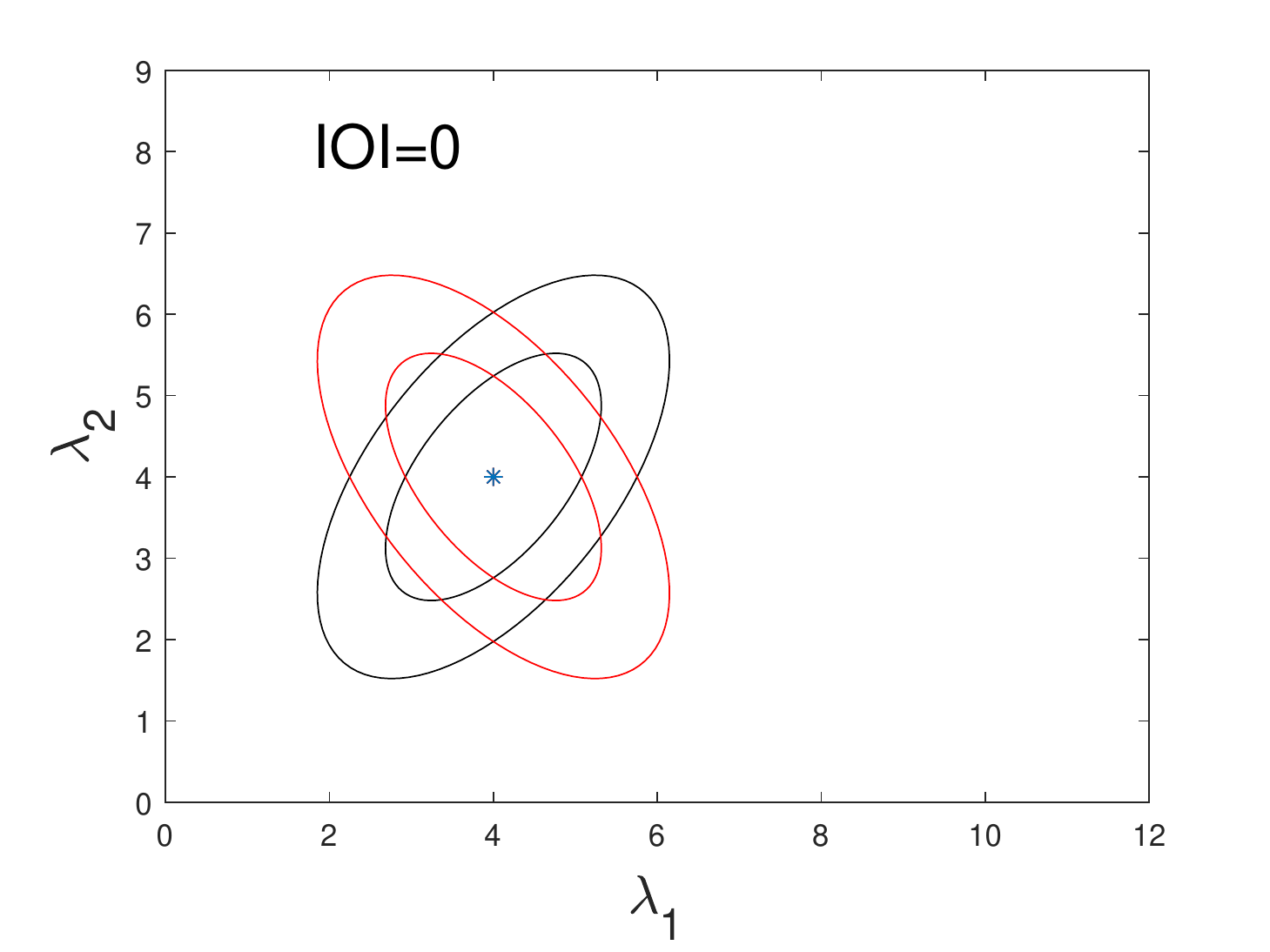}
\includegraphics[width=0.32\textwidth]{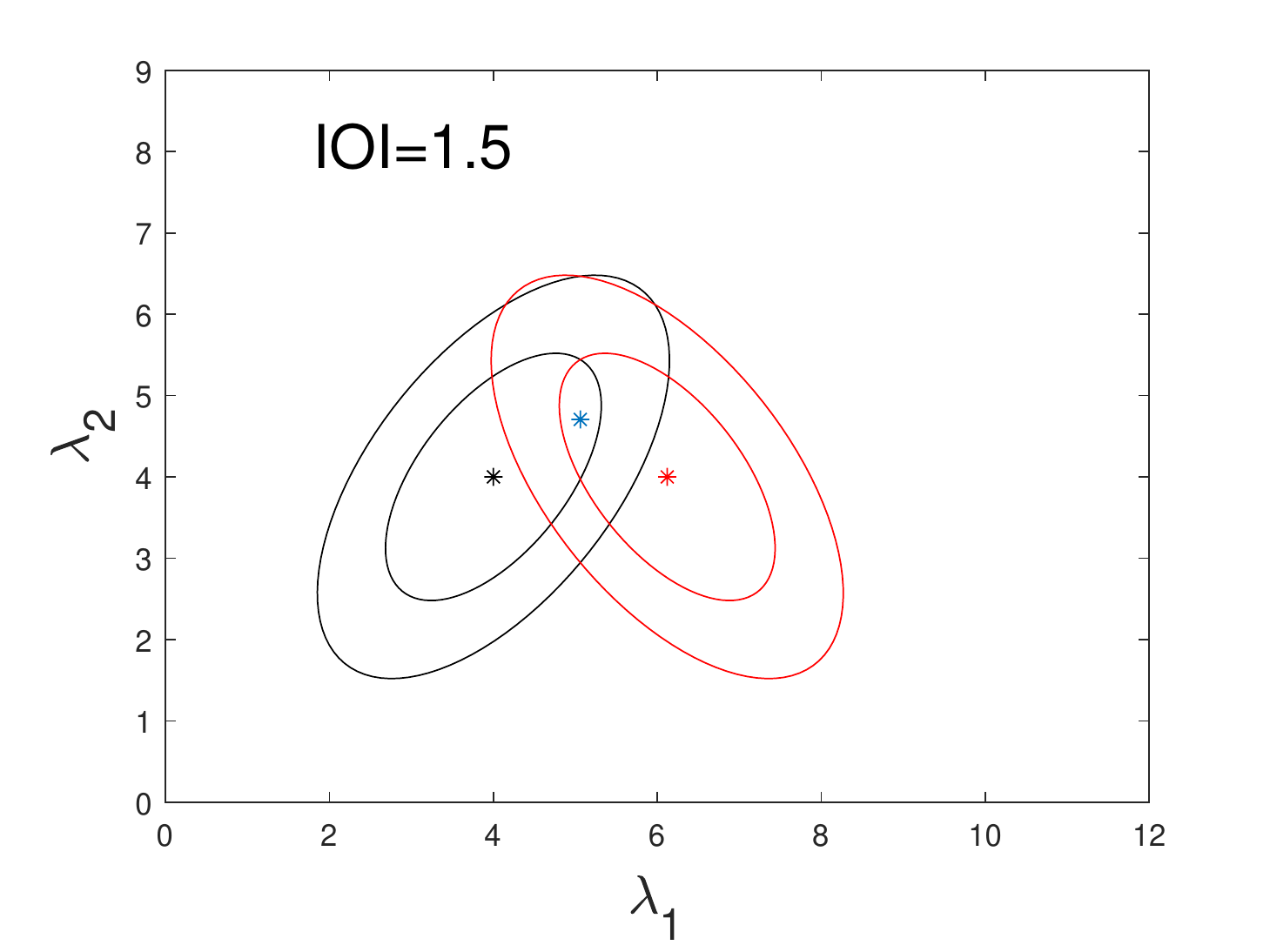}
\includegraphics[width=0.32\textwidth]{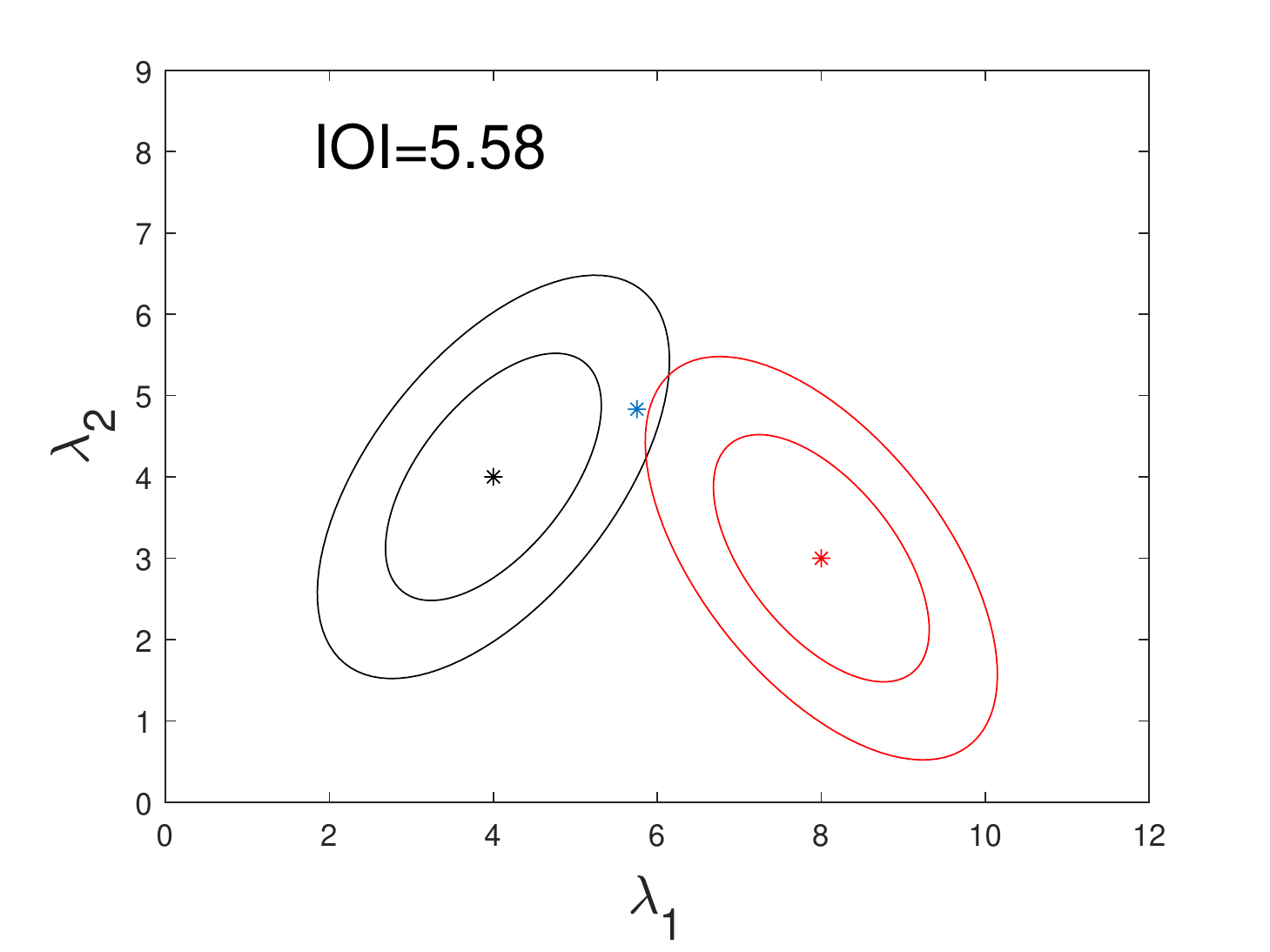}
\includegraphics[width=0.32\textwidth]{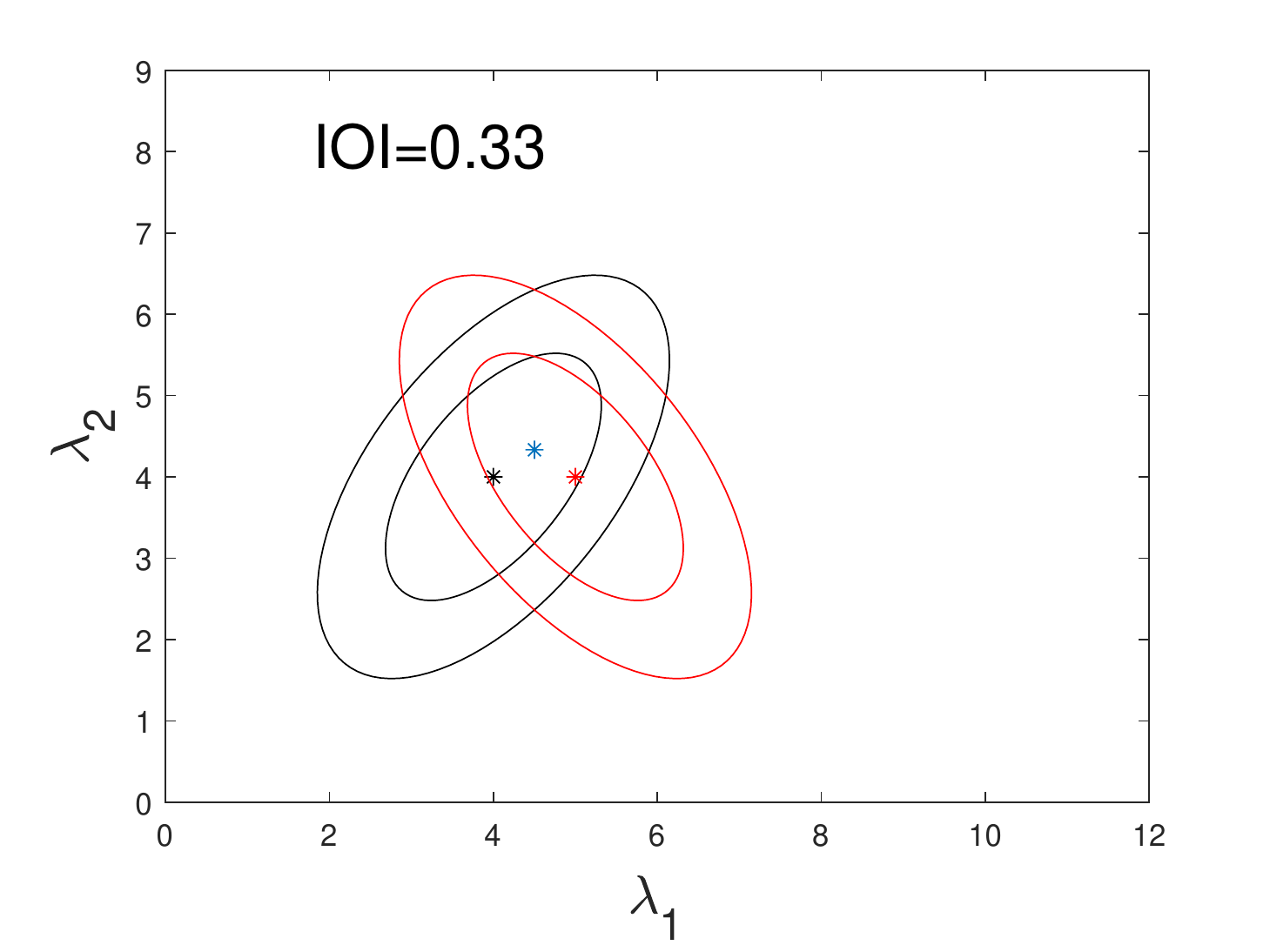}
\includegraphics[width=0.32\textwidth]{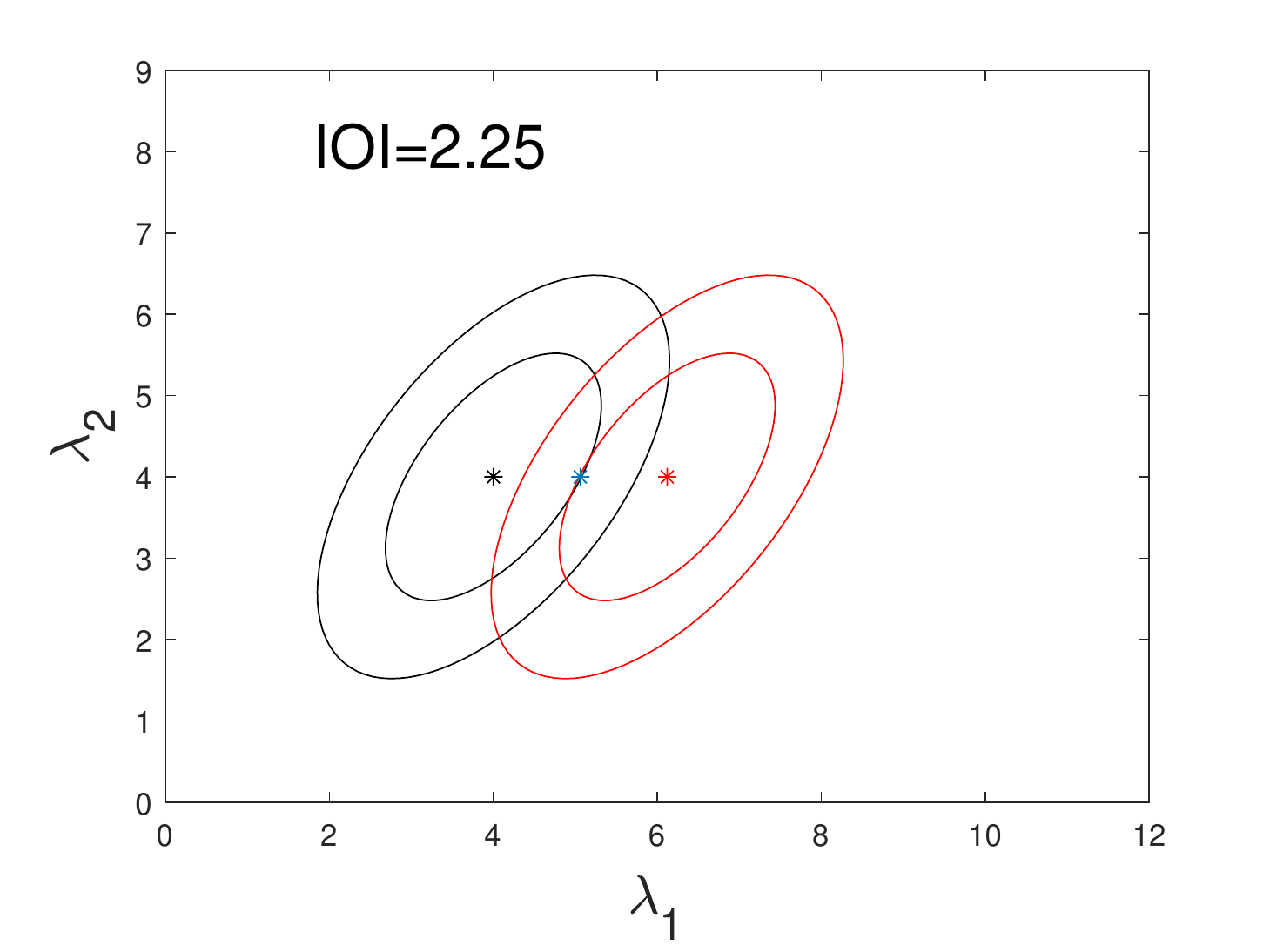}
\includegraphics[width=0.32\textwidth]{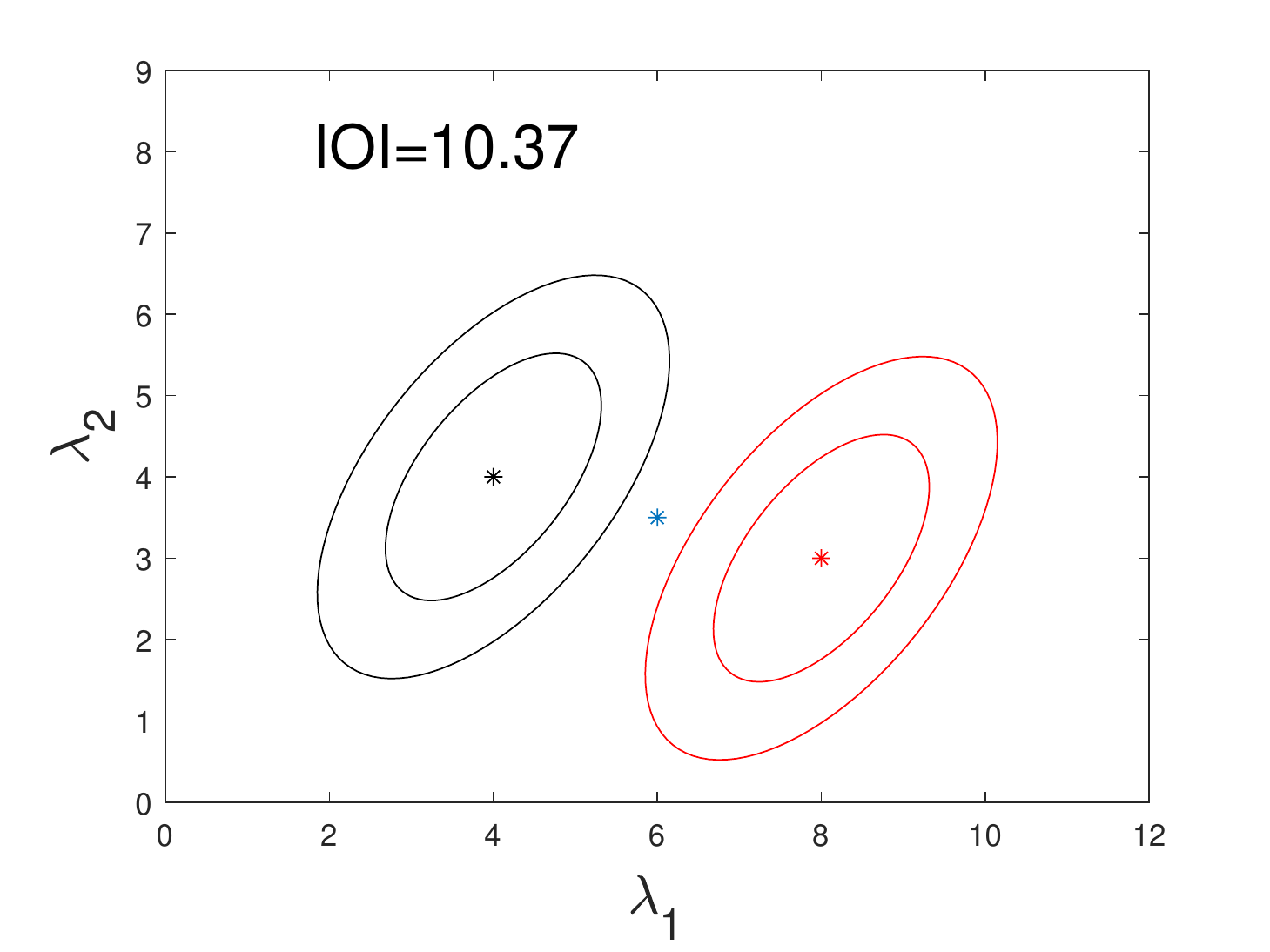}
\caption[Examples of IOI]{\label{fig-examples-of-IOI}Examples of IOIs. The centers of the ellipses are the likelihood maxima. The extra (blue) point is the likelihood maximum of the joint likelihood. The two in the middle have the same deviation of likelihood maxima, but they do not have the same IOI because of the different degenerate directions. Same is also true for the two on the right. }
\end{figure*}

As shown in Appendices \ref{appendix-sub-combine-likes} and \ref{appendix-sub-combine-two-likes}, in the Gaussian limit where $\Delta\chi^2_{(i)}=(\lambdaB-\bm{\mu^{(i)}})^T\bm{L^{(i)}}(\lambdaB-\bm{\mu^{(i)}})$, the mean of the joint likelihood is given by a Fisher-matrix-weighted average,
\begin{equation}\label{eq-Fisher-weighted-average-two}
\muB=\bm{L}^{-1}(\bm{L^{(1)}}\bm{\mu^{(1)}}+\bm{L^{(2)}}\bm{\mu^{(2)}})\,,
\end{equation}
where $\bm{L}=\bm{L^{(1)}}+\bm{L^{(2)}}$. And then $\tfrac{1}{2}\Delta\chi^2$, in this limit, can be computed explicitly as
\begin{equation}\label{eq-DeltaChi2-Gaussian}
\begin{split}
&\tfrac{1}{2}\Delta\chi^2\xrightarrow{\rm Gaussian}\\
&~~~~~\tfrac{1}{2}(\bm{\mu^{(2)}}-\bm{\mu^{(1)}})^T(\bm{C^{(1)}}+\bm{C^{(2)}})^{-1}(\bm{\mu^{(2)}}-\bm{\mu^{(1)}}). \\
\end{split}
\end{equation}
Thus, we define IOI for two experiments to be the Gaussian limit of $\tfrac{1}{2}\Delta\chi^2$ [Eq.\,\eqref{eq-Dchi-square-definition}] and given by
\begin{equation}\label{eq-IOI-definition}
{\rm IOI}\equiv\tfrac{1}{2}\deltaB^T\bm{G}\deltaB\,,
\end{equation}
where $\deltaB\equiv \bm{\mu^{(2)}}-\bm{\mu^{(1)}}$, and $\bm{G}\equiv\big[(\bm{L^{(1)}})^{-1}+(\bm{L^{(2)}})^{-1}\big]^{-1}=(\bm{C^{(1)}}+\bm{C^{(2)}})^{-1}$.
This result is the same as our guided guess in Sec.\,\ref{subsection-motivation-IOI}. In the cases of Gaussian likelihoods, IOI is the same as $\tfrac{1}{2}\Delta\chi^2$. But in general cases, we define IOI as the moment-based quadratic quantity $\tfrac{1}{2}\deltaB^T\bm{G}\deltaB$, with $\bm{L^{(i)}}$ obtained by the inverse of the covariance matrices $(\bm{C^{(i)}})^{-1}$ and $\bm{\mu^{(i)}}$ given by the means. \emph{So the definition of IOI is only motivated by $\tfrac{1}{2}\Delta\chi^2$, but not $\tfrac{1}{2}\Delta\chi^2$ itself.} IOI is a moment-based quadratic quantity.

\begin{table*}[t]
        \caption[Jeffreys' scales for IOI]{\label{table-Jeffrey-scale-IOI} Jeffreys' scales as interpretation of the values of IOI. Jeffreys' scales are empirical scales that originally classify the ranges of the Bayesian evidence ratio. But comparing the numerical values of IOI and the visual separations of likelihood contours in Fig.\,\ref{fig-examples-of-IOI}, we find that Jeffreys' scales are appropriate for the classification and interpretation of IOI. Higher IOI represents higher inconsistency. Since IOI is positive definite, the original interpretation of the negative values of the Bayesian evidence ratio does not apply to IOI (see Table \ref{Jeffrey-scale} for such a difference). }
        \begin{ruledtabular}
        \begin{tabular}{lp{0.15\textwidth}p{0.15\textwidth}p{0.15\textwidth}p{0.15\textwidth}}
        Ranges &IOI$<1$ & $1<$IOI$<2.5$ & $2.5<$IOI$<5$ & IOI$>5$ \\ \hline
Interpretation &No significant\newline inconsistency&Weak\newline inconsistency &Moderate inconsistency&Strong\newline inconsistency
        \end{tabular}
        \end{ruledtabular}
        \end{table*}

If we have priors assigned to the two experiments, we can extend the definition of IOI to include priors in the means and covariance matrices as follows:
\begin{equation}\label{eq-IOI-with-prior}
\begin{split}
\bm{\mu^{(i)}}&\xrightarrow{\rm with~prior}\bm{\bar{\mu}^{(i)}}=\frac{1}{\bm{F^{(i)}}}\big(\bm{L^{(i)}}\bm{\mu^{(i)}}+\bm{P}\bm{\mu^{(p)}}\big)\,,\\
\bm{\delta}&\xrightarrow{\rm with~prior}\bm{\bar{\delta}}=\bm{\bar{\mu}^{(2)}}-\bm{\bar{\mu}^{(1)}}\,,\\
\bm{G}&\xrightarrow{\rm with~prior}\bm{\bar{G}}=\big[(\bm{F^{(1)}})^{-1}+(\bm{F^{(2)}})^{-1}\big]^{-1}\,,\\
{\rm IOI}&\xrightarrow{\rm with~prior} \tfrac{1}{2}\bm{\bar{\delta}}^T\bm{\bar{G}}\bm{\bar{\delta}}\,,
\end{split}
\end{equation}
where $\bm{F^{(i)}}=\bm{L^{(i)}}+\bm{P}$ for Gaussian distributions and is an approximation in mildly non-Gaussian cases. Thus Eq.\,\eqref{eq-IOI-definition} is the weak prior limit ($\bm P\rightarrow 0$) of Eq.\,\eqref{eq-IOI-with-prior}. However, we suggest to use Eq.\,\eqref{eq-IOI-definition} to compare two experiments for the following reasons. First, we usually have only weak priors on parameters. Second, the priors might already be biased or in tension with the experiments considered. Two experiments should be compared with very weak priors. Including strong priors may affect the comparison of two experiments. But, of course, we can use Eq.\,\eqref{eq-IOI-definition} to compare and see if the prior is consistent with the experiments to be compared, by simply replacing $\bm{L^{(2)}}$ and $\bm{\mu^{(2)}}$ with $\bm P$ and $\bm{\mu^{(p)}}$ in Eq.\,\eqref{eq-IOI-definition}.

For illustration, we show in Fig.\,\ref{fig-examples-of-IOI} some examples of toy likelihood contours with their corresponding values of IOI. Those examples show that IOI can properly represent inconsistency between two experiments; i.e., it increases whenever the experimental inconsistency increases graphically. We use Jeffreys' scale (shown in Table \ref{table-Jeffrey-scale-IOI}; see Ref.\,\cite{jeffreys1998theory}) as an interpretation of the values of IOI. Note that Jeffreys' scales were originally empirical scales for the Bayesian evidence ratio. Here IOI is not an evidence ratio but we use this empirical scale as it seems to give sensible meanings for the inconsistencies compared to what is shown in Fig.\,\ref{fig-examples-of-IOI}. Moreover, reference \cite{2013-tension-Verde-etal} proposed a quantity called \textit{tension} to measure experimental inconsistency. They used Jeffreys' scale as the interpretation of the \textit{tension}. We will show further in the paper that the \textit{tension} reduces to IOI in the Gaussian and weak prior limit. So we expect that IOI and the \textit{tension} are similar quantities for nearly Gaussian distributions, and using the same scale is good for comparison. However, we note again that one should not translate IOI into a probability ratio, although we propose Jeffreys' scales as the interpretation of IOI.

The next logical step is to generalize Eq.\,\eqref{eq-IOI-definition} to define IOI for comparison of $N$ experiments as
\begin{equation}\label{eq-DeltaChi2-general-Gaussian}
\frac{1}{N}\sum\limits_i\Delta\chi^2_{(i)}(\mu)
\xrightarrow{\rm Gaussian}\frac{1}{N}\Big(\sum\limits_i\bm{\mu^{(i)}}\,^T\bm{L^{(i)}}\bm{\mu^{(i)}}-\muB^T \bm{L}\muB\Big)\,,
\end{equation}
\begin{equation}\label{eq-IOI-definition-general}
{\rm IOI}\equiv\frac{1}{N}\Big(\sum\limits_{i=1}^{N}\bm{\mu^{(i)}}\,^T\bm{L^{(i)}}\bm{\mu^{(i)}}-\muB^T \bm{L}\muB\Big)\,,
\end{equation}
where $\muB$ is the mean of the joint likelihood combining all experiments and it is also given by the Fisher-matrix-weighted average,
\begin{equation}\label{eq-peak-fisher-weighted}
\muB=\bm{L}^{-1}\Big(\sum\limits_i \bm{L^{(i)}}\bm{\mu^{(i)}}\Big)\,,
\end{equation}
where $\bm{L}=\sum \bm{L^{(i)}}$. Each term in the sum in the left-hand side of Eq.\,\eqref{eq-DeltaChi2-general-Gaussian} has the same interpretation as that in Eq.\,\eqref{eq-Dchi-square-definition}, which is measuring the difficulty for the $i$th experiment to support the mean of the joint analysis. So $\frac{1}{N}\sum\Delta\chi^2_{(i)}(\mu)$ represents the averaged difficulty for the $N$ experiments to support their joint mean. We have again taken the Gaussian limit of Eq.\,\eqref{eq-DeltaChi2-general-Gaussian} and defined IOI for $N$ experimental as the moment-based quantity Eq.\,\eqref{eq-IOI-definition-general}. But since Eq.\,\eqref{eq-IOI-definition-general} contains more than two means, it cannot be reduced into a quadratic form as in the case of two experiments. For the case of two experiments, the two different forms Eq.\,\eqref{eq-IOI-definition} and Eq.\,\eqref{eq-IOI-definition-general} are equivalent; see Appendix \ref{appendix-sub-combine-two-likes}.

Before closing this subsection, we give some remarks. Our IOI and other measures of inconsistency in the literature are a model-dependent description of inconsistency between experiments. For example, in Sec.\,\ref{section-application} we will see that IOI between the geometry experiments growth experiments for the $\Lambda$CDM model is different from that for the $w$CDM model. Reference \cite{2013-tension-Verde-etal} suggested that a better model should provide more consistent explanations on different experiments. With IOI, it means a better model should give a smaller IOI on two or more experiments.

\subsection{Inconsistency measures and Gaussianity}\label{subsection-IOI-non-Gaussian}
Current cosmological experiments usually provide approximately Gaussian probability distributions on parameters. In those cases, it is usual to approximate the actual probability distribution by a Gaussian one based on its moments. The mean and covariance matrix of the approximate Gaussian distribution are taken to be those of the actual distribution. If a measure does not correctly describe inconsistency in Gaussian cases, one would not expect it to correctly describe inconsistency in general. Therefore constructing a correct measure of inconsistency for Gaussian and nearly Gaussian cases is an important step to the measure for general cases, and this is what we aim to accomplish in this work. We argue that IOI works correctly in Gaussian cases, and we expect it to give meaningful information in mildly non-Gaussian cases. So IOI has an important role: one can compare a measure with IOI to see whether that measure properly describes experimental inconsistency in Gaussian cases.

\begin{figure}[h]
\includegraphics[width=\linewidth]{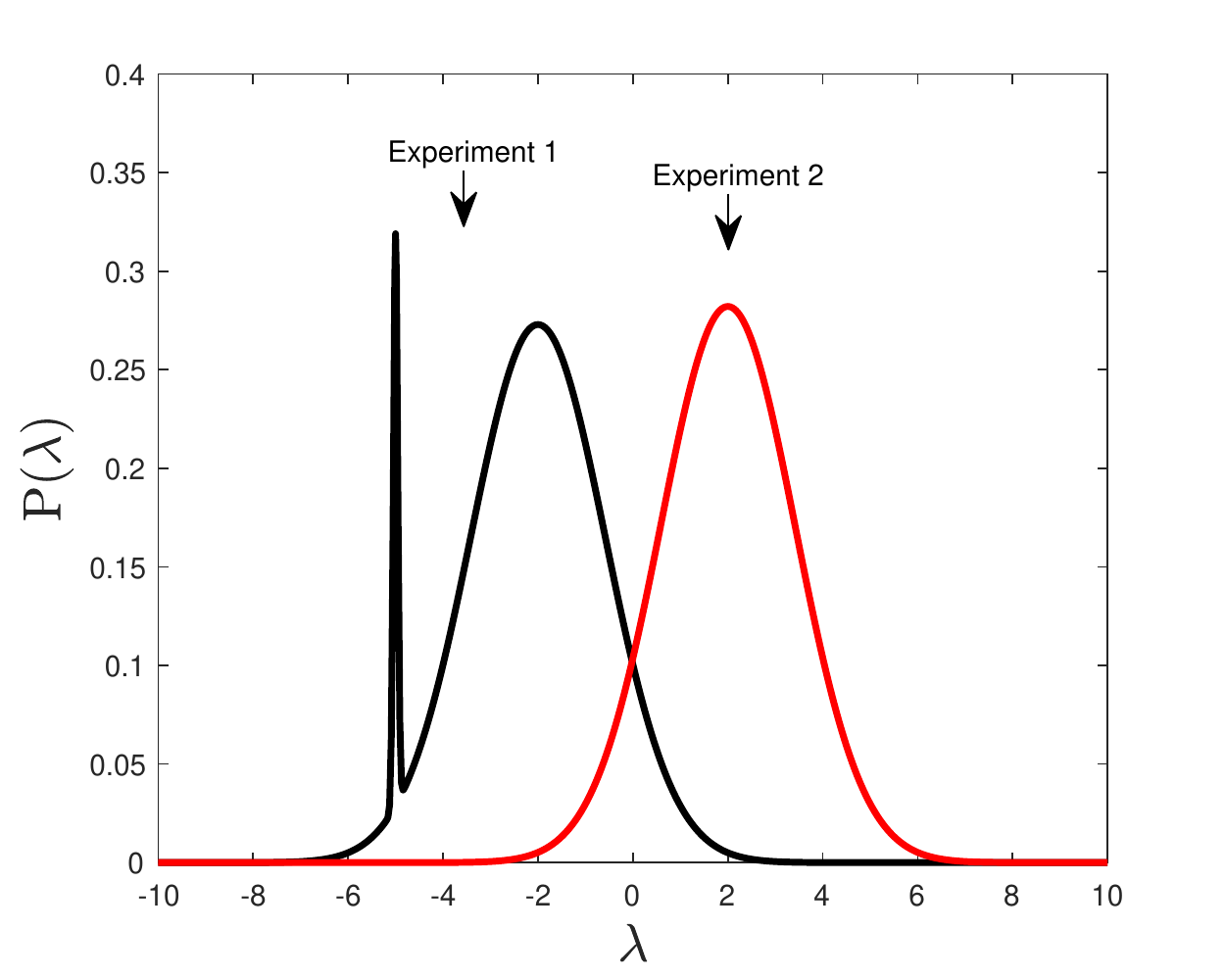}
\caption{\label{fig-non-Gaussian1}{An example of non-Gaussian case. $P_{(2)}(\lambda)$} given by experiment 2 is Gaussian, while $P_{(1)}(\lambda)$ given by experiment 1 is slightly different from Gaussian. The narrow peak of $P_{(1)}(\lambda)$ should be ignored when we compare the two experiments. Using $\Delta\chi^2$ as a measure of inconsistency only considers the maximum that may have insignificant probability weight, but ignores the overall distribution. The moment-based IOI cares about the distribution as a whole.}
\end{figure}

However, we do not expect IOI to describe inconsistency properly in highly non-Gaussian cases. In fact, it is even difficult to define inconsistency when distributions are non-Gaussian. There might be other criteria to consider for general cases, which need to be further explored in future works. We note that even if a measure reduces to IOI in the Gaussian limit, it does not necessarily mean such a measure can describe inconsistency properly. For example, in the Gaussian limit $\Delta\chi^2$ is the same as IOI, but it can give us misleading results in some non-Gaussian cases as we will explain. $\Delta\chi^2$ focuses on its minimum value but ignores the overall distribution. For example, in Fig.\,\ref{fig-non-Gaussian1} the 1D probability distribution $P_{(2)}(\lambda)$ given by experiment 2 is Gaussian, while $P_{(1)}(\lambda)$ given by experiment 1 is not. In fact, $P_{(1)}(\lambda)$ is only slightly different from a Gaussian distribution: there is a narrow local peak at $\lambda=-5$ which is also the global maximum of $P_{(1)}(\lambda)$; the rest of $P_{(1)}(\lambda)$ is overall Gaussian. When we compare the two experiments, intuition tells us we can ignore that narrow peak. That is because the integrated probability around that peak is small compared to the rest, which gives it a negligible probability weight. So even for such a mildly non-Gaussian distribution, $\Delta\chi^2$ does not work. One would expect the situation shown in Fig.\,\ref{fig-non-Gaussian1} can be approximated by two Gaussian distributions. That is exactly what the moment-based IOI does. Compared to the case without the narrow peak, $P_{(1)}(\lambda)$ with that peak only slightly changes the parameters' means and covariance and the value of IOI.

It is worth noting that IOI is invariant under a linear parameter transformation, $\lambdaB_{new}=\bm{M} \lambdaB_{old}$ with a transformation matrix $\bm{M}$. Under such a transformation, the mean difference becomes $\deltaB\rightarrow \bm{M}\deltaB$ and the Fisher matrices become $\bm{L^{(i)}}\rightarrow (\bm{M}^{-1})^T\bm{L^{(i)}}\bm{M}^{-1}$. Then IOI transforms as
\begin{equation}\label{eq-IOI-reparam-invariance}
\begin{split}
  \rm{IOI}_{old}
   \rightarrow& \rm{IOI}_{new}\\
   \rm{IOI}_{new}=&\tfrac{1}{2}\deltaB^T \bm{M}^T\Big[\big((\bm{M}^{-1})^{T}\bm{L^{(1)}}\bm{M}^{-1}\big)^{-1}\\
   &~~~~~~~~~+\big((\bm{M}^{-1})^{T}\bm{L^{(2)}}\bm{M}^{-1}\big)^{-1}\Big]^{-1}\bm{M}\deltaB\\
    =&\tfrac{1}{2}\deltaB^T \bm{M}^T(\bm{M}^T)^{-1}\big[(\bm{L^{(1)}})^{-1}\\
    &~~~~~~~~~~~~~~~~~~+(\bm{L^{(2)}})^{-1}\big]^{-1}\bm{M}^{-1}\bm{M}\deltaB\\
    =&\tfrac{1}{2}\deltaB^T \bm{G} \deltaB=\rm{IOI}_{old}\,.
\end{split}
\end{equation}

However, IOI is not invariant under a general invertible parameter transformation which should be a desirable property for a measure of inconsistency. IOI is only nearly invariant under an invertible parameter transformation that can be approximated by a linear transformation in the parameter region of interest. The parameter region of interest here means the parameter space spanned by several confidence levels around the joint mean. Constructing a measure based on IOI, which can describe inconsistency for general cases and is general parameter invariant, is out of the scope of this paper and is left for future work.

\section{Factors affecting IOI and inconsistency measures}\label{subsection-inconsistency-factors}
Experimental inconsistencies are affected by at least three factors: the likelihood mean deviation, the volume of the covariance matrix (how big the iso-likelihood ellipses are), and the degeneracy directions (orientations of the ellipses). The first two factors are related in an obvious way to degree of inconsistency. Basically a larger mean deviation and a smaller covariance matrix volume give a larger experimental inconsistency. The third factor is a little bit more subtle and can be seen as follows. Let us consider the two right panels in Fig.\,\ref{fig-examples-of-IOI} for example. We can see that those two panels have the same mean deviation and the same covariance matrix volumes, but the upper obviously has a smaller inconsistency due to different constraint orientations. So how does IOI relate to and describe different constraint orientations? We recall that IOI incorporates the joint mean to specify the inconsistency. For a given mean deviation and constraint volumes, the orientations of those ellipses determine the location of the joint mean. If the ellipse orientations make the joint mean to locate closer to the mean of each experiment mean, then IOI will be smaller, and vice versa. A different set of ellipse orientations leads to a different matrix $\bm G$ and a different IOI. So for the case of two experiments, the third factor of inconsistency is reflected by the matrix $\bm{G}$. The values of IOI can then describe such a change of the graphical inconsistency due to the change of ellipse orientations.

\begin{figure}[tbp!]
\includegraphics[width=0.49\textwidth]{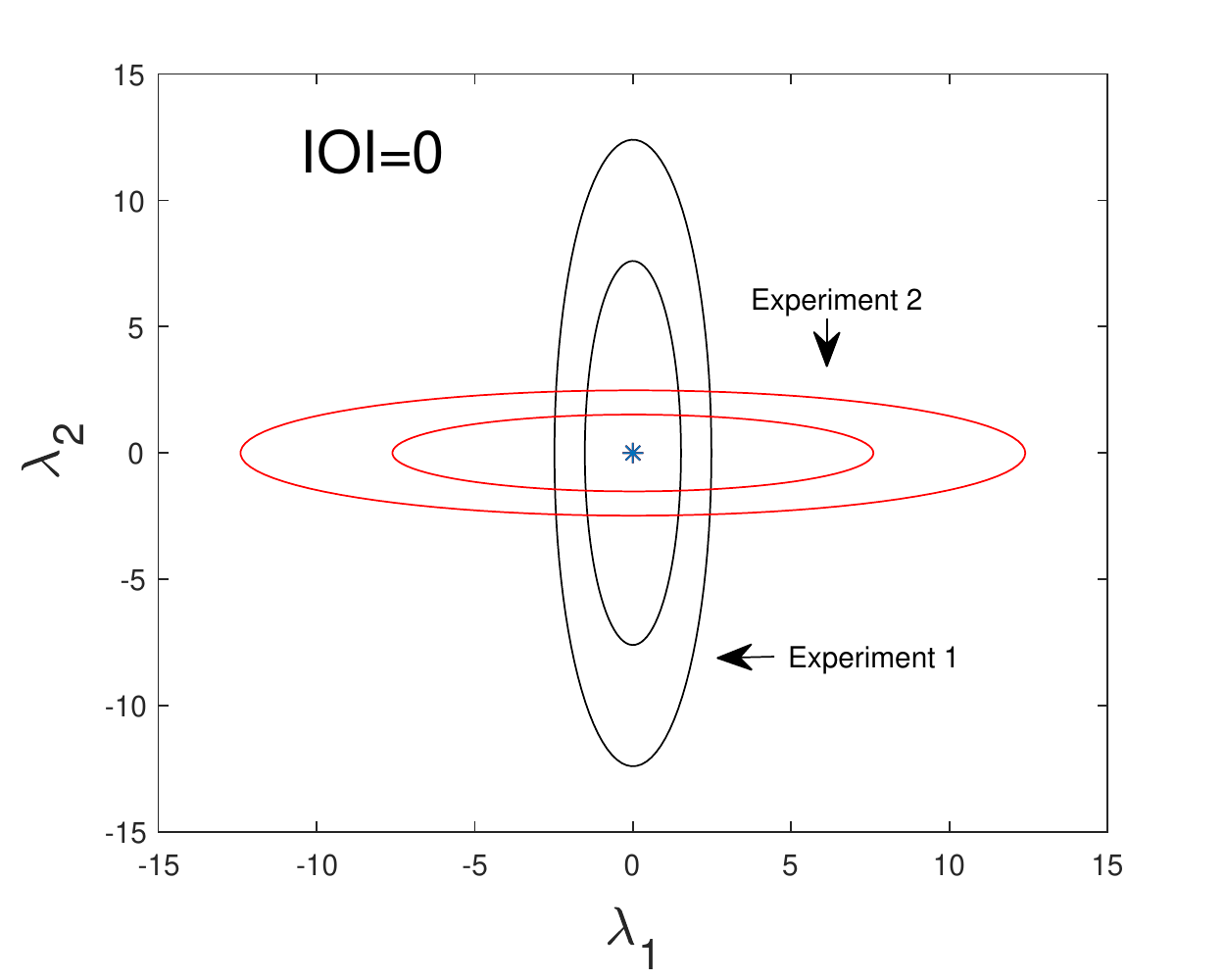}
\caption[A problem of normalized robustness]{\small\label{figure-robustness_normalized-toys} Two toy experiments with different degenerate directions. As we explain in the text, inconsistency between experiments should be treated separately from the small overlapping area caused by degeneracy breaking.}
\end{figure}

We discussed how the lower right panel in Fig.\,\ref{fig-examples-of-IOI} has a larger inconsistency than the upper right panel and explained that this is because the ellipse orientations in the lower panel make the joint mean further away from each individual mean. However, it is not correct to associate the larger inconsistency in the lower panel with the smaller ``overlap'' between the two ellipses compared to the upper panel. A smaller overlap does not necessarily mean a larger inconsistency, as we will explain. We recall that the purpose of a (in)consistency check is to see whether we can jointly study two or more experiments. Two experiments can have a small overlap but still be consistent with each other. Take Fig.\,\ref{figure-robustness_normalized-toys} for example. The two toy experiments have orthogonal degenerate directions, which give them a small overlap. The small overlap here does not prevent us from jointly studying the two experiments, so we say the two experiments are consistent with each other. Thus, this kind of small overlap should not be counted as an increase of inconsistency. Some might worry that the small overlap here means the two distributions are different. But a (in)consistency test is not to look for difference between experiments, but to see if they share anything in common. To see this again, let us assume for a moment that the small overlap does suggest an increase of inconsistency. Imagine that we have a third experiment with a diagonal ($45^\circ$) degenerate direction going through the overlap of the previous two experiments. Now the three experiments also have a small overlap and a lot of differences. But do they have a large inconsistency? Obviously not.

Finally, it is worth clarifying that different ellipse orientations affect simultaneously the inconsistency between two experiments and the constraining power of the joint analysis from these two experiments. The latter is related to the ``overlap area'' of the two experiments. In fact, it can be described by the well-known figure of merit (FOM) (see, e.g., Ref.\,\cite{2010Mortonson-Huterer-Hu-FOM-DE})
\begin{equation}\label{eq-FOM}
{\rm FOM} = |\bm C|^{-1/2}\,,
\end{equation}
where $\bm C$ is the covariance matrix of a distribution. FOM quantifies how good a constraint is. For two distributions with certain constraints powers (i.e. certain FOM$_1$ and FOM$_2$) and a fixed prior, different ellipse orientations change how strong the joint constraint is, i.e. how large FOM$_j$ is. For Gaussian cases (and certain FOM$_1$ and FOM$_2$), FOM$_j$ only depends on ellipse orientations but not on how the ellipses are separated. In contrast, IOI incorporates ellipse orientations and mean separation to specify experimental inconsistency. So while FOM describes how powerful the joint analysis is, IOI tells us whether we can do such a joint analysis.

\section[Parameter marginalization and IOI]{Marginalization can hide inconsistency}\label{subsection-marginalization-and-IOI}
\begin{figure*}[!tbp]
\centering
\includegraphics[width=0.49\textwidth]{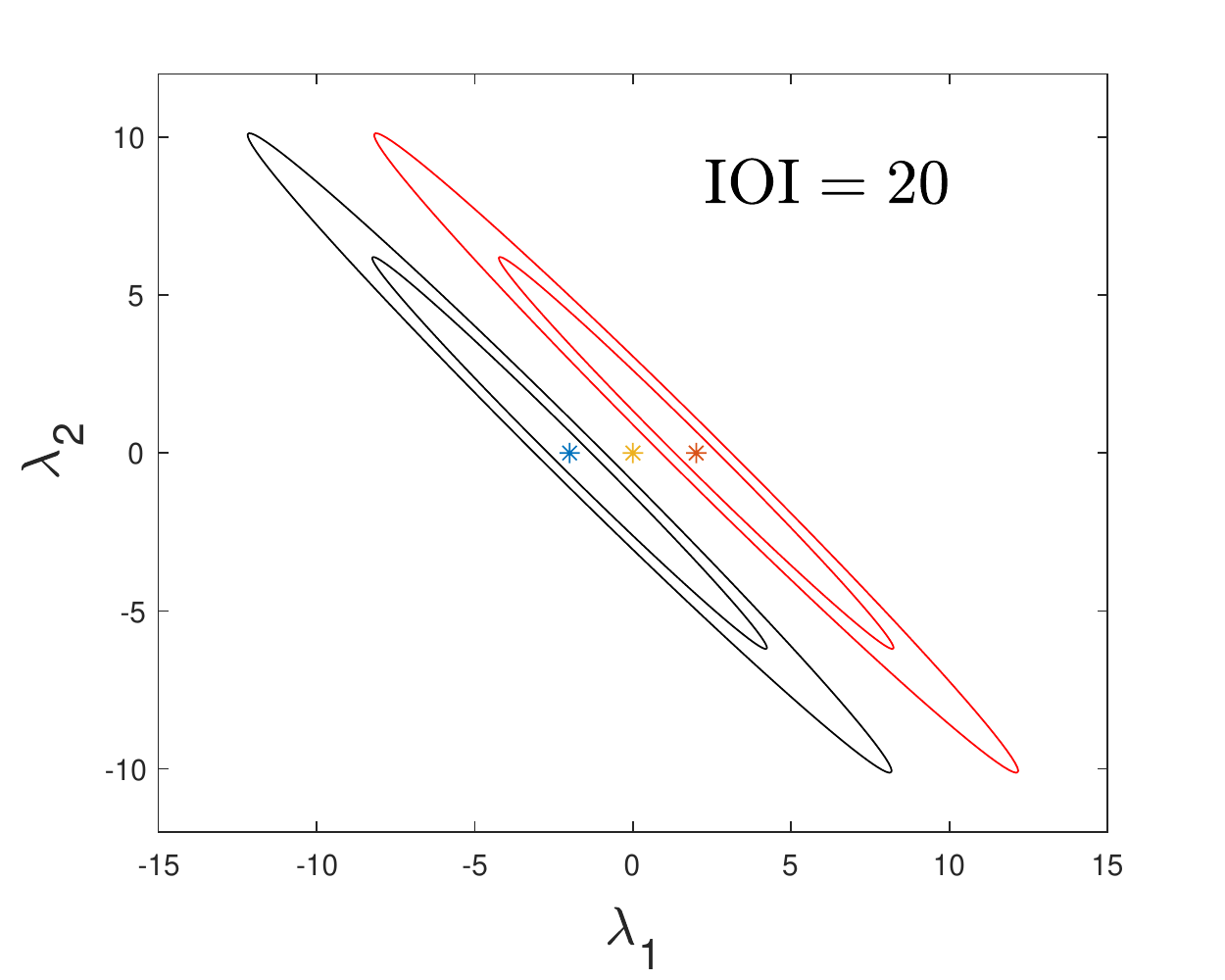}
\includegraphics[width=0.49\textwidth]{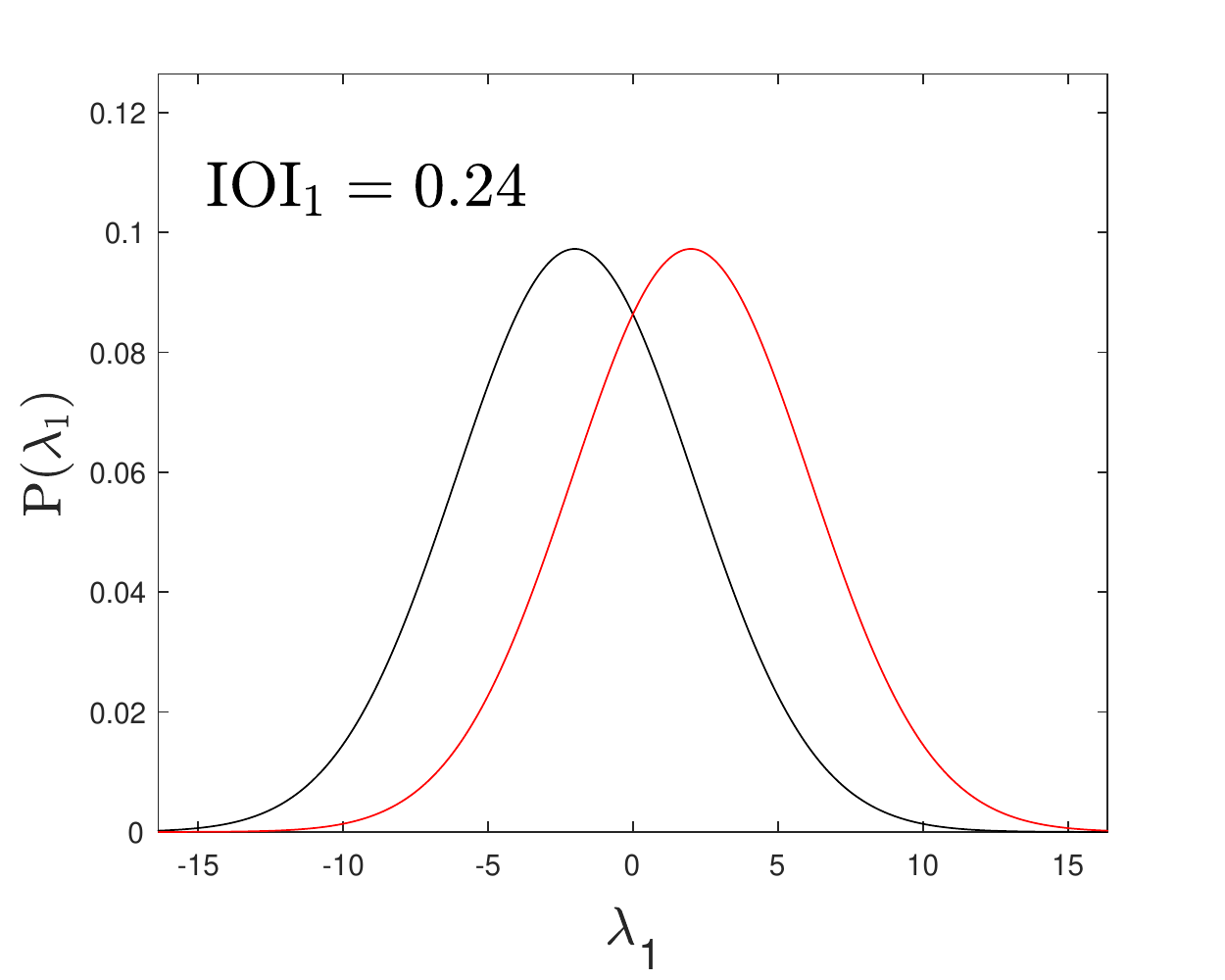}
\caption[Parameter marginalization can hide inconsistency]{\label{fig-marginalization}Toy experiments to show marginalization artificially lowers inconsistency. The left panel shows the 2D contour plot of the posteriors of two toy experiments. The left panel shows the 1D probability distributions of the two experiments marginalized over the second parameter. Clearly the marginalized 1D plots have more overlapped region than the 2D plot, indicating  an artificial decrease of inconsistency between the two experiments. The decrease of the 1D IOI also reflects such a fact but one needs to consider rather the full IOI on the left.}
\end{figure*}

It is usual to investigate the experimental inconsistency in some marginalized 1D or 2D likelihood contour plots, and see if the constraints from different experiments coincide. But there are two problems for this common method. First, if the two experiments have more than three common parameters, we cannot show the inconsistency of them by a plot. Second, since we marginalize over other parameters, we would like to know if the process of marginalization brings up or down the inconsistency between the two experiments. We will show graphically that marginalization hides and lowers artificially the full inconsistency. So after marginalization two experiments will look more consistent one with another but they actually may not be so. Using a measure of discordance solves the first problem, since a measure can be obtained in an arbitrary parameter dimension. But we are going to show that only IOI can correctly track this effect of marginalization.

We recall that marginalization over the $i$th parameter corresponds to deleting the $i$th component from each mean along with the $i$th column and $i$th row from each covariance matrix. Parameter marginalization ignores the detailed information of the probability distribution on the parameters being marginalized over \cite{Cosmomc}. For the inconsistency of two independent and Gaussian distributions, intuitively the loss of information on inconsistency is due to the following two factors. First, in the case where we have more parameters, an unmatched numerical value of any parameter constrained by two experiments introduces some experimental inconsistency. Marginalization over some unmatched parameters obviously hides the inconsistency raised by those parameters. Second, even the mean values of those parameters are the same (i.e., no unmatched numerical values of those parameters), and marginalization can also hide some inconsistency due to the correlations among parameters. One example has already been given by Fig.\,\ref{fig-examples-of-IOI} in our Sec.\,\ref{subsection-IOI-definition}. We can see that the upper right panel in Fig.\,\ref{fig-examples-of-IOI} has a smaller inconsistency than the lower panel. This is because the orientations of the contour ellipses in the upper right panel allow them to overlap more, leading to a smaller inconsistency. However, if we marginalize both panels into 1D plots, they both have identical 1D plots. This example shows that ignoring the orientations of contour ellipses is a reason that marginalization hides inconsistency.

To further demonstrate the second point above, we show in Fig.\,\ref{fig-marginalization} two toy experiments. These two toy experiments have the same mean of $\lambda_2$ (here it is the second element of $\lambdaB$), but they have different means in $\lambda_1$. If we marginalize over $\lambda_2$, we will make the two experiments look more consistent with each other as shown in the right panel of Fig.\,\ref{fig-marginalization}. This is because in this case both experiments have broad marginalized uncertainties in $\lambda_1$, but the actual degeneracy direction of the experiments is not along the direction of $\lambda_1$. Marginalizing over $\lambda_2$ hides the fact that the two experiments actually have little overlap shown in a higher dimension.
This effect is caught by the drop in IOI. The total IOI is $20$ meaning a significant inconsistency between the two toy experiments, but after marginalizing over $\lambda_2$, IOI becomes $0.24$ meaning no significant inconsistency.

From the above discussion, we can see that parameter marginalization hides experimental inconsistency when present. We need to explore the following two questions.  Does IOI always drop after parameter marginalization when it should? When marginalizing over a parameter that has no inconsistency associated with it and it does not correlate with other parameters, then IOI remains the same after marginalizing over that parameter. We find that it is case if those two conditions are satisfied. We prove analytically in Appendix \ref{appendix-sub-the-proof-marg} that IOI never increases after parameter marginalization. The general condition for IOI to remain the same after marginalizing over $\lambda_k$ is,
\begin{equation}\label{eq-IOI-marg-equal-condition}
\sum\limits_{i,j\neq k}\big(\bm{\widetilde{C_{kk}}}^{-1}\big)_{ij}{C_{ki}}\delta_j=\delta_k\,,~\rm{(Equality~Condition)}
\end{equation}
where $C_{ij}$'s are the elements of $\bm{C}=\bm{C^{(1)}}+\bm{C^{(2)}}$, and $\bm{\widetilde{C_{kk}}}$ is a matrix obtained by taking out the \textit{k}th row and \textit{k}th column from $\CB$.  If two experiments give the same mean value of $\lambda_k$ (i.e. $\delta_k=0$), and if $\lambda_k$ does not correlate with other parameters (i.e. $C_{ki}=0$ for $i\neq k$), the equality condition [Eq.\,\eqref{eq-IOI-marg-equal-condition}] is satisfied since both sides are $0$. Then IOI will remain the same after marginalizing over $\lambda_k$. So IOI does track exactly the effect of marginalization. This is a unique feature of IOI among all the measures of inconsistency\footnote{In Gaussian cases, it is also a feature of $\Delta\chi^2$ and the \textit{tension}.}.

However, the condition $\delta_k=0$ and $C_{ki}=0$ is only a subset of the equality condition [Eq.\,\eqref{eq-IOI-marg-equal-condition}] (although such a subset is more common and practical). Equation \eqref{eq-IOI-marg-equal-condition} suggests that there are some other conditions for the experimental inconsistency to be preserved after parameter marginalization. The equality condition \eqref{eq-IOI-marg-equal-condition} can be satisfied without $\delta_k=0$ or $C_{ki}=0$. An example of such situation and a full discussion are provided in Appendix \ref{appendix-sub-discuss-equality-condition}. In other words, if the drop of IOI after marginalizing over one parameter is small, it does not necessarily mean we are safe to marginalize over that parameter. As we discuss in the next section, we should also check whether the residual IOI$_i$ after marginalizing over all the other parameters is also small. If both are small, we can marginalize safely over that parameter to analyze the inconsistency for the remaining parameter space of the model.

\begin{table*}[htbp!]
\begin{ruledtabular}
\caption{\label{table-one-parameter-IOI}Summary of the two measures of one-parameter inconsistency.}
\begin{tabular}{llc}
Notations &\multicolumn{1}{c}{Meanings} & Relative measures\\
\hline
IOI$_i$ & The residual IOI after marginalizing over all the other parameters except $\lambda_i$ & ${{\rm IOI}_i^{\rm Rel}}=\frac{{\rm IOI}_i}{{\rm IOI}}$\\
$\Delta$IOI$_i$ & The drop of IOI after marginalizing over $\lambda_i$ & $\Delta{\rm IOI}_i^{\rm Rel}=\frac{\Delta{\rm IOI}_i}{{\rm IOI}}$
\end{tabular}
\end{ruledtabular}
\end{table*}

\section[Two single-parameter measures]{Zooming on one parameter: relative drop in IOI and relative residual IOI}\label{subsection-supplemental-measure}
The usual way to zoom the experimental inconsistency of a given parameter is to plot and compare the two marginalized 1D likelihoods in the same graph. Examples are the marginalized 1D plots in Fig.\,\ref{fig-gg-l-triagular} and Fig.\,\ref{fig-gg-w-triagular} in Sec.\,\ref{section-application}. As we show in Sec.\,\ref{subsection-marginalization-and-IOI}, this is not an accurate description of experimental inconsistency. An important point is that parameters are correlated with each other (in the language of IOI, the correlation is specified by the combined covariance matrix $\bm{C}=\bm{C^{(1)}}+\bm{C^{(2)}}$ instead of $\bm{C^{(1)}}$, $\bm{C^{(2)}}$ or the joint covariance matrix $\bm{C}_{joint}=\big[(\bm{C^{(1)}})^{-1}+(\bm{C^{(2)}})^{-1}\big]^{-1}$), but marginalized 1D plots do not reflect this fact. So if the two experiments yield a consistent result on one parameter in this way, it does not mean it is safe to marginalize such parameter and analyze the inconsistency of the other parameters. Even all marginalized 1D plots show quite consistent constraints on all parameters; it does not mean the two experiments are actually consistent with each other. Adding a measure to a 1D marginalized plot does not improve the situation at all, because a numerical value associated with a 1D plot does not reflect the effect of marginalization either. All the above points suggest that the usual way of viewing the experimental inconsistency of parameters in marginalized plots is not accurate, and a measure of inconsistency for the full parameter space should be used.

On the other hand, we still want to see if two experiments agree on one particular parameter. For example, we want to see if the geometry and the growth sets of experiments give consistent properties of dark energy as its equation of state $w$. For IOI, an (obvious) one-parameter inconsistency IOI$_i$ can be defined as the residual IOI of parameter $\lambda_i$ after marginalizing over all the other parameters in the model. But as we show below, this IOI$_i$ does not fully specify the inconsistency due to parameter $\lambda_i$. Indeed, to better specify the inconsistency due to one particular parameter in a model, we propose a supplemental measure based on IOI. We  define the drop of IOI after marginalizing over $\lambda_i$ itself, that is
\begin{equation}\label{eq-subpplemental-measure}
\Delta{\rm{IOI}}_i= {\rm{IOI}}-{\rm{IOI}}_{marg,i}\,.
\end{equation}
This definition is motivated by the fact that IOI never increases whenever we perform a parameter marginalization. Since other measures may increase in general, this supplemental measure of one-parameter inconsistency is unique among the other measures in the literature.

With the two measures of one-parameter inconsistency, we can see if it is safe to marginalize over a parameter to analyze the experimental inconsistency. The residual IOI$_i$ can be directly associated with the corresponding marginalized 1D plot, while the drop $\Delta$IOI$_i$ can tells us the degree of inconsistency hidden by marginalizing over $\lambda_i$. If two experiments do not have inconsistency due to $\lambda_i$, both IOI$_i$ and $\Delta{\rm{IOI}}_i$ should be small on Jeffreys' scales. On the other hand, if one of them is large, it means there exists inconsistency associated with that parameter, and it is not safe to marginalize over it to analyze the inconsistency of the other parameters. In Sec.\,\ref{appendix-sub-discuss-equality-condition} we describe in detail different situations regarding different magnitudes of IOI$_i$ and $\Delta{\rm{IOI}}_i$, and we summarize them in Table \ref{table-marginalization-situations}. In sum, marginalizing over a parameter that has a large $\Delta$IOI$_i$ or IOI$_i$ will either hide the inconsistency due to correlation or ignore the shape of the original distributions. The requirement is that both measures must be small.

The two measures of one-parameter inconsistency IOI$_i$ and $\Delta$IOI$_i$ can be compared to Jeffreys' scales (Table \ref{table-Jeffrey-scale-IOI}). However, we often have situations where all IOI$_i$'s and $\Delta$IOI$_i$'s are not significant on Jeffreys' scales but they are still not insignificant compared to the full IOI. In this situation, it is not good to marginalize over any parameter to analyze the experimental inconsistency. Therefore, when we decide whether we can marginalize over a parameter, we can compare IOI$_i$ and $\Delta$IOI$_i$ to the full IOI. For that, we define the relative residual ${\rm IOI}_i^{\rm Rel}$ and relative drop $\Delta{\rm IOI}_i^{\rm Rel}$ for the parameter $\lambda_i$,
\begin{equation}\label{eq-relative-IOI}
{\rm IOI}_i^{\rm Rel}=\frac{{\rm IOI}_i}{{\rm IOI}}\,,~~~~~
\Delta{\rm IOI}_i^{\rm Rel}=\frac{\Delta{\rm IOI}_i}{{\rm IOI}}\,.
\end{equation}
Both ${\rm IOI}_i^{\rm Rel}$ and $\Delta{\rm IOI}_i^{\rm Rel}$ need to be much smaller than 1 for us to safely marginalize over $\lambda_i$. The definitions of the two measures of one-parameter inconsistency are summarized in Table \ref{table-one-parameter-IOI}.

Note that the summation of all IOI$_i$'s or $\Delta$IOI$_i$'s is not the full IOI. That is
\begin{equation}\label{eq-summation-IOIi-unequal}
\sum\limits_i{\rm IOI}_i\neq {\rm IOI}\,,~~~~\sum\limits_i\Delta{\rm IOI}_i\neq {\rm IOI}\,.
\end{equation}
Consequently,
\begin{equation}\label{eq-summation-rel-IOIi-unequal}
\sum\limits_i{\rm IOI}_i^{\rm Rel}\neq 1\,,~~~~\sum\limits_i\Delta{\rm IOI}_i^{\rm Rel}\neq 1\,.
\end{equation}
Thus we cannot interpret ${\rm IOI}_i^{\rm Rel}$ or $\Delta{\rm IOI}_i^{\rm Rel}$ as the fractional inconsistency for $\lambda_i$. Normalizing ${\rm IOI}_i$ and $\Delta{\rm IOI}_i$ by deriving the full IOI is only a way to see whether the inconsistency due to $\lambda_i$ is significant to the full inconsistency of a model.

The analytic expression for $\Delta{\rm{IOI}}_i$ is given by (see Appendix \ref{appendix-sub-IOI-drop})
\begin{equation}\label{eq-DeltaIOI-general-repeat}
\Delta{\rm IOI}_i=\frac{\Big[\delta_i-\sum\limits_{m,n\neq i}\Big(\mathbf{\widetilde{C_{ii}}}^{-1}\Big)_{mn}C_{im}\delta_n\Big]^2} {C_{ii}-\sum\limits_{m,n\neq i}\big(\mathbf{\widetilde{C_{ii}}}^{-1}\big)_{mn}C_{im}C_{in}}\,.
\end{equation}
However, despite its analytical neatness, the above expression is not as practically useful  as the definition of $\Delta{\rm{IOI}}_i$ itself as given by [Eq.\,\eqref{eq-subpplemental-measure}].

\section[IOI eigenmodes of inconsistency]{IOI eigen-mode decomposition}\label{subsection-eigen-modes}
Since IOI is in a quadratic form, it possible to define eigenmodes of inconsistency as the eigenmodes of the matrix $\bm{G}$. By diagonalizing $\bm{G}$ we obtain $N$ eigenmodes with eigenvalues $g_i$. Note that the eigenmodes are characterized by $\bm{G}$, instead of $\bm{C^{(i)}}$ in the usual principal component analysis. Suppose $\bm{M}$ is the orthogonal transformation matrix from the original parameter space to the eigenmodes, i.e. $\bm{\xi}=\bm{M}\lambdaB$ and $\bm{\delta^\xi}=\bm{M}\bm{\delta}$, then Eq.\,\eqref{eq-IOI-definition} in the eigenmodes of $\bm{G}$ reads,
\begin{equation}\label{eq-eigen-sum-1}
{\rm{IOI}}=\sum\limits_i\tfrac{1}{2}g_i(\delta^\xi_{~i})^2\,.
\end{equation}
From the above equation, we can see that the full IOI is a summation of $N$ terms, each of which involves the mean difference of one and only one eigenmode parameter. We can then identify each term as the inconsistency of each eigenmode parameter, i.e. IOI$^{\rm{eig}}_i=\tfrac{1}{2}g_i(\delta^\xi_{~i})^2$ (assumed in a descending order), and we have
\begin{equation}\label{eq-eigen-sum}
{\rm IOI}=\sum\limits_i {\rm IOI}^{\rm{eig}}_i\,.
\end{equation}
A nice feature of the inconsistency eigenmodes is that these IOI$^{\rm{eig}}_i$'s are equal to the drop $\Delta$IOI$_i$ after marginalizing over the corresponding parameter, and also are equal to the residual IOI$_i$ of $\lambda_i$ after marginalizing over all the other parameters, that is
\begin{equation}\label{eq-eigen-three-equal}
\Delta{\rm{IOI}}_i={\rm{IOI}}_i={\rm{IOI}}^{\rm{eig}}_i\,.
\end{equation}
Recall Eqs.\,\eqref{eq-eigen-sum} and \eqref{eq-eigen-three-equal} are only valid for eigenmodes of inconsistency. In general cases, the summation of $\Delta{\rm{IOI}}_i$'s or ${\rm{IOI}}_i$'s does not equal IOI of the full parameter space.

\begin{figure*}[!tbp]
\centering
\includegraphics[width=0.49\textwidth]{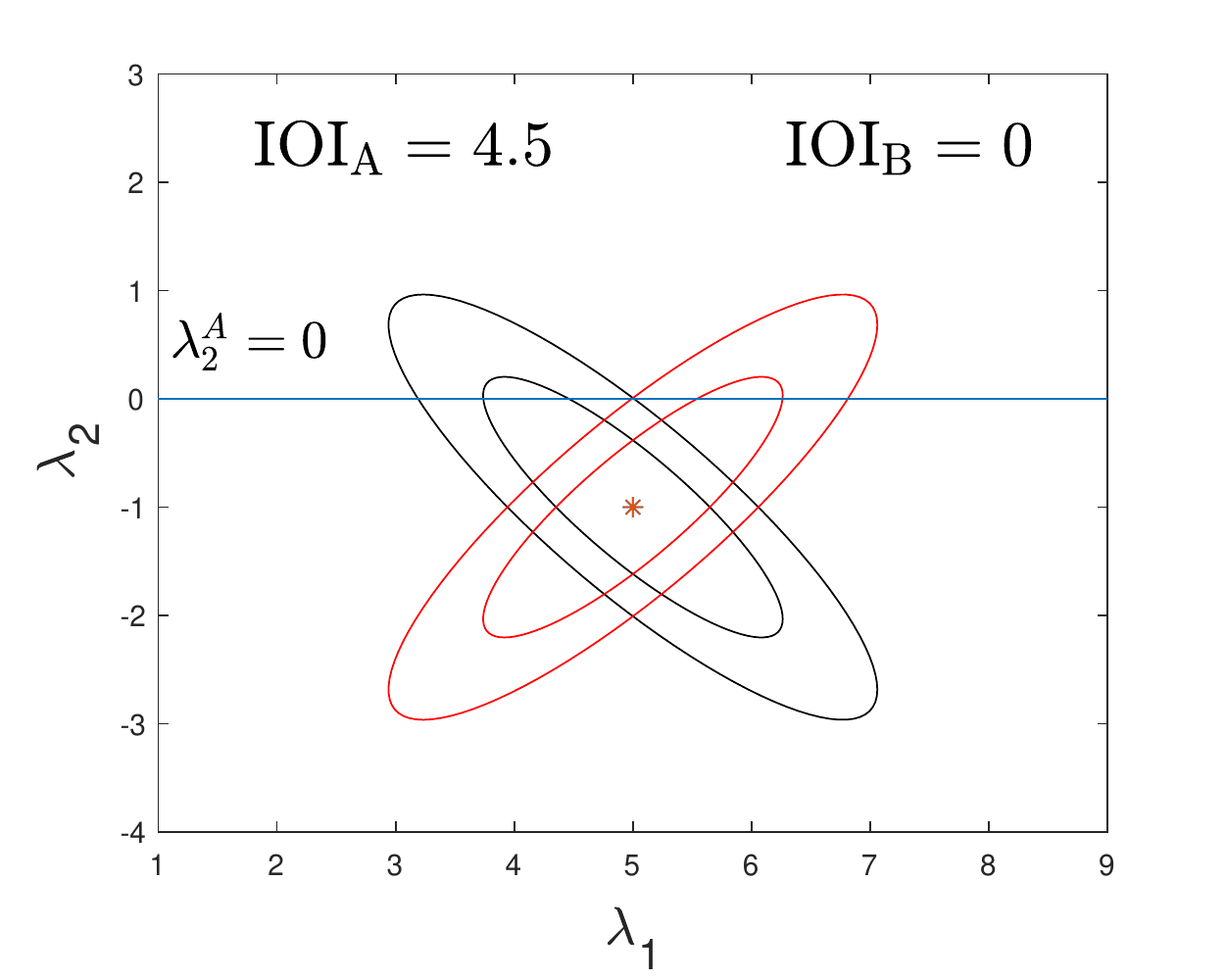}
\includegraphics[width=0.49\textwidth]{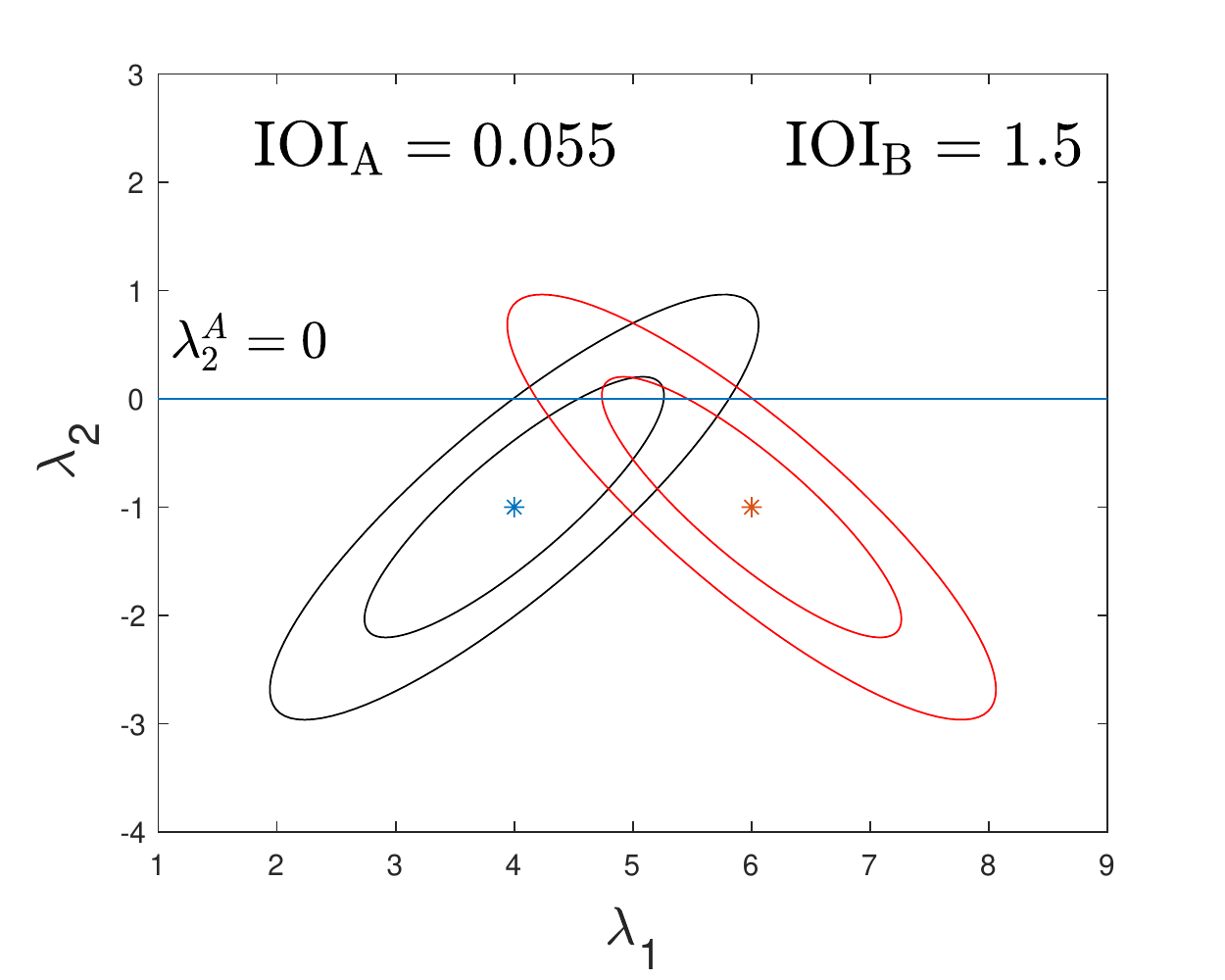}
\caption[Converge and diverge model extensions]{\label{fig-nest-converge-diverge}Toy experiments to show IOI can be smaller or larger for model extension. Model A is nested in model B, and model A has a fixed value of $\lambda_2=\lambda_2^{A}=0$. In the left (right) panel model B has a larger (smaller) IOI than model A. See the text for a detailed explanation.}
\end{figure*}

Let us further clarify the meaning of inconsistency eigenmodes. First, this is not like the principal component analysis in which the uncertainties of eigenmodes are specified by the eigenvalues of a Fisher matrix. Here the eigenmode inconsistency is specified by IOI$^{\rm{eig}}_i=\tfrac{1}{2}g_i(\delta^\xi_{~i})^2$, rather than $g_i$ alone. A larger $g_i$ does not necessarily mean a larger inconsistency. Second, if we rotate the parameter space and make one parameter along the deviation vector $\deltaB$, we can express the full IOI as $\tfrac{1}{2}G_{\delta\delta}\delta^2$, where  $G_{\delta\delta}$ (or $(\bm{G})_{\delta\delta}$) is the diagonal element of the rotated $\bm{G}$ corresponding to the deviation direction, and $\delta=|\deltaB|$ is the magnitude of the deviation vector. So the largest inconsistency ``mode'' seems to be just along the deviation vector $\deltaB$. However, the numerical value of $\tfrac{1}{2}G_{\delta\delta}\delta^2$ is not preserved after marginalizing over other parameters. It is the element $(\bm{G}^{-1})_{\delta\delta}$ that is preserved after marginalization. And since $(\bm{G})_{\delta\delta}$ is not $\big[(\bm{G}^{-1})_{\delta\delta}\big]^{-1}$ in general,  $\tfrac{1}{2}G_{\delta\delta}\delta^2$ changes its value after marginalizing over any other parameter. In contrast, the eigenmodes of inconsistency here have a feature that each IOI$^{\rm{eig}}_i$ is preserved after marginalizing over any other parameter.

\section[IOI in nested models]{IOI in nested models (fixing parameters)}\label{subsection-IOI-nested}
We have shown that marginalization can hide experimental inconsistency and lowers IOI, but it does not mean that models with more parameters will necessarily have a higher IOI. Especially if model A is nested in model B (or alternatively said, model B is an extension to model A), it does not necessarily mean model B will have a higher IOI than model A. It was shown in Ref.\,\cite{2013-tension-Verde-etal} with their own measure called the \textit{tension} that extending a model does not necessarily increase or decrease experimental inconsistency. Here we use IOI to show this point in an explicit way. If model A is nested in model B, one (some) of the parameters in model B is (are) fixed in model A. This actually has two effects on the likelihoods. First, fixing a parameter will take out the corresponding elements from the Fisher matrices of model B, and the covariance matrices and consequently the matrix $\bm{G}$ will be changed. Second, the mean values are forced to be shifted onto a parameter subspace of one (or some) constant parameter(s). These two effects can be shown more explicitly with an example in two dimensions. We assume that the extension model B has means $\bm{\mu^{(1)}}$ and $\bm{\mu^{(2)}}$ from two experiments), and Fisher matrices $\bm{L^{(1)}}$ and $\bm{L^{(2)}}$. When we fix, e.g., $\lambda_2$ to $\lambda_2^{A}$, for each likelihood we can manipulate the term $(\lambdaB-\bm{\mu^{(i)}})^T\bm{L^{(i)}}(\lambdaB-\bm{\mu^{(i)}})$ as follows:
\begin{equation}\label{eq-fixing-parameter}
\begin{split}
&(\lambdaB-\bm{\mu^{(i)}})^T\bm{L^{(i)}}(\lambdaB-\bm{\mu^{(i)}})\\
=&L^{(i)}_{~11}(\lambda_1-\mu^{(i)}_{~1})^2+2L^{(i)}_{~12}(\lambda_1-\mu^{(i)}_{~1})(\lambda_2^{A}-\mu^{(i)}_{~2}) \\ &~~~~~~~~~~~~~~~~~~~~~~~~~+L^{(i)}_{~11}(\lambda_2^{A}-\mu^{(i)}_{~2})^2 \\
=&L^{(i)}_{~11}\big[\lambda_1-\mu^{(i)}_{~1}-\frac{L^{(i)}_{~12}}{L^{(i)}_{~11}}(\mu^{(i)}_{~2}-\lambda_2^{A})\big]^2\\
&~~~~~~~~~~~~~~~~~~~~~~~~~+{\rm{terms~independent~of~\lambda_1}}\,,
\end{split}
\end{equation}
where $L^{(i)}_{~jk}$ is the $(j,k)$ element of $\bm{L^{(i)}}$. After taking the exponential of the above, the terms independent of $\lambda_1$ become  only a multiplying factor that can be taken care of by normalization. So fixing parameter $\lambda_2$ to $\lambda_2^{A}$ changes the squared uncertainty of $\lambda_1$ from $C^{(i)}_{~11}$ to $1/L^{(i)}_{~11}$, and shifts the means of $\lambda_1$ from $\mu^{(i)}_{~1}$ to $\mu^{(i),A}_{~1}=\mu^{(i)}_{~1}+\tfrac{L^{(i)}_{~12}}{L^{(i)}_{~11}}(\mu^{(i)}_{~2}-\lambda_2^{A})$.

The shifts of means change IOI the most: it may either bring the means closer or push them fartheraway. As a result, an extension may have a larger or smaller IOI than the model being extended. Figure \ref{fig-nest-converge-diverge} shows the two possibilities. In both panels in Fig.\,\ref{fig-nest-converge-diverge}, model B is an extension to model A. Model B has two parameters, $\lambda_1$ and $\lambda_2$. Parameter $\lambda_2$ is fixed to $0$ in model A. In the left panel, the two toy experiments give two consistent distributions for model B, and IOI$_B$ is $0$. But since we fix $\lambda_2$ to $\lambda_2^{A}=0$, the distributions are now on the straight line. The two means are now pushed farther apart in model A. The extension model B is more consistent than model A when comparing the two experiments, and accordingly IOI$_B$ is smaller than IOI$_A$. On the other hand, in the right panel the means are brought closer when fixing $\lambda_2$ to $\lambda_2^{A}=0$. The extension model B is then less consistent than model A, and IOI$_B$ is accordingly larger than IOI$_A$. Therefore, an extension may have larger or smaller IOI.

We classify model B to be either a more consistent extension to model A (if IOI$_{B}$ is smaller than IOI$_{A}$), or a less consistent extension (if IOI$_{B}$ is larger than IOI$_{A}$). But since IOI depends on particular experiments, whether an extension is more or less consistent is relative and depends on the experiments under consideration.

\section{Nuisance and unshared parameters}\label{subsection-IOI-uncommon-param}
When we compare two experiments, we often come across situations where one experiment can constrain some parameters in a model but the other cannot, or each experiment has its own systematic/nuisance parameters. In those situations, the two experiments give two probability distributions with different parameter dimensions. It is then necessary to marginalize over the uncommon parameters to get two distributions with the same dimension. Does parameter marginalization in these cases hide some experimental inconsistencies? The answer is no. Marginalizing over any systematic parameters or unshared parameters does not affect IOI even in the case where the inconsistencies can be caused by those parameters.

For a real example, in Sec.\,\ref{section-application} we will compare the geometry experiments and the growth experiments in the $\Lambda$CDM model. Since the geometry experiments do not constrain the parameter $\tau$ (optical depth), we need to fix it when fitting the $\Lambda$CDM model to the geometry experiments. We then marginalize over $\tau$ for the growth experiments to compare the two sets of experiments. If the geometry experiments do not constrain or very poorly constrain $\tau$, it is equivalent to the situation as follows: (1) where it gives very wide (weak) uncertainty on $\tau$, and $\big(\bm{L^{geom}}\big)_{\tau\tau}\rightarrow0$; (2) where $\tau$ has no correlation with other parameters in the geometry set, and $\big(\bm{L^{geom}}\big)_{\tau i}\rightarrow0$ for $i\neq\tau$. No matter what constraint of $\tau$ is given by the growth experiments, the geometry experiments do not ``object'' to such a constraint, and there is no conflict between the two sets on the value of $\tau$. This does not necessarily mean that the joint mean of $\tau$ would be the same as the mean given by the growth experiments. In fact they are different in general, and the joint mean is given by the Fisher-matrix-weighted average [Eq.\,\eqref{eq-Fisher-weighted-average-two}]. However, the form of $\bm{L^{geom}}$ now implies that the joint mean of $\tau$ does not depends on the fixed value of $\tau$ for in geometry set, or the mean given by the geometry set (if it gave one).

Mathematically, we can see more clearly from the analytical expression of $\Delta$IOI$_{\tau}$ given by Eq.\,\eqref{eq-DeltaIOI-general}. If $\big(\bm{L^{geom}}\big)_{\tau i}\rightarrow0$ for $i\neq\tau$ and $\big(\bm{L^{geom}}\big)_{\tau\tau}\rightarrow0$, we have $C^{geom}_{\tau \tau}\rightarrow\infty$. Every term in Eq.\,\eqref{eq-DeltaIOI-general} is finite except for $C_{\tau\tau}=C^{geom}_{\tau\tau}+C^{grow}_{\tau\tau}\rightarrow\infty$. So the denominator of Eq.\,\eqref{eq-DeltaIOI-general} goes to $\infty$ and $\Delta$IOI$_{\tau}\rightarrow0$. In another word, the drop of IOI after marginalizing over $\tau$ is $0$. Also IOI$_{\tau}\rightarrow0$ as well since $C_{\tau\tau}\rightarrow\infty$ and IOI$_{\tau}=\tfrac{1}{2}\delta_\tau^2/C_{\tau\tau}\rightarrow0$ for finite $\delta_\tau$. Since both $\Delta$IOI$_\tau$ and IOI$_\tau$ are zero, there is no problem to marginalize over $\tau$ to analyze the inconsistency of the whole model. For the same reason, it is also safe to marginalize over all the nuisance parameters of each experiment.

\section[Other measures of inconsistency in the literature]{Comparison with other measures of inconsistency in the literature}\label{section-comparison}
In this section we compare IOI to some other measures of inconsistency/consistency in the literature. Their Gaussian and weak prior limits will be derived for the comparison. We also examine how such measures track or not inconsistencies in various particular cases.

\subsection{The \textit{robustness}}\label{subsection-robustness}
The \textit{robustness} was introduced in Refs.\,\cite{2006-Marshall-etal-Bayesian,2011Robustness-March-etal}, and has been applied, e.g., in Ref.\,\cite{2013Amendola-Marra-Quartin-Internal-robustness} to investigate the internal inconsistency of type Ia supernovae data set. It is defined as the Bayesian evidence ratio between the combined model and the split model in order to quantify the consistency between two experiments as follows. Combined model: the two experiments are jointly fit by a set of parameters; Split model: each experiment is individually fitted by one set of parameters. The robustness is then defined as
\begin{equation}\label{robustness-definition}
R=\frac{E_{comb}}{E_{split}}\,, 
\end{equation}
where $E$ is the corresponding Bayesian evidence, so
\begin{align}
\begin{split}\label{eq-E-combined}
E_{comb}&=\int d^N\lambda~\mathcal{L}(\bm{Q};\lambdaB)\mathcal{P}(\lambdaB)\\
        &=\int d^N\lambda~\mathcal{L}^{(1)}(\bm{Q^{(1)}};\lambdaB)\mathcal{L}^{(2)}(\bm{Q^{(2)}};\lambdaB)\mathcal{P}(\lambdaB)
\end{split}\\
\begin{split}\label{eq-E-split}
E_{split}&=E_1\times E_2\\
&=\int d^N\lambda~\mathcal{L}^{(1)}(\bm{Q^{(1)}};\lambdaB)\mathcal{P}(\lambdaB)\\
&~~~~~~~~~~~~\times\int d^N\lambda'~\mathcal{L}^{(2)}(\bm{Q^{(2)}};\bm{\lambda'})\mathcal{P}(\bm{\lambda'})\,.\end{split}
\end{align}
The values of $R$ are interpreted by Jeffreys' scales listed in Table \ref{Jeffrey-scale}.

        \begin{table*}[!t]
        \caption[Jeffreys' scales for Baysian evidence ratio]{\label{Jeffrey-scale}Jeffreys' scales - empirical scales that classify the ranges of the Bayesian evidence ratio. Positive $\ln R$ ($R>1$) favors the combined model while negative $\ln R$ favors the split model. If $|\ln R|<1$, empirically it means there is not enough evidence to tell which model is supported by the data. A higher value of $|\ln R|$ has a higher preference.}
        \begin{ruledtabular}
        \begin{tabular}{lcccc}
        Ranges & $|\ln R|<1$ & $1<|\ln R|<2.5$ & $2.5<|\ln R|<5$ & $|\ln R|>5$ \\
        Meanings &No preference & Weak evidence & Moderate evidence& Strong evidence\\
        \end{tabular}
        \end{ruledtabular}
        \end{table*}

        In the Gaussian limit, we can calculate the \textit{robustness} analytically. Different from the treatment in Ref.\,\cite{2011Robustness-March-etal}, we do not set one of the means to be $\bm{0}$ to find
        \begin{equation}\label{robustness-Gaussian}
        \begin{split}
        R=&\sqrt{\frac{|\bm{F^{(1)}}\bm{F^{(2)}}|}{|\bm{F}\bm{P}|}} \exp\Big[-\frac{1}{2}\big(\bm{\bar{\mu}^{(1)}}\,^T\bm{F^{(1)}}\bm{\bar{\mu}^{(1)}}\\
        &+ \bm{\bar{\mu}^{(2)}}\,^T\bm{F^{(2)}}\bm{\bar{\mu}^{(2)}}-\bm{\mu^{(p)}}\,^T\bm{P}\bm{\mu^{(p)}}-\mubar^T\bm{F}\mubar\big)\Big]\,,
        \end{split}
        \end{equation}
        where,
        \begin{equation}
        \begin{split}
        \bm{F^{(i)}}&=\bm{L^{(i)}}+\bm{P}\,,\\
        \bm{F}~&=\bm{L^{(1)}}+\bm{L^{(2)}}+\bm{P}\,, \\ \bm{\bar{\mu}^{(i)}}&=\tfrac{1}{\bm{F^{(i)}}}\big(\bm{L^{(i)}}\bm{\mu^{(i)}}+\bm{P} \bm{\mu^{(p)}}\big)\,,\\
        \mubar~&=\tfrac{1}{\bm{F}}\big(\bm{L^{(1)}}\bm{\mu^{(1)}}+\bm{L^{(2)}} \bm{\mu^{(2)}}+\bm{P}\bm{\mu^{(p)}}\big)\,.
        \end{split}
        \end{equation}
The major steps of the above calculation can be found in our Appendix \ref{appendix-integral-likelihood}. In the weak prior limit Eq.\,\eqref{robustness-Gaussian} can be further reduced to,
        \begin{equation}\label{robustness-Gaussian-weak}
        \ln R=\tfrac{1}{2}\ln\big(\frac{|\bm{L^{(1)}}\bm{L^{(2)}}|}{|\bm{L}\bm{P}|} \big)-{\rm IOI}\,.
        \end{equation}
       So the \textit{robustness} in the case of Gaussian and weak prior is related to the negative of IOI with an additional term $\tfrac{1}{2}\ln\big(\frac{|\bm{L^{(1)}}\bm{L^{(2)}}|}{|\bm{L^{(2)}}\bm{P}|}\big)$ that accounts for the Occam's razor factor (favoring the model with fewer assumptions). A similar relation [to Eq.,\eqref{robustness-Gaussian-weak}] is pointed out in Ref.\,\cite{2006-Marshall-etal-Bayesian} where $\ln R$ is related to $\tfrac{1}{2}\Delta\hat{\chi}^2$ (same as $\tfrac{1}{2}\Delta\chi^2$ with our notation) which in the Gaussian case is defined as IOI in this work. Ref.\,\cite{2016-quantify-concor-Seehars-etal}
noted that the dependence of prior does not go away in the weak prior limit and $\ln R$ diverges as $|\bm{P}|\rightarrow0$. This seems to imply that two experiments are always consistent with each other when a very weak prior is used, which cannot be true.

        The dependence of $|\bm{P}|$ can be eliminated by a normalization. So for example, the authors of Ref.\,\cite{2011Robustness-March-etal} define $R_N=R/R_*$ where $R_*=\frac{|\bm{F^{(1)}}|}{(|2\bm{L^{(1)}}+\bm{P}||\bm{P}|)}$, so that $R_N=1$ if the second experiment gives the same likelihood as the first one. Then from Eq.\,\eqref{robustness-Gaussian-weak}, $R_N$ in the Gaussian and weak prior limit reads
        \begin{equation}\label{robustness-normalized}
        \ln R_N=\tfrac{1}{2}\ln\big(\frac{|2\bm{L^{(2)}}|}{|\bm{L^{(1)}}+\bm{L^{(2)}}|} \big)-{\rm IOI}\,,
        \end{equation}
        and the dependence on $|\bm{P}|$ is eliminated. But other problems seem to persist. For example, consider two Gaussian likelihoods with the same mean, i.e., $\bm{\delta}=\bm{\mu^{(2)}}-\bm{\mu^{(1)}}=0$. And suppose their Fisher matrices are given by
       \begin{equation}\label{eq-robustness_normalized_thought_experiments}
       \bm{L^{(1)}}=\begin{pmatrix}1 & ~0\\0 &~ \epsilon \end{pmatrix}\,,~~~\bm{L^{(2)}}=\begin{pmatrix}\epsilon & ~0\\0 & ~1 \end{pmatrix}\,,~~{\rm{with}}~\epsilon\ll1\,.
       \end{equation}
       Similar likelihood contours of these two toy experiments are shown in Fig.\,\ref{figure-robustness_normalized-toys}, but now the contours are much narrower. The first constraint ellipse is very elongated along the $y$ direction, while the other is very elongated along the $x$ direction. Although there is little overlap between the two experiments, they have the same parameter mean so there should be no tension between them\footnote{A real and similar example of this situation is the experiments of BAO and distances to supernova.}. But the normalized \textit{robustness} turns out to be (to the leading order of $\epsilon$)
       \begin{equation}
       R_N\approx2\sqrt{\epsilon}\ll1\,.
       \end{equation}
       So $R_N$ implies these two toy experiments are not ``robust'' or not consistent with each other, which seems to be in contradiction with the fact that they are actually consistent with each other.

The term $\tfrac{|\bm{L^{(1)}}\bm{L^{(2)}}|}{|\bm{L}||P|}$ in the original \textit{robustness} also disfavors orthogonal constraints. It may be easier to see this if we relate the \textit{robustness} to IOI and FOM. From the Gaussian and weak prior limit of the \textit{robustness} [Eq.\,\eqref{robustness-Gaussian-weak}] and the definition of FOM [Eq.\,\eqref{eq-FOM}], we can see that
\begin{equation}\label{eq-lnR-IOI-FOM}
-\ln R = \ln\Big(\frac{{\rm FOM}_j\times{\rm FOM}_p}{{\rm FOM}_1\times{\rm FOM}_2}\Big)+{\rm IOI}\,,
\end{equation}
where the subscripts``$1$'', ``$2$'', ``$j$'' and ``$p$'' stand for ``first'', ``second'', ``joint'' and ``prior''. In Sec.\,\ref{subsection-inconsistency-factors} we argued that different constraint orientations do lead to different experimental inconsistencies, which have already been taken care of by IOI. For a fixed FOM$_p$, the term $\frac{{\rm FOM}_j\times{\rm FOM}_p}{{\rm FOM}_1\times{\rm FOM}_2}$ can describe how powerful the joint constraint is compared to the two individual constraints. It is true that this term does also depend on the constraint orientations, and it is larger for orthogonal constraints than parallel constraints. But it should not be taken into accounted as indication of consistency or inconsistency, because it is independent of how far away the two individual constraints are separated from each other.

    \subsection{The \textit{Tension}}\label{subsection-tension}
     Based on Bayesian evidence ratio, a quantity called the \textit{tension} $\mathcal{T}$ was introduced in Ref.\,\cite{2013-tension-Verde-etal}. This measure was used in Ref.\,\cite{2016-Bernal-Verde-Riess-H0}, e.g., to study the tension between the direct measurement of $H_0$ and the one derived from Planck CMB observation. The $\mathcal{T}$ is defined as follows. For two posteriors $\mathscr{P}^{(1)}$ and $\mathscr{P}^{(2)}$, we transfer them while keeping their shapes to get two new posteriors $\mathscr{P}^{(1)}_{sh}$ and $\mathscr{P}^{(2)}_{sh}$. Peaks of $\mathscr{P}^{(1)}_{sh}$ and $\mathscr{P}^{(2)}_{sh}$ coincide at the same location. Then the \textit{tension} $\mathcal{T}$ is defined as
        \begin{equation}\label{eq-tension-definition}
        \mathcal{T}\equiv\frac{\int \mathscr{P}^{(1)}_{sh}\mathscr{P}^{(2)}_{sh} d^N\lambda}{\int \mathscr{P}^{(1)}\mathscr{P}^{(2)} d^N\lambda}\,.
        \end{equation}
         As Ref.\,\cite{2013-tension-Verde-etal} explained, the \textit{tension} $\mathcal{T}$ is defined as the joint Bayesian evidence ratio between a virtual hypothesis and an actual hypothesis. It was also proposed in Ref.\,\cite{2013-tension-Verde-etal} to interpret $\ln\mathcal{T}$ using Jeffreys' scales.

         In the case of Gaussian likelihoods and a Gaussian posterior, the integral of the posterior product in Eq.\,\eqref{eq-tension-definition} can be calculated analytically (see our Appendix \ref{appendix-integral-likelihood} for details)
        \begin{equation}
        \int \mathscr{P}^{(1)}\mathscr{P}^{(2)} d^N\lambda=|\bm{\bar{G}}|^{1/2}\exp\big(-\tfrac{1}{2}\bar{\deltaB}\bm{\bar{G}}\bar{\deltaB}\big)\,,
        \end{equation}
        where (again) $\bm{\bar{G}}=\big[(\bm{F^{(1)}})^{-1}+(\bm{F^{(2)}})^{-1}\big]^{-1}$,  $\bm{F^{(i)}}=\bm{L^{(i)}}+\bm{P}$, $\bar{\deltaB}=\bm{\bar{\mu}^{(1)}}-\bm{\bar{\mu}^{(2)}}$, and $\bm{\bar{\mu}^{(i)}}=(\bm{F^{(i)}})^{-1}(\bm{L^{(i)}}\bm{\mu^{(i)}}+\bm{P}\bm{\mu^{(p)}})$. Note that $\bm{\bar{G}}$ and $\bm{\bar{\mu}^{(i)}}$ are different from $\bm{G}$ and $\bm{\mu^{(i)}}$, and the former ones have the prior involved. In the weak prior limit, $\bm{\bar{G}}\xrightarrow{\rm weak~prior}\bm{G}$ and $\bm{\bar{\mu}^{(i)}}\xrightarrow{\rm weak~prior}\bm{\mu^{(i)}}$. The shifted-posterior-product integral is similar except that $\bar{\deltaB}=\bm{\bar{\mu}^{(1)}}-\bm{\bar{\mu}^{(2)}}=0$, so $\int \mathscr{P}^{(1)}_{sh}\mathscr{P}^{(2)}_{sh} d^N\lambda=|\bm{\bar{G}}|^{1/2}$. Therefore, the \textit{tension} $\mathcal{T}$ becomes
        \begin{equation}\label{eq-tension-Gaussian}
        \ln\mathcal{T}=\tfrac{1}{2}\bar{\deltaB}^T\bm{\bar{G}}\bar{\deltaB}\xrightarrow{\rm weak~prior}\tfrac{1}{2}\deltaB^T\bm{G}\deltaB=\rm{IOI}\,.
        \end{equation}
        So $\ln\mathcal{T}$ reduces to IOI in the Gaussian and weak prior limit. Since the shifted posteriors always have a larger overlapping region for Gaussian cases, $\mathcal{T}>1$, which can also be seen from Eq.\,\eqref{eq-tension-Gaussian}.

        It was also shown in Ref.\,\cite{2013-tension-Verde-etal} that the \textit{tension} is another information for the problem of model selection. For example,  Table 2 in Ref.\,\cite{2013-tension-Verde-etal} shows that while there is not enough Bayesian evidence favoring the $\Lambda$CDM model over the $w$CDM model, the latter significantly reduces the tension on $H_0$ between the Planck and the local measurements.

        Authors in Ref.\,\cite{2016-quantify-concor-Seehars-etal} commented that the \textit{tension} $\ln\mathcal{T}$ is of the order $N$ (dimension of the parameter space) and could overestimate the inconsistency between two experiments. Since $\ln\mathcal{T}$ and IOI are the same in the Gaussian and weak prior limit, that comment would also apply to IOI. But we find here otherwise. First, Table 2 in Ref.\,\cite{2013-tension-Verde-etal} showed that some extensions of the $\Lambda$CDM model increase the \textit{tension} while the others decrease it. So increasing the degree of freedoms does not necessarily increases or decrease $\ln\mathcal{T}$. Second, we can always rotate the parameter space and let the first axis be along the deviation vector $\deltaB$. Then $\ln\mathcal{T}$ consists of only one term $\tfrac{1}{2}g_\delta\delta^2$ in the Gaussian and weak prior limit (see Sec.\,\ref{subsection-eigen-modes}). So it should be of the order unity.

        \begin{figure}[t]
        \includegraphics[width=0.95\linewidth,height=0.5\linewidth]{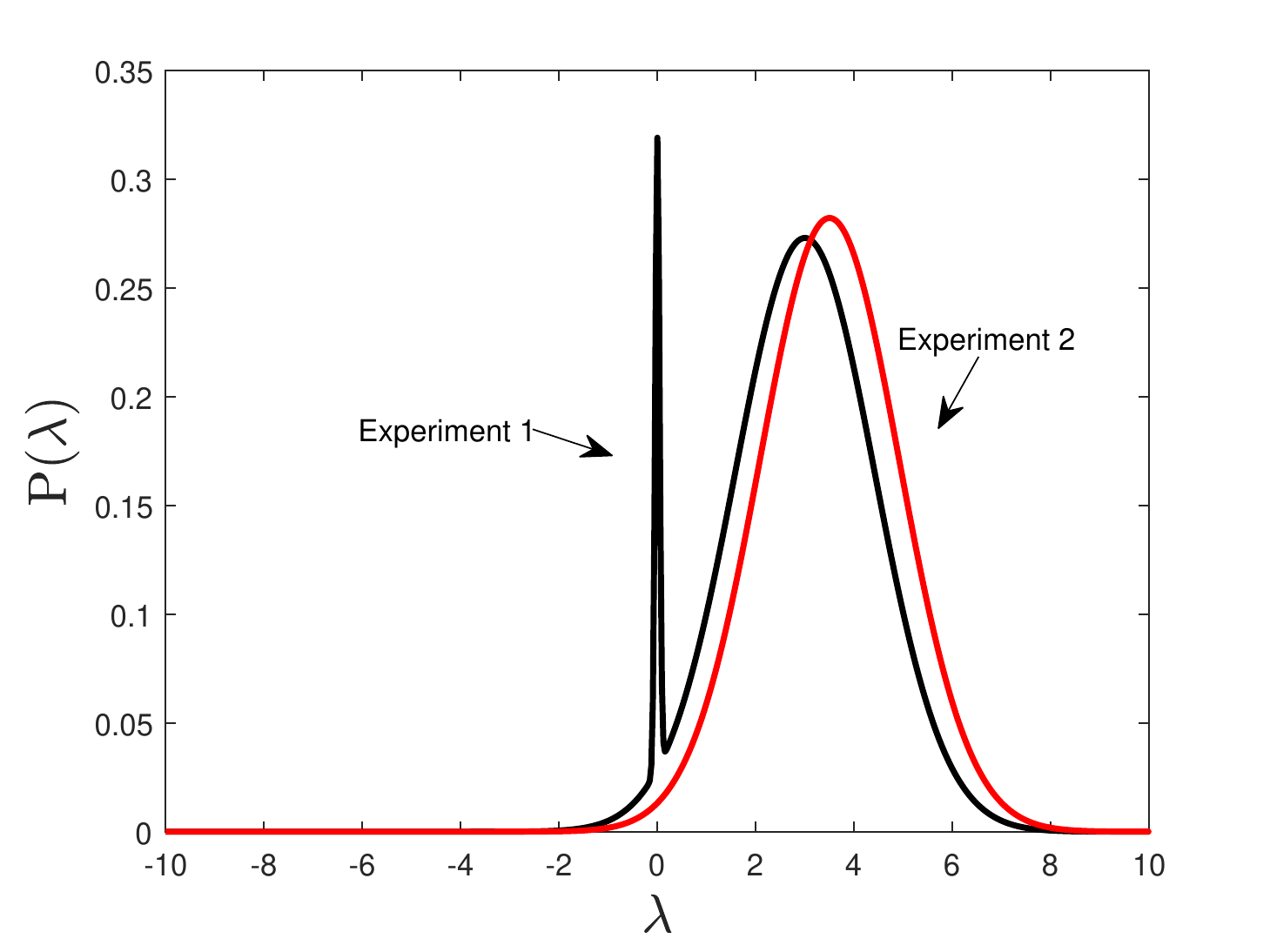}
        \includegraphics[width=0.95\linewidth,height=0.5\linewidth]{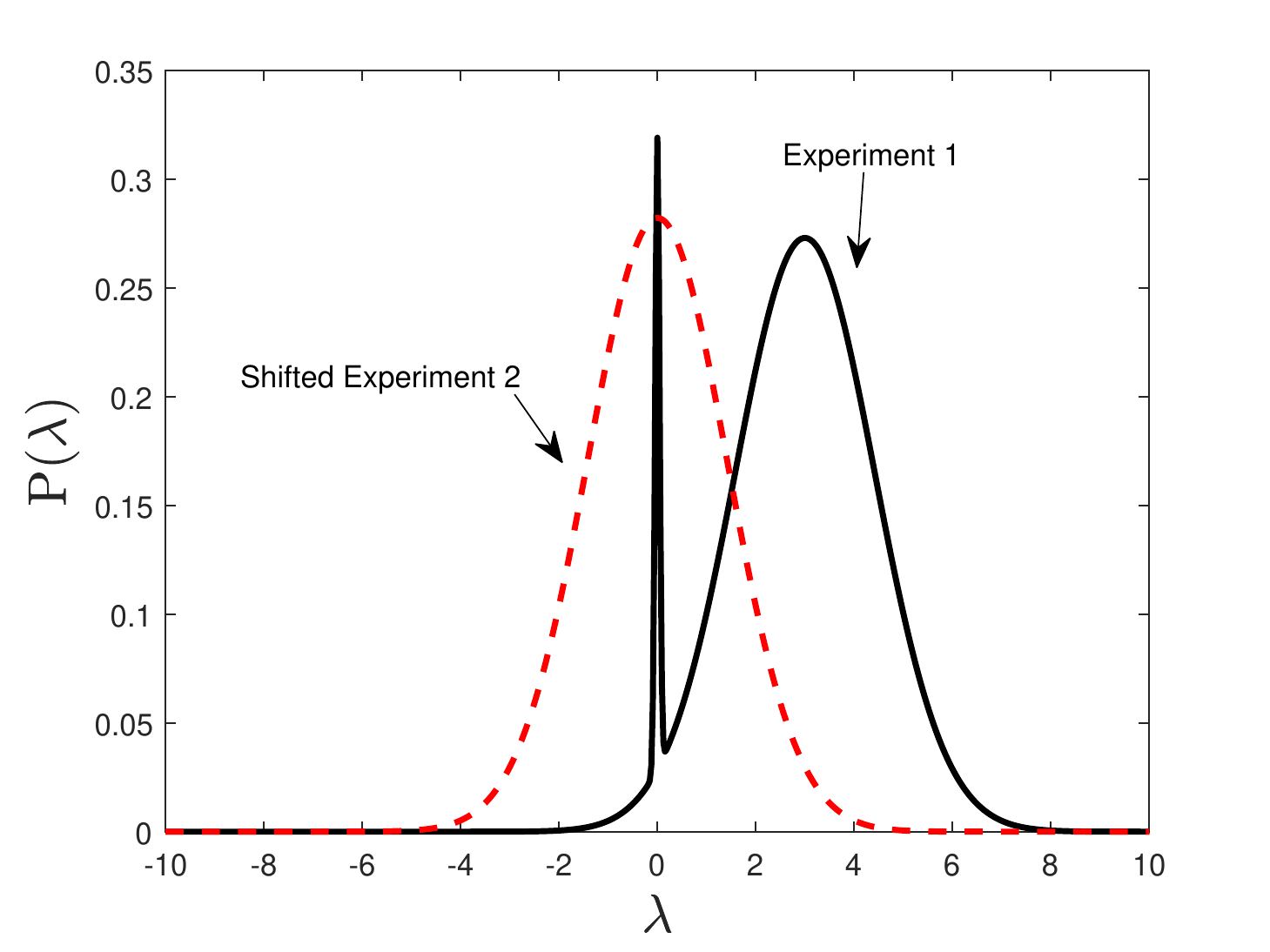}
        \caption{\label{fig-nonGaussian-tension-problem}Demonstration of the problem of the \textit{tension} in a non-Gaussian case. Upper panel: two original one-dimensional probability distributions. The one [$P_{(1)}(\lambda)$] given by experiment 1 is non-Gaussian, while the other ($P_{(2)}(\lambda)$) is Gaussian. Lower panel: the distribution $P_{(2)}(\lambda)$ is shifted to coincide its maximum with that of $P_{(1)}(\lambda)$. Because of the small overlap of the shifted distributions in the lower panel, the numerator in Eq.\,\eqref{eq-tension-definition} is smaller than the denominator, which makes $\ln\mathcal{T}$ negative.}
        \end{figure}

        While the \textit{tension} reduces to IOI in the Gaussian and weak prior limit, the difference between them lies in non-Gaussian cases and in the fact that they originate from distinct concepts (see Table \ref{table-measures}). One advantage of the \textit{tension} compared to IOI is that the \textit{tension} is defined for general distributions. The \textit{tension} may give us a better sense of inconsistency in general cases (although we do not know yet what a good measure is in non-Gaussian cases), while IOI may not work in highly non-Gaussian cases. However, in some cases, it may be better to use IOI. Recall that the definition of the \textit{tension} involves the shifted distributions such that they have coinciding maxima. This may give to the \textit{tension} an ambiguity in choosing the shift in general cases. Consider the two 1D distributions shown in the upper panel of Fig.\,\ref{fig-nonGaussian-tension-problem}. The two experiments give two similar distributions. But $P_{(1)}(\lambda)$ is not Gaussian, and it has a narrow peak at $\lambda=0$ which is unfortunately also the maximum of $P_{(1)}(\lambda)$. If we shift the two distributions to make their maxima coincide, we will get the lower panel in Fig.\,\ref{fig-nonGaussian-tension-problem}. Since the shifted distributions have less overlap than the original distributions, the numerator in Eq.\,\ref{eq-tension-definition} can be smaller than the denominator. We can then get $\mathcal{T}<1$ and $\ln\mathcal{T}<0$. The negative value of $\ln\mathcal{T}$ here certainly cannot be interpreted as `no tension'. In fact, the `majorities' of the two distribution may be very different and exhibit a large tension, but $\ln\mathcal{T}$ may still be negative due to the possibly small overlap of the shifted distributions. The problem is that the numerator in Eq.\,\eqref{eq-tension-definition} is not guaranteed to be larger than the denominator. Better shifted distributions can be defined as those who maximize the numerator in Eq.\,\eqref{eq-tension-definition}, though such definition may not be as practical as the original one. However, distributions in cosmology are usually unimodal, making two maxima to coincide usually also maximize the numerator in Eq.\,\eqref{eq-tension-definition}. Our extreme example here is only to show the possible ambiguity in the choice of distribution shift for general cases, but the measure \textit{tension} should work just fine in the usual and common cases.

\subsection{The \textit{discordance}}\label{subsection-discordance}
The \textit{discordance} was constructed from overlapping iso-likelihood contours and was proposed in Ref.\,\cite{2015-discordance-MacCrann-etal} to measure the inconsistency between two distributions in the following way. First, for a given specific percentage, we can draw two iso-likelihood contours of the two distributions. Depending on how far away the two distributions are separated from each other, the two iso-likelihood contours may or may not overlap. If the two distributions are very far away from each other, the two iso-likelihood contours need to be large (corresponding to a large included percentage) in order to overlap with each other. So the percentage within each overlapping iso-likelihood contours traces the distance of two distributions. The \textit{discordance} is defined as the minimum percentage $\sigma_{eq}$ of the overlapping iso-likelihoods that include equal probability. A larger $\sigma_{eq}$ corresponds to a greater distance between the two distributions, and a larger inconsistency. If the maxima of two distributions coincide at the same location, $\sigma_{eq}$ will be $0$; on the other hand if they are separated very far away and very inconsistent with each other, $\sigma_{eq}\rightarrow1$. The range of \textit{discordance} is $[0,1]$, and it is compared to $1$ to measure the level of inconsistency.

Reference \cite{2015-discordance-MacCrann-etal} also discussed the effect of marginalization on the value of \textit{discordance}. But instead of decreasing after marginalization, the \textit{discordance} tends to increase. In their Appendix A (in Ref.\,\cite{2015-discordance-MacCrann-etal}) they showed a simple example of 2D toy experiments. Both of their toy experiments provide uncorrelated parameter constraints. While the two experiments give different mean values of parameter $x$, the means of $y$ are the same. After they marginalize over $y$, the \textit{discordance} increases. This is different from the behavior of IOI and what we have shown graphically in Sec.\,\ref{subsection-marginalization-and-IOI}. We have argued that marginalizing over inconsistency-irrelevant parameters\footnote{Parameters that have the same mean values and do not correlate with other parameters, or unshared parameters} does not affect the study of inconsistency between two experiments. In their example the two mean values of the $y$ are the same, and there is no correlation between $x$ and $y$, marginalizing over $y$ should not change the inconsistency between the two toy experiments or the measure representing it. It was suggested in Ref.\,\cite{2015-discordance-MacCrann-etal} that, for a model with $N$ dimensions, it is the full $N$-dimensional $\sigma_{eq}$ that should be used for a conservative measure of discordance. However, if two experiments have inconsistent results on only one parameter and if all parameters are uncorrelated, a model with a higher dimension will have a smaller \textit{discordance}. In that case, a very high-dimensional model will have a very low \textit{discordance}. The full-parameter \textit{discordance} actually underestimates the inconsistency in that case.

Unlike for the other measures, we could not obtain an explicit expression of the \textit{discordance}. However, there is a simple algebraic procedure to calculate it in Gaussian and weak prior limit that we provide below. When $|P|\rightarrow0$ so that the prior can be treated as a constant in the parameter range of interest, the integrated probability included within the iso-likelihood contour of the same $\Delta\chi^2$ is given by
\begin{equation}\label{eq-P-DeltaChi2}
P_N(<\Delta\chi^2)=\frac{\gamma(\frac{N}{2},\frac{\Delta\chi^2}{2})}{\Gamma(\frac{N}{2})}\,,
\end{equation}
where $\Gamma$ is the gamma function and $\gamma$ is the lower incomplete gamma function. The overlapping iso-likelihoods with minimum $\sigma_{eq}$ will be tangentially touching with each other at point $\lambdaB_{eq}$. The gradients of two iso-likelihoods at $\lambdaB_{eq}$ must be of opposite directions, that is
\begin{equation}\label{eq-iso-likelihood-gradient}
(1-x)\frac{\partial\Delta\chi^2_{(1)}}{\partial\lambdaB_{eq}}=-x\frac{\partial\Delta\chi^2_{(2)}}{\partial\lambdaB_{eq}}\,,
\end{equation}
where $x\in[0,1]$ is a number we need to solve for. In Gaussian cases with $\Delta\chi^2_{(i)}=(\lambdaB-\bm{\mu^{(i)}})^T\bm{L^{(i)}}(\lambdaB-\bm{\mu^{(i)}})$, we can get $\lambdaB_{eq}$ as a function of $x$ from Eq.\,\eqref{eq-iso-likelihood-gradient}, that is
\begin{equation}\label{eq-iso-likelihood-gradient-reduce}
\lambdaB_{eq}=\frac{1}{(1-x)\bm{L^{(1)}}+x\bm{L^{(2)}}}\big((1-x)\bm{L^{(1)}}\bm{\mu^{(1)}}+x\bm{L^{(2)}}\bm{\mu^{(2)}}\big)\,.
\end{equation}
Equation \eqref{eq-iso-likelihood-gradient-reduce} can be viewed as a modified-Fisher-matrix-weighted average. There are three special cases: $\lambdaB_{eq}=\bm{\mu^{(1)}}$ when $x=0$; $\lambdaB_{eq}=\muB$ when $x=0.5$; and $\lambdaB_{eq}=\bm{\mu^{(2)}}$ when $x=1$. By definition we have
\begin{equation}\label{eq-lambda-eq-definition}
\Delta\chi^2_{(1)}(\lambdaB_{eq})=\Delta\chi^2_{(2)}(\lambdaB_{eq})\,.
\end{equation}
We can first solve Eq.\,\eqref{eq-lambda-eq-definition} for $x$, substitute $x$ in Eq.\,\eqref{eq-iso-likelihood-gradient-reduce} to get $\lambdaB_{eq}$, and substitute $\lambdaB_{eq}$ in Eq.\,\eqref{eq-P-DeltaChi2} to get
\begin{equation}\label{eq-sigma-eq}
\sigma_{eq}=P_N(<\Delta\chi^2(\lambdaB_{eq}))\,.
\end{equation}
We use this algorithm to make corresponding calculations further in the paper.

\begin{figure*}[t]
\centering
\includegraphics[width=0.49\textwidth]{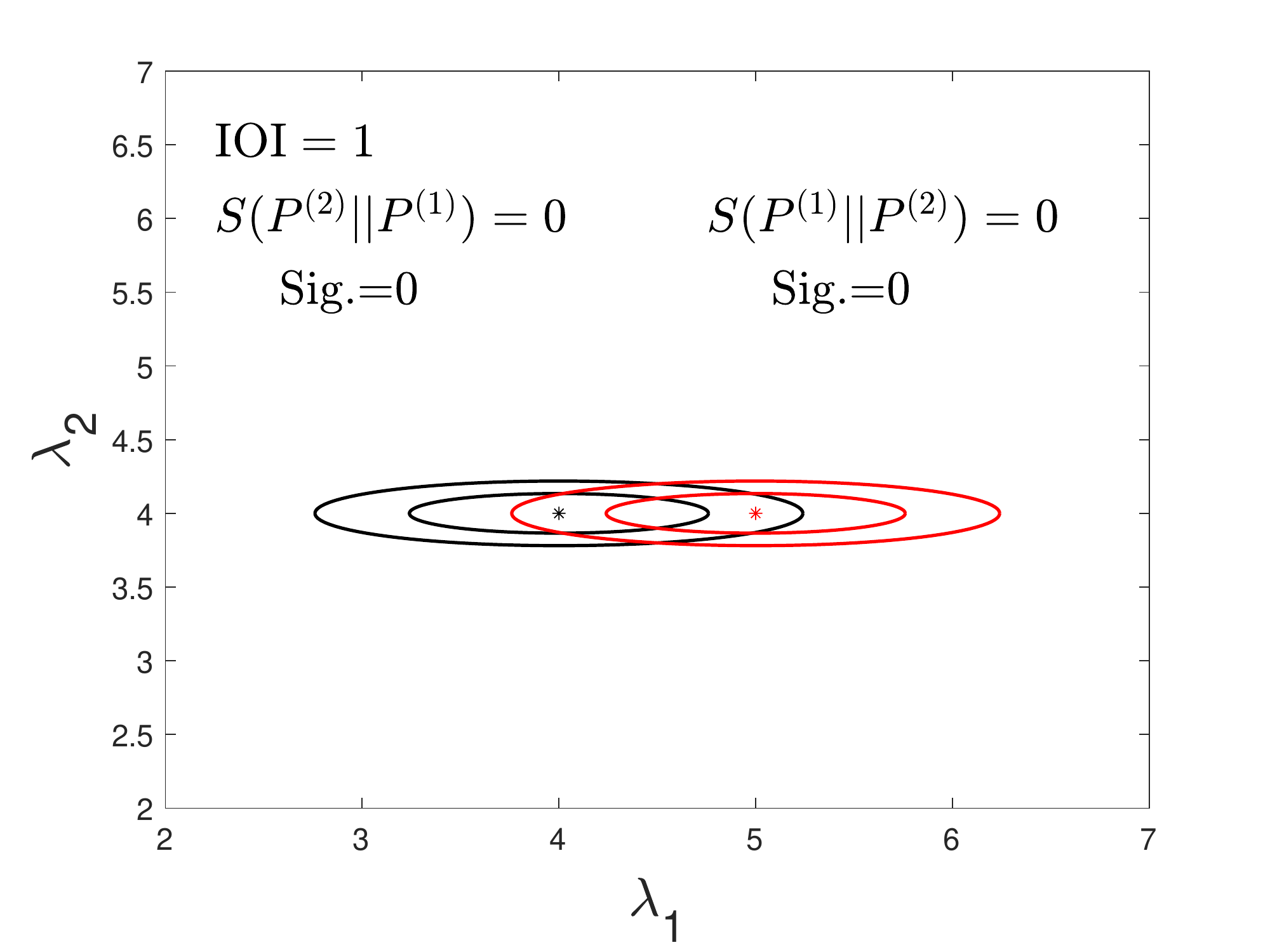}
\includegraphics[width=0.49\textwidth]{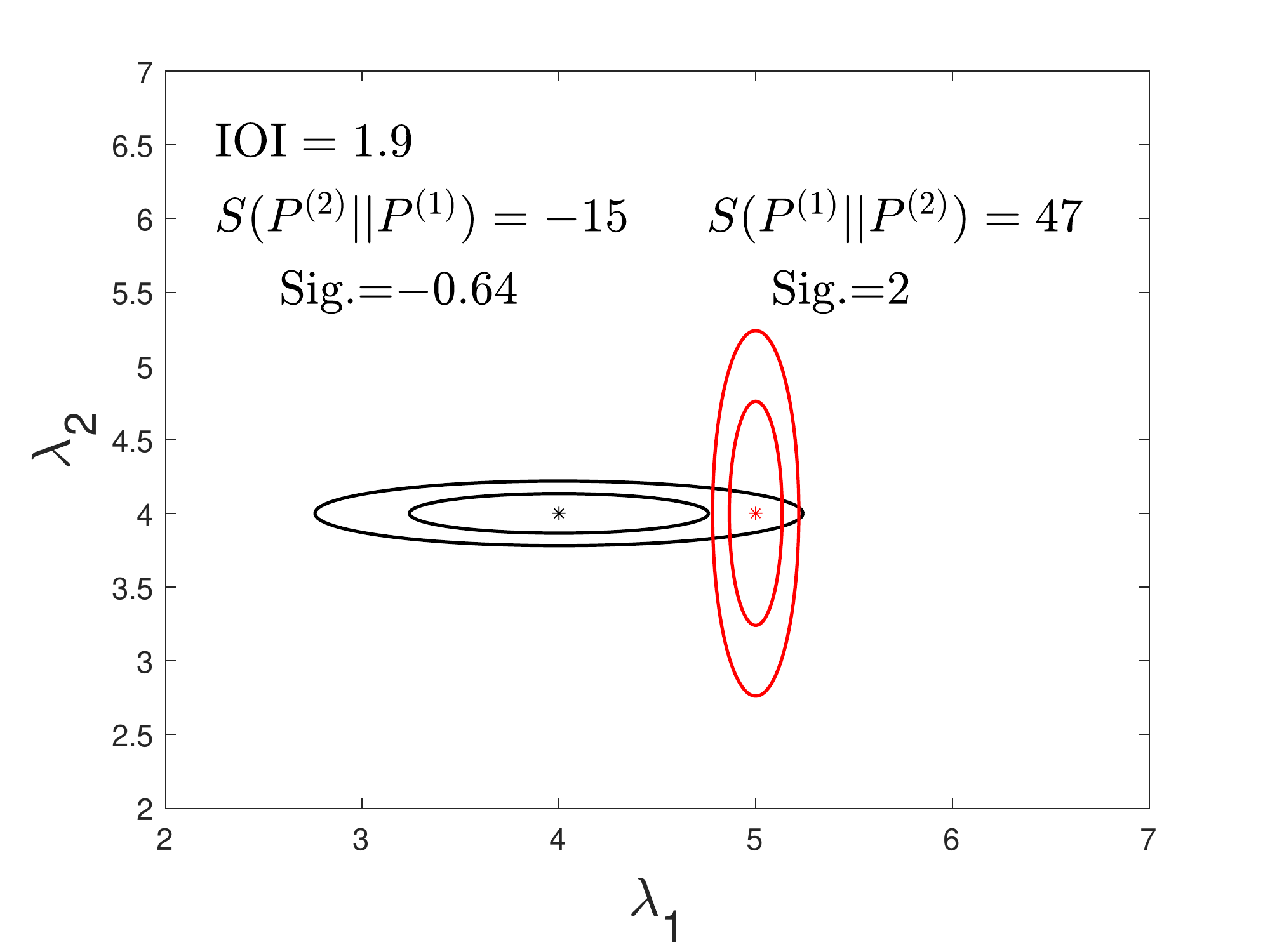}
\caption[A problem of the \textit{surprise} - incorrect description]{\label{figure-surprise-toys}Two sets of toy experiments described by Eqs. \eqref{surprise-toy-experiment-1-mean-diff}-\eqref{surprise-toy-experiment-1} (left panel) and Eqs. \eqref{surprise-toy-experiment-2-mean-diff}-\eqref{surprise-toy-experiment-2} (right panel). Experiment 1 is described by contours in black, and experiment two in red. The two experiments in the right panel seems to have a larger inconsistency, which is correctly described by the increase of IOI.  Using different argument orders in the \textit{surprise} leads to opposite conclusions about the inconsistency in the right panel.}
\end{figure*}

\subsection{The \textit{surprise}}\label{subsection-surprise}

Based on relative entropy (or Kullback-Leibler divergence) in information theory, a measure, called \textit{surprise}, was introduced in Refs. \cite{2014-rel-entropy-Seehars-etal,2016-quantify-concor-Seehars-etal,2016-Grandis-information-Gains}. Reference \cite{2017-GB-Zhao-etal-BAO-surprise} used this measure to study the tension between different distance indicators. For two posterior probability distributions, this quantity $S(\mathscr{P}^{(2)}||\mathscr{P}^{(1)})$ is defined as the difference between the relative entropy $D(\mathscr{P}^{(2)}||\mathscr{P}^{(1)})$ and its average $\langle D\rangle_{Q_2}$
\begin{align}
D(\mathscr{P}^{(2)}||\mathscr{P}^{(1)})&=\int d^N\lambda \mathscr{P}^{(2)}(\lambdaB)\ln\big(\tfrac{\mathscr{P}^{(2)}(\lambdaB)}{\mathscr{P}^{(1)}(\lambdaB)}\big)\,,\\
S(\mathscr{P}^{(2)}||\mathscr{P}^{(1)})&=D(\mathscr{P}^{(2)}||\mathscr{P}^{(1)})-\langle D\rangle_{Q_2}\,.\label{surprise-definition}
\end{align}
The average above is taken over a probability distribution $P(\bm{Q^{(2)}}|\bm{Q^{(1)}})$ of the second data $\bm{Q^{(2)}}$ given the first data $\bm{Q^{(1)}}$:
        \begin{align}
        \langle D\rangle_{Q_2}&=\int d \bm{Q^{(2)}}~D\times P(\bm{Q^{(2)}}|\bm{Q^{(1)}})\,,\label{eq-def-average-over-data}\\
        P(\bm{Q^{(2)}}|\bm{Q^{(1)}})&=\int d^N\lambda \mathcal{L}(\bm{Q^{(2)}}|\lambdaB)\mathscr{P}(\lambdaB|\bm{Q^{(1)}})\,.
        \end{align}
The uncertainty of $D$ can also be calculated,
\begin{equation}
\sigma^2(D)=\langle D^2\rangle_{Q_2}-\langle D\rangle^2_{Q_2}\,.
\end{equation}
And the significance of $S(\mathscr{P}^{(2)}||\mathscr{P}^{(1)})$ is defined as
\begin{equation}\label{eq-surprise-significance-definition}
{\rm significance~of~}S=\frac{S(\mathscr{P}^{(2)}||\mathscr{P}^{(1)})}{\sigma(D)}\,.
\end{equation}

\begin{figure*}[t]
       \centering
       \includegraphics[width=0.49\textwidth]{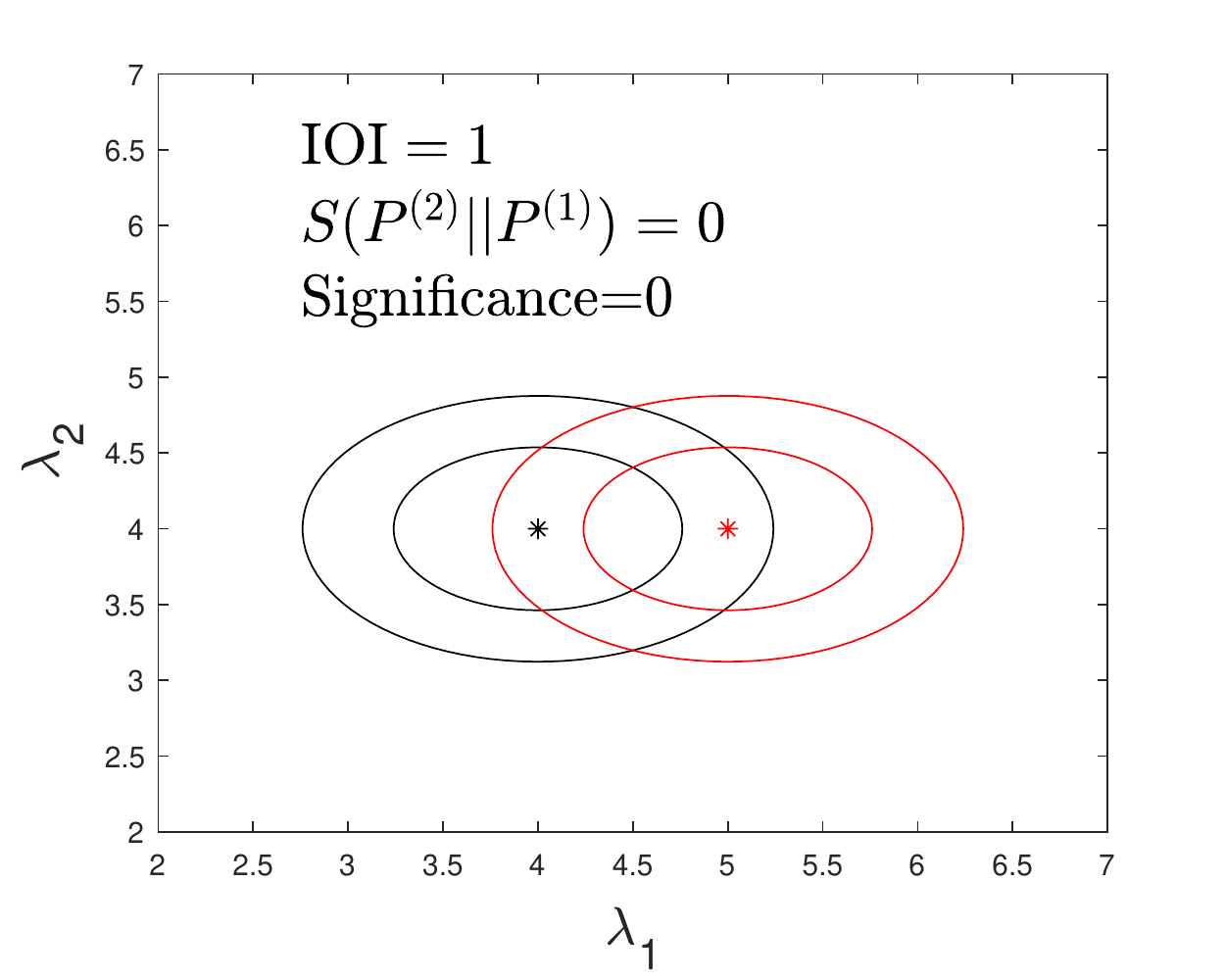}
       \includegraphics[width=0.49\textwidth]{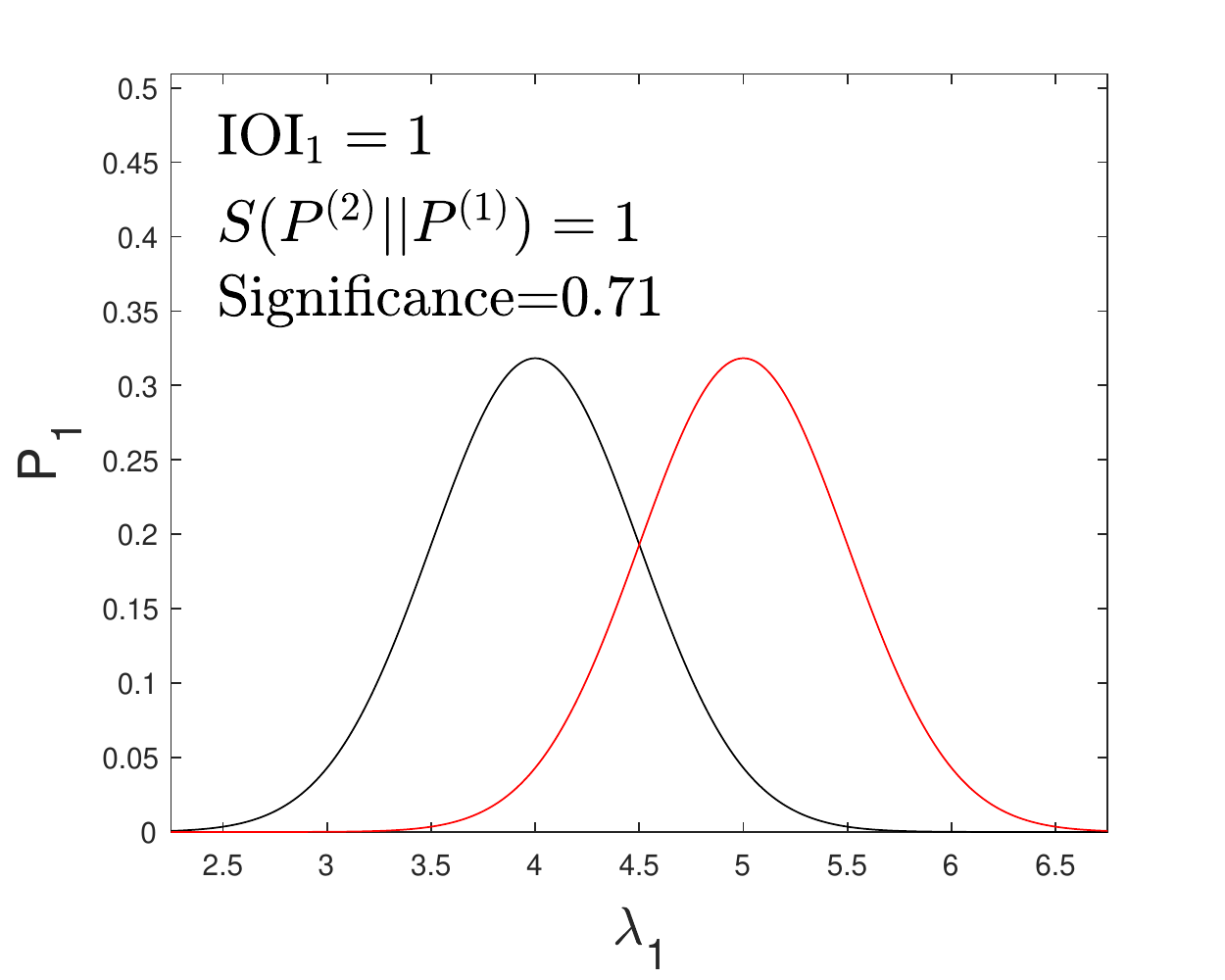}
       \caption[A problem of the \textit{surprise} - marginalization]{\label{surprise-marginalization} The posterior probability contours of the two experiments described by \eqref{surprise-toy-experiment-3} (left panel) and their marginalized $1$D distribution over $\lambda_2$ (right panel). The \textit{surprise} before marginalization is $S(\mathscr{P}^{(2)}||\mathscr{P}^{(1)})=0$, but after marginalization the \textit{surprise} is $S_1=1$ with a significance $0.7$. However, we argued in Sec.\,\ref{subsection-marginalization-and-IOI} that $\lambda_2$ should not be relevant to the inconsistency, a measure of the inconsistency should be preserved after marginalization over $\lambda_2$.}
        \end{figure*}

        Two different cases are discussed in Ref.\,\cite{2014-rel-entropy-Seehars-etal}: one is called ``updating data'' where the second experiment is used to updated the constraints from the first experiments as a prior; the other is called ``replacing data'' where two experiments are used separately to set constraints on parameters. Since we are considering the inconsistency between two experiments with similar constraining powers, it is the second case that is more relevant to our work. The \textit{surprise} and $\sigma^2(D)$ for the second case in the Gaussian limit is given in Ref.\,\cite{2016-quantify-concor-Seehars-etal}, which in our notation read
        \begin{equation}\label{eq-surprise-Gaussian}
        \begin{split}
        &S(\mathscr{P}^{(2)}||\mathscr{P}^{(1)})\\&\xrightarrow{\rm Gaussian}\tfrac{1}{2}\bar{\deltaB}^T\bm{F^{(1)}} \bar{\deltaB}-\tfrac{1}{2}{\rm{tr}}(\bm{I_N}+(\bm{F^{(2)}})^{-1}\bm{F^{(1)}}) \\
        &\xrightarrow{\rm{weak ~ prior}}\tfrac{1}{2}\deltaB^T \bm{L^{(1)}}\deltaB-\tfrac{1}{2}{\rm{tr}}(\bm{I_N}+(\bm{L^{(2)}})^{-1}\bm{L^{(1)}})\,,
        \end{split}
        \end{equation}
        \begin{equation}\label{eq-sigma-surprise-Gaussian}
           \sigma^2(D)=\tfrac{1}{2}{\rm{tr}}\big[(\bm{I_N}+(\bm{F^{(2)}})^{-1}\bm{F^{(1)}})^2\big]\,,
         \end{equation}
        where $\bm{I_N}$ is the $N$-dimensional identity matrix. The relative entropy (and the \textit{surprise}) is divided by $\ln(2)$ when being measured in the unit of bits. For example, if the second experiment only gives a narrower constraint than the first one, $D(\mathscr{P}^{(2)}||\mathscr{P}^{(1)})=1~{\rm{bit}}$ roughly means that the second experiment improves and shrinks the constraint to half compared to the first experiment .

        Despite its interesting concept, the outcome of the \textit{surprise} $S$ seems to depend on the choice of its argument order leading to some ambiguity. For example, let us consider two sets of toy experiments described by
        \begin{align}
                 &\rm{Left~panel~in~Fig.~\ref{figure-surprise-toys}:}~ \bm{\bar{\mu}^{(2)}}-\bm{\bar{\mu}^{(1)}}=\begin{pmatrix} 1\\0\end{pmatrix}\,,\label{surprise-toy-experiment-1-mean-diff}\\
                 &\qquad\bm{F^{(1)}}=\begin{pmatrix} 4 & 0 \\ 0 & 128\end{pmatrix}\,,~\bm{F^{(2)}}=\begin{pmatrix} 4 & 0 \\ 0 & 128\end{pmatrix}\,,\label{surprise-toy-experiment-1}\\
                 &\rm{Right~panel~in~Fig.~\ref{figure-surprise-toys}:}~\bm{\bar{\mu}^{(2)}}-\bm{\bar{\mu}^{(1)}}=\begin{pmatrix} 1\\0\end{pmatrix}\,,\label{surprise-toy-experiment-2-mean-diff}\\
                 &\qquad\bm{F^{(1)}}=\begin{pmatrix} 4 & 0 \\ 0 & 128\end{pmatrix}\,,~\bm{F^{(2)}}=\begin{pmatrix} 128 & 0 \\ 0 & 4\end{pmatrix}\,. \label{surprise-toy-experiment-2}
        \end{align}
These two sets of experiments are shown in Fig.\,\ref{figure-surprise-toys}.

The contour plots indicate that the two experiments shown on the left panel seem to have a smaller inconsistency than those on the right panel. This is, for example, reflected quantitatively on the respective values of IOI with the right value (i.e. 1.9) being larger than the left one (i.e. 1.0). However, if $S(\mathscr{P}^{(2)}||\mathscr{P}^{(1)})$ is used, it decreases from $0$ on the left to $-15$ (or $-21.6$ bits) on the right, implying a decrease of inconsistency. The significance of $S(\mathscr{P}^{(2)}||\mathscr{P}^{(1)})$ also has the same implication: $0$ for the left panel and $-0.64$ for the right one. Now, since the \textit{surprise} is not symmetric about the two experiments, one may choose a different argument order in $S$. So, if we instead use $S(\mathscr{P}^{(1)}||\mathscr{P}^{(2)})$ (i.e., switching the argument order), the \textit{surprise} and its significance changes both from $0$ on the left panel to $\sim47$ and $\sim2$ respectively on the right panel. It is unclear what should be a good rationalization to choose the order of arguments in $S$.
It could be argued that the choice should be physically motivated favoring the more trusted data set with less expected systematic uncertainties, thus with a prior preference of a given data set which could be viewed as consistent with a Bayesian statistics approach \cite{private1}. It remains unclear though what is the right choice if both data sets have their own significant systematic errors or when the inconsistency is rooted into a problem with the model or the theory itself.

Finally, the \textit{surprise} and its significance are found to increase after parameter marginalization in, for example, the following toy experiments:
       \begin{equation}\label{surprise-toy-experiment-3}
                 \bm{\bar{\mu}^{(2)}}-\bm{\bar{\mu}^{(1)}}=\begin{pmatrix} 1\\0\end{pmatrix}\,,~~~ \bm{F^{(1)}}=\bm{F^{(2)}}=\begin{pmatrix} 4 & 0 \\ 0 & 8\end{pmatrix}\,,
        \end{equation}
which are shown also in Fig.\,\ref{surprise-marginalization}. We argued in Sec.\,\ref{subsection-marginalization-and-IOI} that $\lambda_2$ should not be relevant to the inconsistency, because the two experiments have the same mean of $\lambda_2$ and there is no correlation between $\lambda_1$ and $\lambda_2$. So after marginalizing over $\lambda_2$ a measure of inconsistency should be preserved, as is the case for IOI. As shown in Fig.\,\ref{surprise-marginalization}, before marginalizing over $\lambda_2$, the \textit{surprise} $S(\mathscr{P}^{(2)}||\mathscr{P}^{(1)})$ (and its significance) is $0$. After marginalizing over $\lambda_2$, it becomes $1$ with a significance of $0.7$.
It can be argued this is a normal behaviour for the \textit{surprise} because the results of $\lambda_2$ are more consistent than expected \cite{private1}. So marginalizing over $\lambda_2$ emphacises the inconsistency in $\lambda_1$ and that is reflected by the increase of the marginalized \textit{surprise} \cite{private1}.
But then when going from the right to the left panel, the inconsistencies as indicated by the values of the \textit{surprise} and its significance on the left seem to go away.
It is unclear if this is really the case since the inconsistency in $\lambda_1$ is still present.

    \subsection{The calibrated evidence ratio}\label{subsection-CER}
Reference\,\cite{2016-tensions-Grandis-etal} introduced the calibrated evidence ratio (CER hereafter) by calibrating the \textit{robustness} using its average. The level of CER can also be described by its significance. But different from the case of the \textit{surprise}, the average here is taken over a probability distribution of both the first and the second data $\bm{Q^{(1)}}$ and $\bm{Q^{(2)}}$, which is given by the joint evidence $E(\bm{Q^{(1)}},\bm{Q^{(2)}})$,
        \begin{align}
        {\rm{CER}}&=\ln R-\langle\ln R\rangle_{Q_1,Q_2}\,,\label{CER-definition}\\
        \sigma^2(R)&=\big\langle\big(\ln R-\langle \ln R\rangle\big)^2 \big\rangle_{Q_1,Q_2}\,.\label{CER-sigma}
        \end{align}
        In Gaussian cases, CER in $N$ dimensions is given by
        \begin{align}
        \begin{split}
        -{\rm{CER}}&=\deltaGhalfbar -\tfrac{N}{2}\\
        &\xrightarrow{\rm weak~prior}\deltaGhalf -\tfrac{N}{2}\\
        &={\rm{IOI}}-\tfrac{N}{2}\,,
        \end{split}\label{eq-CER-Gaussian}\\
        \sigma(R)&=\sqrt{N/2}\,.\label{eq-sigma-CER}
        \end{align}
        So the opposite of CER is related to IOI in the Gaussian and weak prior limit by an additive constant, $-N/2$.

The CER measure did address some of the issues with its predecessors. It is worth noting though that since IOI is positive definite, $-N/2$ is the minimum value of $-$CER. Because of the term $-N/2$, $-$CER can be be more negative with higher dimensional parameter spaces. The significance defined as $-$CER$/\sigma(R)$ goes as $-\sqrt{N/2}$ when IOI $=0$. The term $-N/2$ (and $-\sqrt{N/2}$ for its significance) seems to suggest that a model with a higher dimension tends to have a smaller inconsistency. It can be argued that this is normal because a higher dimensional model allows larger scatter of the data, which leads to a larger expected $- \ln R$ \cite{private1}.
Consider though the following situation: if two experiments have inconsistency on only one parameter (e.g., $\lambda_1$) and if all parameters are uncorrelated, a model with a higher dimension will have a smaller $-$CER. This seems to underestimate the inconsistency that is due to one particular parameter.

\begin{table*}[tp]
\caption[Measures of discordance in Gaussian and weak prior limit]{\label{table-measures}Table of measures of discordance in the Gaussian and weak prior limit considered in this work. $\deltaB=\bm{\mu^{(1)}}-\bm{\mu^{(2)}}$ and $\bm{G}=\big((\bm{L^{(1)}})^{-1}+(\bm{L^{(2)}}\big)^{-1}\big)^{-1}$. We take the opposite of the \textit{robustness}, the normalized \textit{robustness} and the CER to be the measures of inconsistency, since they are originally defined to be measures of consistency.}
    \begin{ruledtabular}
    \begin{tabular}{lccc}
    Quantities & Symbols  & Relevant concept & Gaussian and weak prior limit\\
    \hline
    Index of inconsistency & IOI & $\Delta\chi^2_{(1)}+\Delta\chi_{(2)}^2$ & $\tfrac{1}{2}\deltaB^T\bm{G}\deltaB$ ~~(definition)\\
    \textit{Robustness} \cite{2006-Marshall-etal-Bayesian,2011Robustness-March-etal} & $-\ln R$ & Bayesian evidence ratio& $\rm{IOI}+\tfrac{1}{2}\ln\Big(\frac{|\mathbf{L}\mathbf{P}|}{|\mathbf{L^{(1)}}\mathbf{L^{(2)}}|}\Big)$\\
    Normalized \textit{robustness} \cite{2011Robustness-March-etal} & $-\ln R_N$ & Bayesian evidence ratio & $\rm{IOI}+\tfrac{1}{2}\ln\Big(\frac{|\mathbf{L^{(1)}}+\mathbf{L^{(2)}}|}{|2\mathbf{L^{(2)}}|}\Big)$\\
    \textit{Tension} \cite{2013-tension-Verde-etal} & $\ln\mathcal{T}$ & Bayesian evidence ratio & $\rm{IOI}$\\
    \textit{Discordance} \cite{2015-discordance-MacCrann-etal}  & $\sigma_{eq}$ & Iso-likelihood contour & Algorithm provided by Eqs.\,\eqref{eq-iso-likelihood-gradient-reduce}-\eqref{eq-sigma-eq}\\
    \textit{Surprise} \footnote{The case of replacing data described in Ref.\,\cite{2014-rel-entropy-Seehars-etal} is used in this work.}  \cite{2014-rel-entropy-Seehars-etal,2016-quantify-concor-Seehars-etal,2016-Grandis-information-Gains} & $S(\mathscr{P}^{(2)}||\mathscr{P}^{(1)})$  & Relative entropy & $\tfrac{1}{2}\deltaB^T\bm{L^{(1)}}\deltaB-\tfrac{1}{2}{\rm{tr}}\big(\bm{I_N}+(\bm{L^{(2)}})^{-1}\bm{L^{(1)}}\big)$\\
    \multicolumn{1}{r}{and its deviation}& $\sigma(D)$ & & $\sqrt{\tfrac{1}{2}{\rm{tr}}\big[\big(\bm{I_N}+(\bm{L^{(2)}})^{-1}\bm{L^{(1)}}\big)^2\big]}$\\
    Calibrated evid. ratio \cite{2016-tensions-Grandis-etal} & $-$CER & Bayesian evidence ratio & ${\rm{IOI}}-N/2$\\
    \multicolumn{1}{r}{and its deviation}&$\sigma(R)$& & $\sqrt{N/2}$\\
    Level of concordance \cite{2017MNRAS-Joudaki-etal-DIC-inconsistency} & $\mathcal{G}(\bm{Q^{(1)}}, \bm{Q^{(2)}})$ & Deviance information criterion & ${\rm IOI}-N/2$
    \end{tabular}
    \end{ruledtabular}

    \caption[Comparison of Measures]{\label{table-Comparison-of-measures}Comparison of properties of the measures of discordance considered in this work in Gaussian and weak prior limit. In Gaussian and weak prior limit, $\ln \mathcal{T}$ = IOI and $\tfrac{1}{2}\mathcal{G}(\bm{Q^{(1)}},\bm{Q^{(2)}})$ = $-$CER. By saying whether or not a measure reflects the effects of marginalization, we mean if the measure can describe the hidden inconsistency after parameter marginalization. For measures like \textit{robustness} and normalized \textit{robustness} $-$CER, this column does not apply (N/A). }
      \begin{ruledtabular}
\begin{tabular}{lccccc }
        Quantities & Symmetric? & \multicolumn{1}{b{0.1\textwidth}}{Follow \newline inconsistency?} & \multicolumn{1}{b{0.07\textwidth}}{Give 0\newline if $\deltaB=\bm{0}$?} & \multicolumn{1}{b{0.13\textwidth}}{Reflect effects of\newline marginalization?} & Proposed interpretation\\ \hline
        IOI & \cmark & \cmark & \cmark & \cmark  & Jeffreys' scales\\
        \textit{Robustness} & \cmark & \xmark & \xmark & N/A &  Jeffreys' scales\\
        Normalized $\ln R_N$ & \xmark & \xmark & \xmark & \xmark &  Not specified\\
        \textit{Discordance} & \cmark & \cmark & \cmark & \xmark &  Compared to $100\%$\\
        \textit{Surprise} & \xmark & \textbf{?}\footnote{Due to the asymmetric behavior, if the \textit{surprise} has the wrong choice of argument order, the trend will be incorrect.} & \xmark & \xmark &  Bits $|$ significance\\
        $-$CER & \cmark & \cmark & \xmark & N/A &  Significance\\
      \end{tabular}
      \end{ruledtabular}
\end{table*}

        It is also worth mentioning that from Eq.\,\eqref{eq-CER-Gaussian} we can see, in the Gaussian and weak prior limit, $-$CER has a close relation with the Akaike information criterion \cite{Akaike-criterion} for the following reason. The Akaike information criterion (AIC) is to select a model with a smaller AIC$=-2\ln \mathcal{L}+2N$. For a model with $N$ parameters, the corresponding splitting model has $2N$ parameters. The AIC difference between the combined and the split models is
\begin{equation}\label{eq-AIC-CER}
\begin{split}
\tfrac{1}{2}\Delta{\rm AIC}&=\tfrac{1}{2}{\rm AIC_{comb}}-\tfrac{1}{2}{\rm AIC_{spl}}\\
&={\rm IOI}-\tfrac{1}{2}N\\
&=-{\rm CER}\,.
\end{split}
\end{equation}

\subsection[Level of concordance]{Level of concordance}\label{subsection-DIC-inconsistency}
Analogous to the \textit{robustness}, Ref.\,\cite{2017MNRAS-Joudaki-etal-DIC-inconsistency} most recently introduced the \textit{level of concordance}, defined as $\exp\big[-\tfrac{1}{2}\mathcal{G}(\bm{Q^{(1)}},\bm{Q^{(2)}})\big]$, to measure experimental (in)consistency. Instead of using the Bayesian evidence ratio between the combined and the split models, $\mathcal{G}(\bm{Q^{(1)}},\bm{Q^{(2)}})$ uses the deviance information criterion (DIC) (see Refs.\,\cite{2002-Spiegelhalter-etal-DIC,2007-Liddle-IC-for-astroph}), and it is defined as
\begin{equation}\label{eq-G-Delta-DIC}
\mathcal{G}(\bm{Q^{(1)}},\bm{Q^{(2)}})\equiv{\rm DIC_{comb}}-{\rm DIC_{spl}}\,,
\end{equation}
where DIC$= \chi^2_{min}+2p_D$, and $p_D$ is the effective number of parameters given by $p_D=\langle\chi^2\rangle-\chi^2_{min}$. Jeffreys' scale was used to interpret this measure.

Since DIC approaches AIC in the Gaussian limit (when parameters are well constrained) \cite{2007-Liddle-IC-for-astroph}, $\mathcal{G}$ is equivalent to $-$CER (see Sec.\,\ref{subsection-CER} for the connection between $-$CER and AIC) in Gaussian limit, i.e.,
\begin{equation}\label{eq-G-CER-IOI}
\tfrac{1}{2}\mathcal{G}(\bm{Q^{(1)}},\bm{Q^{(2)}})=-{\rm CER}={\rm IOI}-\tfrac{N}{2}\,.~~(\rm Gaussian)
\end{equation}
Therefore, \textit{level of concordance} measure seams to share the points mentioned above for the $-$CER.

\subsection[Relation between IOI and the \textit{Surprise}]{Is there a connection between IOI, \textit{Surprise} and $-$CER}\label{subsection-connection-S-IOI}
The definitions of the \textit{robustness} $\ln(R)$, \textit{tension} $\mathcal{T}$, \textit{surprise} $S$, $-$CER and IOI are quite different. But we have seen in the previous sections that $\ln(R)$, $\mathcal{T}$ and $-$CRE are related to IOI by equations \eqref{robustness-Gaussian-weak} \& \eqref{robustness-normalized}, \eqref{eq-tension-Gaussian} and \eqref{eq-CER-Gaussian}. Is there a relation between the \textit{surprise} $S$ and IOI? We find that only in one dimension the significance of the \textit{surprise} is simply related to IOI, and to the significance of $-$CER.

In one dimension with Gaussian and weak prior, the \textit{surprise} given by Eq.\,\eqref{eq-surprise-Gaussian} reduces to
\begin{equation}
S(\mathscr{P}^{(2)}||\mathscr{P}^{(1)})=\tfrac{1}{2}\delta^2/{\sigma_{(1)}^2}-\tfrac{1}{2}(1+\sigma_{(2)}^2/\sigma_{(1)}^2),
\end{equation}
the deviation of reduces to
\begin{equation}
\sigma(D(\mathscr{P}^{(2)}||\mathscr{P}^{(1)}))=\sqrt{\tfrac{1}{2}}(1+\sigma_{(2)}^2/\sigma_{(1)}^2),
\end{equation}
and the significance is given by
\begin{equation}\label{eq-significance-S-1D}
\begin{split}
{\rm{significance~of~}}S&=\frac{S(\mathscr{P}^{(2)}||\mathscr{P}^{(1)})}{\sigma(D(\mathscr{P}^{(2)}||\mathscr{P}^{(1)}))}\\
&= \sqrt{2}\Big[\frac{1}{2}\frac{\delta^2}{(\sigma_{(1)}^2+\sigma_{(2)}^2)}-\frac{1}{2}\Big]\,.
\end{split}
\end{equation}
In one dimension, the significance of the \textit{surprise} becomes symmetric about two experiments. IOI in this case is given by $\frac{1}{2}\frac{\delta^2}{(\sigma_{(1)}^2+\sigma_{(2)}^2)}$, and  from \eqref{eq-CER-Gaussian} and \eqref{CER-sigma} we can obtain the significance of $-$CER in this case as $\sqrt{2}\big({\rm{IOI}}-\frac{1}{2}\big)$, for $N=1$. Therefore the three quantities are related by
\begin{equation}\label{eq-S-CER-IOI}
{\rm{signif.~of~}}S={\rm{signif.~of~(-CER)}}=\sqrt{2}\big({\rm{IOI}}-\tfrac{1}{2}\big)\,.
\end{equation}

In $N$ dimensions, due to the connection of $-$CER to AIC, the relation between IOI and $-$CER (and its significance) remains simple in the Gaussian and weak prior limit, and it is given by Eq.\,\eqref{eq-CER-Gaussian}, and
\begin{equation}\label{eq-signif-cer-ioi}
{\rm significance~of~(-CER)}=\sqrt{\tfrac{2}{N}}\big({\rm{IOI}}-\tfrac{N}{2}\big)\,.
\end{equation}
But it is not obvious to see the relation between IOI and \textit{surprise}. The \textit{surprise} and its significance become asymmetric for two- or multiple-dimensional parameter spaces, while IOI is always symmetrical.

Finally, we summarize the measures of inconsistency surveyed in this section in Table \ref{table-measures} and Table \ref{table-Comparison-of-measures}.

\section[Application of IOI]{Application of IOI to cosmological experiments: geometry vs growth}\label{section-application}
To test the internal consistency of the $\Lambda$CDM and $w$CDM models, we fit each of them separately to a set of geometry experiments and a set of growth experiments. This is partly motivated by the fact that different theories (e.g. modified gravity) have  relations between the cosmic expansion background and the large scale structure growth, e.g. \cite{2006-Ishak-splitting}. We use the publicly available Markov Chain Monte Carlo code \textsc{CosmoMC} \cite{Cosmomc} for parameter constraints. Note that, in this work, we will not use IOI for multiple experiments [see Ref.\,\eqref{eq-IOI-definition-general} to study the full experimental inconsistency among them. Rather, we will treat the geometry experiments as one set of experiments and the growth experiments as the other set, and then apply IOI to measure the (in)consistency between the two set of experiments [Eq.\,\eqref{eq-IOI-definition}].

\begin{table*}[htp!]
\caption[List of geometry and growth experiments]{\label{table-experiments} Summary of geometry and growth experiments/data used in testing the model consistency. }
\begin{ruledtabular}
\begin{tabular}{|c|c|c|c|}
\multicolumn{2}{|c|}{Geometry}&\multicolumn{2}{c|}{Growth}\\
\hline
\multicolumn{2}{|c|}{Supernovae Type Ia \cite{2014supernova740} }& \multicolumn{2}{c|}{ Planck 2015 low-$\ell$ CMB temperature and polarization \cite{Planck2015XIII-Cos.Param.} }\\
\hline
\multirow{3}{*}{~BAO~} & Six Degree Field Glactic Survey (6dF) $(z_{eff}=0.106)$ \cite{2011BAO-6df}& \multicolumn{2}{c|}{Planck 2015 CMB lensing \cite{Planck2015XV-lensing}}\\
\cline{2-4}
 & SDSS main galaxy sample (MGS) $(z_{eff}=0.15)$ \cite{2015BAO-sdss-mgs} &  \multicolumn{2}{c|}{Sunyaev--Zel'dovich effect \cite{2015Planck-SZ-cluster-count}}  \\
\cline{2-4}
 &SDSS quasar-Lyman-$\alpha$ forest $(z_{eff}=2.34)$ \cite{2015bao-sdssIII} & \multicolumn{2}{c|}{CFHTlens galaxy weak lensing \cite{2013CFHTlens}}\\
\hline
\multicolumn{2}{|c|}{Planck 2015 High-$\ell$ CMB temperature \cite{Planck2015XIII-Cos.Param.}} &\multirow{2}{*}{~RSD~} & WiggleZ Dark Energy Survey \cite{2010WiggleZ-MPK,2012WiggleZ-MPK}\\
\cline{4-4}
\multicolumn{2}{|c|}{} & &SDSS DR12 CMASS and LOWZ catalogs \cite{2015sdss-dr12}~ \\
\end{tabular}
\end{ruledtabular}
\end{table*}

\begin{figure*}[!tbp]
\includegraphics[width=\textwidth,height=0.8\textwidth]{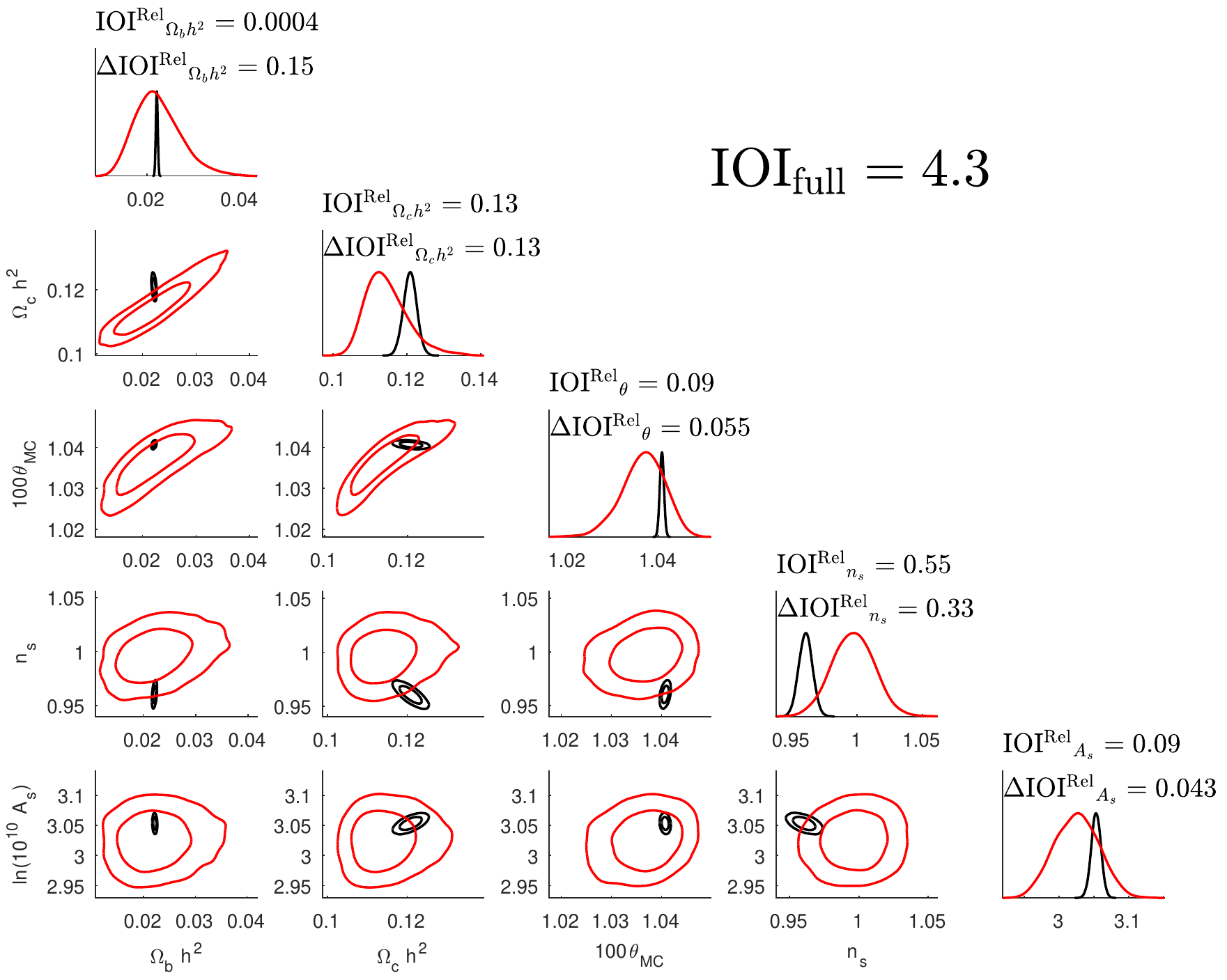}
\caption[Geometry and growth constraints for $\Lambda$CDM]{\label{fig-gg-l-triagular} The marginalized 1D probability distributions and 2D likelihood contour plots for the $\Lambda$CDM model constrained by the geometry versus the growth data sets. Black for the geometry set and red for the growth set. The two relative measures of one-parameter inconsistency (IOI$_i^{\rm Rel}$ and $\Delta$IOI$_i^{\rm Rel}$, see Eq.\,\eqref{eq-relative-IOI} for their definitions) are shown on top of each 1D marginalized plot. \textit{While these relative IOI measures can indicate the parameter(s) most associated with the inconsistency, \textbf{the important number here is the value of the full IOI} shown at the upper right corner of the figure}. Note that summation of all individual IOI$^{\rm Rel}_i$'s or $\Delta$IOI$^{\rm Rel}_i$'s is generally not $1$. The full IOI between the two data sets is 4.34 and indicates a moderate inconsistency on Jeffreys' scales (see Table III).}
\end{figure*}

{
\begin{table*}[!tbp]
\caption[Measures of inconsistency for $\Lambda$CDM - geometry vs growth]{\label{table-Measures-values} Measures of inconsistency between geometry and growth experiments for $\Lambda$CDM and $w$CDM models. Measures of inconsistency are obtained by using the Gaussian and weak prior limits listed in Table \ref{table-measures}. The weak prior limit of the \textit{robustness} diverges, so we do not calculate it in this table. Quantities $\ln\mathcal{T}$ and $\mathcal{G}(\bm{Q^{(1)}},\bm{Q^{(2)}})$ are not listed here, since they equal IOI and $-$CER respectively in the Gaussian and weak prior limit. The IOI between the two data sets is 4.34 and 5.41 for the $\Lambda$CDM and $w$CDM, respectively. These indicate  moderate and strong inconsistency,  respectively, on Jeffreys' scales. The other measures also indicate an overall moderate to strong  [except $S(\rm gm||gr)$].}
\begin{ruledtabular}
\begin{tabular}{ld|d|d|dd|dd|dd}
Measures & \multicolumn{1}{r}{IOI} & \multicolumn{1}{r}{$-\ln R_N$} & \multicolumn{1}{r}{$\sigma_{eq}$} & \multicolumn{1}{r}{$S({\rm{gr}}||{\rm{gm}})$\footnote{``gm'' $=$ ``geometry'', ``gr'' $=$ ``growth''.}}&\multicolumn{1}{r}{Significance} & \multicolumn{1}{r}{$S({\rm{gm}}||{\rm{gr}})$}& \multicolumn{1}{c}{Sig.} & \multicolumn{1}{r}{$-$CER} & \multicolumn{1}{c}{Sig.} \\
 \hline
 $\Lambda$CDM & \IOInumggl & \RNnumggl & \sigmaeqggl & \Snumggl &\SigSnumggl & \SnumgglP & \SigSnumgglP & \CERnumggl & \SigCERnumggl\\
 $w$CDM & \IOInumggw & \RNnumggw & \sigmaeqggw & \Snumggw & \SigSnumggw & \SnumggwP & \SigSnumggwP & \CERnumggw & \SigCERnumggw
\end{tabular}
\end{ruledtabular}
\end{table*}}

\begin{figure*}[!tbp]
\includegraphics[width=\textwidth,height=0.9\textwidth]{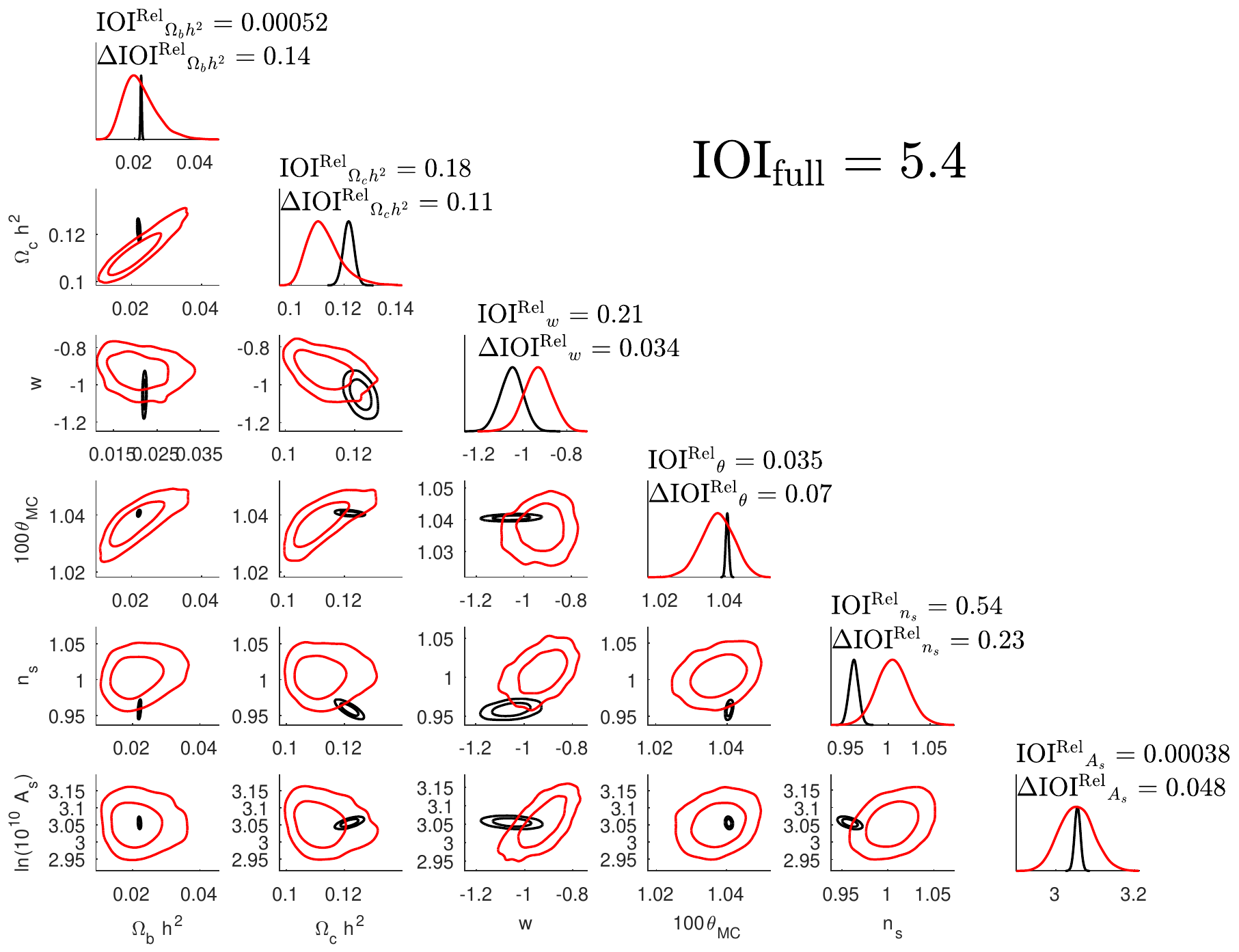}
\caption[Geometry and growth constraints of $w$CDM]{\label{fig-gg-w-triagular} Marginalized 1D probability distributions and 2D likelihood contour plots in a triangular displacing diagram for the $w$CDM model constrained by the geometry and growth sets of experiments. The structure of this figure is the same as Fig.\,\ref{fig-gg-l-triagular}.
\textit{Again, the important number here is the value of the full IOI} shown at the upper right corner of the figure. Note that summation of all individual IOI$^{\rm Rel}_i$'s or $\Delta$IOI$^{\rm Rel}_i$'s is generally not $1$. The full IOI between the two data sets is 5.34 for $w$CDM and indicates a strong inconsistency on Jeffreys' scales (see Table III).
}
\end{figure*}

\subsection{Data sets}
For the geometry set of experiments, we use the $740$ type Ia supernovae catalog compiled in Ref.\,\cite{2014supernova740}, the baryon acoustic oscillation (BAO) standard ruler from the Six Degree Field Glactic Survey (6dF) \cite{2011BAO-6df}, the main galaxy sample (MGS) of data release seven (DR7) from the Sloan Digital Sky Survey (SDSS) and the SDSS quasar-Lyman-$\alpha$ forest \cite{2015bao-sdssIII}, and the high-$\ell$ CMB temperature data from Planck 2015 \cite{Planck2015XIII-Cos.Param.}. For the growth set of experiments, we use the low-$\ell$ CMB temperature and polarization data from Planck 2015 \cite{Planck2015XV-lensing}, CMB lensing \cite{Planck2015XV-lensing}, thermal Sunyaev-Zel'dovich effect \cite{2015Planck-SZ-cluster-count}, galaxy weak lensing cosmic shear from the Canada France Hawaii Lensing Survey (CFHTlens) \cite{2013CFHTlens}, and the redshift space distortion from the WiggleZ Dark Energy Survey (WiggleZ MPK) \cite{2010WiggleZ-MPK,2012WiggleZ-MPK} and the SDSS DR12 CMASS and LOWZ catalogs \cite{2015sdss-dr12}. Those two sets of experiments are summarized in Table \ref{table-experiments}. Note that we put high-$\ell$ CMB temperature data in the geometry set as in Ref.\,\cite{2016Bernal-etal-param-splitting}, since its acoustic peaks strongly constrain the expansion history of the Universe.

From Table \ref{table-experiments} we can see that there are some overlaps between the data sets used in this work and those used in Refs. \cite{2016Bernal-etal-param-splitting,2015Ruiz-etal-param-splitting}. But there are some differences. First, for the RSD data from SDSS, while Ref.\,\cite{2016Bernal-etal-param-splitting} separately used the measurements of $Hr_s$ and $f\sigma_8$ as geometry and growth sets, we use them all in the growth set because these two measurements (as well as $D_A/rs$) are correlated. Second, we use the full data catalog of RSD from WiggleZ, while Ref.\,\cite{2016Bernal-etal-param-splitting} used its compressed information on BAO and growth. Third, we use the combined Lyman-$\alpha$ BAO data at $z_{eff}=2.34$ from Ly$\alpha$ forest autocorrelation and the quasar-Ly$\alpha$ cross-correlation \cite{2015bao-sdssIII}, while Ref.\,\cite{2016Bernal-etal-param-splitting} only used the data from cross-correlation.

\subsection{Results from application to data: A moderate to strong inconsistency}

For the $\Lambda$CDM model, we use the following parameters: $\Omega_bh^2$, $\Omega_ch^2$, $\theta$, $\ln(10^{10}A_s)$, $n_s$, and $\tau$ (for the ($w$CDM) we add $w$). We fit each model separately into the geometry set and growth set of experiments listed in Table \ref{table-experiments}. Since the geometry experiments poorly constrain $\tau$, we set it to the mean value obtained in Ref.\,\cite{Planck2016-reionization} $\tau=0.058$, while we allow it to vary in the fitting of the growth experiments. The constraints from the two sets of experiments are shown in Figs.\,\ref{fig-gg-l-triagular} and \ref{fig-gg-w-triagular}. To quantify the inconsistency between the geometry and the growth experiments, we calculate IOI for the full parameter space (marginalized over $\tau$). Additionally, to compare IOI to the other measures of inconsistency, we also calculate other measures in the Gaussian and weak prior limit, like the normalized \textit{robustness} $\ln R_N$, the \textit{discordance}, the \textit{surprise} $S({\rm{grow}}||{\rm{geom}})$ and $S({\rm{geom}||{\rm{grow}}})$ (and their significance) and the $-$CER (and its significance). All results of inconsistency between the geometry and the growth experiments  for the $\Lambda$CDM and $w$CDM models are listed in Table\,\ref{table-Measures-values}.

We can see from Table \ref{table-Measures-values} that all the measures (except for $S({\rm{grow}}||{\rm{geom}})$) overall suggest there are inconsistencies between geometry and growth data sets for $\Lambda$CDM and $w$CDM models. However, different measures provide different levels of inconsistency. Take, for example, the $\Lambda$CDM model. IOI suggests a moderate experimental inconsistency (in fact not too far from a significant one on Jeffreys' scales). We note that the normalized \textit{robustness} gives an inconsistency much higher than other measures and ranked as very strong (decisive) if we insist to interpret it in Jeffreys' scales. However, this high inconsistency using the normalized \textit{robustness} is artificial as we explain. The geometry experiments give much stronger constraints than the growth ones, so that $|\bm{L^{geom}}|\gg |\bm{L^{grow}}|$, $|\bm{L^{geom}}+\bm{L^{grow}}|\simeq|\bm{L^{geom}}|$ and the term $-\ln(\tfrac{2|\bm{L^{geom}}\bm{L^{grow}}|}{\bm{L^{geom}}(|\bm{L^{geom}}+\bm{L^{grow}}|)})\gg1$. Therefore the apparent inconsistency suggested by $\ln R_N$ is only due to the unmatched constraining powers of the two data sets. The quantity $-$CER is simply related to IOI by Eq.\,\eqref{eq-CER-Gaussian}. It is suggested in Ref.\,\cite{2016-tensions-Grandis-etal} that $-$CER should be interpreted in terms of its significance. The significance of $-$CER gives a confidence level of inconsistency of $\SigCERnumggl\sigma$. For the \textit{surprise}, different orders of experiments in its argument lead to different results. While $S({\rm{geom}}||{\rm{grow}})$ is positive and finite, $S({\rm{grow}}||{\rm{geom}})$ is negative and large. The significance of the \textit{surprise} also has such asymmetric problem. While the significance of $S({\rm{grow}}||{\rm{geom}})$ gives no inconsistency, the significance of $S({\rm{grow}}||{\rm{geom}})$ suggests $\SigSnumgglP\sigma$ inconsistency, which seems more consistent with the one suggested by IOI and Fig.\,\ref{fig-gg-l-triagular}. The value of \textit{discordance} is $56.9\%$. Compared with $100\%$, the \textit{discordance} also seems to suggest some inconsistencies.

Comparing the results between $\Lambda$CDM and $w$CDM models, we can see that all measures (except for $S({\rm{grow}}||{\rm{geom}})$) suggest that the $w$CDM model has a larger inconsistency regarding the geometry and the growth experiments. In the language of IOI in nested models (see Sec.\,\ref{subsection-IOI-nested}), we say that the $w$CDM model is a less consistent extension to the $\Lambda$CDM model when comparing growth versus geometry. The result that the $w$CDM model has a larger experimental inconsistency is different from what was found in Ref.\,\cite{2013-tension-Verde-etal}. This may not be surprising, because Ref\,\cite{2013-tension-Verde-etal} were comparing Planck with local measurement on the Hubble constant and the age of the Universe, which are different experiments from what we are considering. The values of inconsistency measures are experiment dependent. Reference \cite{2017-Valentino-etal-DEdynamics-constraints} also found that the $w$CDM model is more favored than $\Lambda$CDM model when Planck, the local measurement of $H_0$ and BAO datasets are analyzed jointly. Indeed, when IOI is applied to investigate the inconsistency between Planck and local measurement on $H_0$, we find IOI$=4.0$ for $\Lambda$CDM and $0.59$ for $w$CDM. These numbers are consistent with those provided in Table 2 in Ref.\,\cite{2013-tension-Verde-etal} (recall that IOI and the \textit{tension} are the same in the Gaussian limit). So $w$CDM does have a smaller IOI in this case. However, the reason might be simply that Planck data have a poor constraint on $H_0$ if $w$ is allowed to vary; $\sigma_{H_0}\sim10$ km/s/Mpc. Recall that one of the three factors affecting IOI is the constraint volume: a larger constraint volume leads to a smaller IOI.

\begin{table*}[tp]
\caption[One-parameter measures for $\Lambda$CDM - geometry vs growth]{\label{table-IOI-one-param-lcdm} Two measures of one-parameter inconsistency for the $\Lambda$CDM model. The first one is the residual IOI$_i$ (and the relative residual between parentheses) for $\lambda_i$ after marginalizing over all the other parameters. The second one is the inconsistency drop $\Delta$IOI$_i$ (and the relative drop between parentheses) after marginalizing over $\lambda_i$. Both IOI$_i^{\rm Rel}$ and $\Delta$IOI$_i^{\rm Rel}$ need to be much smaller than unity for us to safely marginalize over $\lambda_i$ to analyze inconsistency. Four different situations are listed in Table \ref{table-marginalization-situations} regarding different values of IOI$_i$ and $\Delta$IOI$_i$. It seems that none of the parameters can be safely marginalized over, although $\theta$ and $A_s$ seem to be least responsible for the inconsistency. The geometry and the growth experiments give a quite inconsistent result on $n_s$ since both IOI$_{n_s}^{\rm{Ref}}$ and $\Delta$IOI$_{n_s}^{\rm{Ref}}$ are large. The bottom three rows: drops of $\sigma_{eq}$, $S({\rm{geom}}||{\rm{grow}})$ and its significance evaluated after marginalizing over one parameter. Note that their drops are not always positive after parameter marginalization as one would expect.}
\begin{ruledtabular}
\begin{tabular}{lccccc}
$\Lambda$CDM parameters & $\Omega_bh^2$ & $\Omega_ch^2$  & $\theta$ & $A_s$ & $n_s$ \\
\hline
IOI$_i$ ~(IOI$_i^{\rm Rel}$) & $0.0017$ ~($0.0004$) & $0.55$ ~($0.13$) & $0.39$ ~($0.09$) & $0.39$ ~($0.09$) & $2.4$ ~($0.55$)  \\
$\Delta{\rm{IOI}}_i$ ~($\Delta{\rm{IOI}}_i^{\rm Rel}$)  & $0.64$ ~($0.15$) & $0.55$ ~($0.13$) & $0.24$ ~($0.06$) & $0.19$ ~($0.04$) & $1.5$ ~($0.33$)  \\
\hline
$\Delta\sigma_{eq}$ & $-10.7\%$ & $-14.5\%$ & $-6.6\%$ & $-13.5\%$ & $10.5\%$ \\
$\Delta S({\rm{geom}}||{\rm{grow}})$ & $0.97$ & $1.03$ & $-0.7$ & $-0.13$ & $1.05$\\
$\Delta$(significance of $S$) & $0.62$ & $0.47$ & $-0.32$ & $0.19$ & $0.81$\\
\end{tabular}
\end{ruledtabular}
\caption[One-parameter measures for $w$CDM - geometry vs growth]{\label{table-IOI-one-param-wcdm} Two measures of one-parameter inconsistency for the $w$CDM model. This is similar to Table \ref{table-IOI-one-param-lcdm}. For the $w$CDM model, it seems that $\theta$ and $A_s$ are least responsible for the inconsistency, and we can marginalize over them. The geometry and the growth experiments give a quite inconsistent result on $n_s$, as in the case of the $\Lambda$CDM model. The bottom three rows are similar to those in Table \ref{table-IOI-one-param-lcdm}.}
\begin{ruledtabular}
\begin{tabular}{lcccccc}
$w$CDM parameters & $\Omega_bh^2$ & $\Omega_ch^2$  & $\theta$ & $w$ & $A_s$ & $n_s$ \\
\hline
IOI$_i$ ~(IOI$_i^{\rm Rel}$) & $0.0028$ ~($0.0005$)  & $1.00$ ~($0.18$) & $0.19$ ~($0.035$) & $1.12$ ~($0.21$)& $0.0021$ ~($0.0004$) & $2.93$ ~($0.54$)  \\
$\Delta{\rm{IOI}}_i$ ~($\Delta{\rm{IOI}}_i^{\rm Rel}$) & $0.77$ ~($0.14$) & $0.58$ ~($0.11$) & $0.38$ ~($0.07$) & $0.18$ ~($0.034$) & $0.26$ ~($0.048$) & $1.27$ ~($0.23$) \\
\hline
$\Delta\sigma_{eq}$ & $-4.2\%$ & $-7.1\%$ & $-2.0\%$ & $-8.1\%$ & $-8.7\%$ & $11.4\%$\\
$\Delta S({\rm{geom}}||{\rm{grow}})$ & $1.39$ & $0.66$ & $-0.57$ & $-0.63$ & $0.65$ & $0.2$\\
$\Delta$(significance of $S$) & $0.41$ & $0.03$ & $-0.4$ & $-0.81$ & $0.07$ & $-0.01$\\
\end{tabular}
\end{ruledtabular}
\end{table*}

Finally in this subsection, we briefly investigate the roles of CFHTlens data and SDSS RSD data on the value of IOI for $\Lambda$CDM model. If we remove CFHTlens data alone, IOI drops from $4.3$ to $3.2$. If we remove SDSS RSD data alone, IOI drops from $4.3$ to $3.8$. Although the CFHTlenS data seem to cause more inconsistency, neither removing the CFHTlenS data nor the SDSS RSD data dramatically drops IOI.  Therefore, the inconsistency between the geometry and the growth experiments does not seem to be caused by one particular data set. Furthermore, if we remove both CFHTlens and SDSS RSD data, IOI drops from $4.3$ to $2.5$, but a weak inconsistency still persists. In a follow-up work, we will systematically explore the internal and cross inconsistencies between a number of data sets and more applications of IOI.

\subsection[two measures in $w$CDM]{Relative drop of IOI and relative residual IOI in the geometry vs growth application}\label{subsection-supplemental-measure-wCDM}
We have shown that parameter marginalization can hide experimental inconsistency in Sec.\,\ref{subsection-marginalization-and-IOI}. This is reflected by the drop of IOI after marginalizing over each parameter, which is shown in Tables \ref{table-IOI-one-param-lcdm} and \ref{table-IOI-one-param-wcdm}. The fact that IOI never increases offers us two ways to see the experimental inconsistency due to a particular parameter, which we apply here to the geometry and growth experiments.

The two measures of one-parameter inconsistency IOI$_i$ and $\Delta{\rm{IOI}}_i$ described in Sec.\,\ref{subsection-supplemental-measure} can help us zoom into the parameter space, to see experimental inconsistency due to each parameter. For a given model, if two experiments do not have inconsistency due to $\lambda_i$, both IOI$_i$ and $\Delta$IOI$_i$ are small in Jeffreys' scales. But in some cases, even though IOI$_i$ and $\Delta$IOI$_i$ are small in Jeffreys' scales, they are not negligible compared to the full IOI. So more conveniently, we can use the relative quantities IOI$_i^{\rm Rel}$ and $\Delta$IOI$_i^{\rm Rel}$, and compare them to unity to see the role of each parameter on experimental inconsistency. Only when both IOI$_i^{\rm Rel}$ and $\Delta$IOI$_i^{\rm Rel}$ are much smaller than $1$, can we safely marginalize over $\lambda_i$ to analyze the full inconsistency. We listed the two measures and their relative measures in Tables \ref{table-IOI-one-param-lcdm} and \ref{table-IOI-one-param-wcdm} for the $\Lambda$CDM model and the $w$CDM model, respectively.

Take the $w$CDM model for example, the geometry and the growth experiments have a significant (full) inconsistency (IOI=5.4). For individual parameters, the 1D marginalized plots in Fig.\,\ref{fig-gg-w-triagular} seem to show little inconsistency in $\Omega_bh^2$, $\theta$ and $A_s$, and the corresponding IOI$_i$'s (and IOI$_i^{\rm Rel}$) are small. But it does not mean it is safe to marginalize over these three parameters. For example, $\Delta$IOI$_{bh^2}^{\rm Rel}$ is not much smaller than $1$, so marginalizing over $\Omega_bh^2$ will hide some experimental inconsistency due to its correlation with the other parameters (specified by $\bm{C}=\bm{C^{(1)}}+\bm{C^{(2)}}$). On the other hand, since both IOI$_{A_s}^{\rm Rel}$ and $\Delta$IOI$_{A_s}^{\rm Rel}$ are small, the geometry and the growth experiments give a consistent result on $A_s$, and it is safe to marginalize over it to study the full inconsistency. Similarly, it is also safe to marginalize over $\theta$. For the $\Lambda$CDM model, from Table \ref{table-IOI-one-param-lcdm} it seems inappropriate to marginalize any parameter to analyze inconsistency, although $\theta$ and $A_s$ seem to be least responsible for the inconsistency.

Which parameter have the most inconsistent results given by the geometry and the growth experiments? From Tables \ref{table-IOI-one-param-lcdm} and \ref{table-IOI-one-param-wcdm}, we can see that for both the $\Lambda$CDM model and the $w$CDM model, the results of $n_s$ are the most inconsistent. Both IOI$_{n_s}^{\rm Rel}$ and $\Delta$IOI$_{n_s}^{\rm Rel}$ are large. The next inconsistent parameter should be $\Omega_c h^2$, but the inconsistency caused by it is quite small compared to that due to $n_s$.

The definition of $\Delta{\rm{IOI}}_i$ is based on the fact that IOI never increases after a parameter marginalization. This drop of IOI represents the hidden inconsistency due to a parameter after we marginalize over it. This is a unique feature for IOI. To demonstrate it, we show as examples the drops of $\sigma_{eq}$, $\Delta S({\rm{geom}}||{\rm{grow}})$ and its significance after marginalizing over each parameter in the last three rows of Tables \ref{table-IOI-one-param-lcdm} and \ref{table-IOI-one-param-wcdm}. We can see that their drops are not always positive, suggesting that they are not able to represent the hidden inconsistency after a parameter marginalization.

It is, however, important to emphasize that although combining $\Delta{\rm{IOI}}_i$'s and IOI$_i$'s can give us more insights into the inconsistency due to each parameter, they cannot fully describe the experimental inconsistency. It is the full IOI (recall this is not the summation of $\Delta{\rm{IOI}}_i$'s or IOI$_i$'s) for the whole parameter space that should be calculated and used.

\vspace{-1em}
\section{Summary}\label{section-summary}
The $\Lambda$CDM model is considered to be the standard and concordance model in cosmology, but different experiments have been yielding constraints of parameters that do not perfectly agree with each other. Those inconsistencies could be caused by either systematics or the breakdown of the underlying model and theory. It is timely and important to develop tools to quantify the degree of inconsistency between different experiments.

We defined a moment-based measure that we call IOI. We suggested to interpret IOI using Jeffreys' scales listed in Table \ref{table-Jeffrey-scale-IOI}. We showed that IOI tracks well the separation of the means, the volume of the covariance ellipsoids, and the orientations of those ellipsoids. We discussed criteria that must be fulfilled by measures of inconsistency. For example, the numerical value of the measure must follow the graphical inconsistency of likelihood contour plots, represent experiments with definite consistency, and correctly reflect the effect of parameter marginalization. We find that IOI satisfies all those criteria. We provided an eigenmode decomposition of IOI and also discussed the application of IOI for nested models.

We showed how marginalizing over a parameter can hide inconsistency between experiments when present. This is because every parameter can induce its own experimental inconsistency and also has correlations with other parameters. In fact, even if all marginalized distribution plots look consistent for two experiments, they might still be inconsistent. These facts show that descriptions of experimental inconsistency involving parameter marginalization is not accurate. We provided an analytical proof that IOI never increases whenever a parameter marginalization is performed but rather decreases in most cases. This work seems to be the first to quantify the loss of information on cosmological experimental inconsistency due to parameter marginalization.

In order to zoom on the inconsistency residing in a particular parameter, we defined the relative drop in IOI and  the relative residual IOI for a given parameter, IOI$^{\rm Rel}_i=\tfrac{{\rm IOI}_i}{\rm IOI}$ and $\Delta$IOI$^{\rm Rel}_i=\tfrac{\Delta{\rm IOI}_i}{\rm IOI}$. We explained that only when both relative measures are small can one safely marginalize over such a parameter when considering inconsistency measures. We also explained that in order to correctly quantify the inconsistency we should use the full IOI of all parameters in a model that are commonly constrained by two experiments.

We provided a comparative survey of other measures of inconsistency and their relationships, if any, with IOI.

We then applied IOI and other measures in the Gaussian and weak prior limit to quantify the inconsistency of the $\Lambda$CDM and $w$CDM models between geometry and growth experiments. As discussed in the previous section, IOI indicates a moderate to strong inconsistency between the two data sets when the $\Lambda$CDM and $w$CDM models are used with values of 4.3 and 5.4 on Jeffreys' scales. Overall, although different measures use different scales to quantity inconsistency, most of the measures (except for $S(\rm geom||grow)$) also suggest that there are moderate to strong experimental inconsistencies in $\Lambda$CDM and $w$CDM models with the $w$CDM model having a larger inconsistency between geometry and growth experiments.
As we discussed in the previous section, this comparison and its results are different from two other works that compared Planck with local measurement of the Hubble constant and the age of the Universe, finding that $w$CDM brings some reconciliation. As we explained, the values of inconsistency measures are experiment dependent, as they should.

Finally, we used the IOI measure and found that CFTlenS data, SDSS RSD data and other growth data all contribute to the inconsistency. So this seems to indicate that the inconsistency cannot be associated with a single data set. In future work, we will explore in more detail the role of each data set in the experimental inconsistencies.

\begin{acknowledgments}
We would like to thank L. Amendola, S. Grandis, M. Kesden, L. King, E. Linder, L. Pogosian, D. Rapetti, L. Verde and J. Zuntz for useful comments. M.I. acknowledges that this material is based upon work supported in part by NSF under Grant No. AST-1517768 and an award from the John Templeton Foundation.
\end{acknowledgments}

\appendix
\section{Derivations and proofs}\label{appendix-derivations}
This appendix provides the calculations and derivations used in this work, including those used in the definition of IOI and its relationship to other measures of inconsistency.
\subsection{IOI for arbitrary numbers of experiments}\label{appendix-sub-combine-likes}
We consider combination of likelihoods that do not correlate with each other. A Gaussian likelihood for the $i$th experiment reads
\begin{equation}\label{eq-ith-likelihood}
\begin{split}
\mathcal{L}^{(i)}&=\exp\big[-\tfrac{1}{2}\chi^2_{(i)}(\lambdaB)\big]\\&=\mathcal{L}^{(i)}_{max}\exp\big[-\tfrac{1}{2}\Delta\chi^2_{(i)}(\lambdaB)\big]\\
&=\mathcal{L}^{(i)}_{max}\exp\left[-\tfrac{1}{2}(\lambdaB-\bm{\mu^{(i)}})^T\bm{L^{(i)}}(\lambdaB-\bm{\mu^{(i)}})\right]\,.
\end{split}
\end{equation}
For arbitrary number of experiments, the likelihood of a joint analysis is defined as
\begin{equation}\label{combine-likelihood-def}
\begin{split}
\mathcal{L}&=\prod_i\mathcal{L}^{(i)}\\
&=\Big(\prod_i\mathcal{L}^{(i)}_{max}\Big)\exp\Big[\sum\limits_i-\frac{1}{2}(\lambdaB-\bm{\mu^{(i)}})^T\bm{L^{(i)}}(\lambdaB-\bm{\mu^{(i)}})\Big]\,.
\end{split}
\end{equation}
Note that $\bm{\mu^{(i)}}$ is the mean of the $i$th likelihood, not the $i$th component of $\muB$. The sum in Eq.\,\eqref{combine-likelihood-def} can be reduced to
\begin{equation}\label{eq-exponential-reduced}
\begin{split}
&~~~\sum\limits_i(\lambdaB-\bm{\mu^{(i)}})^T\bm{L^{(i)}}(\lambdaB-\bm{\mu^{(i)}})\\
&=\lambdaB^T\bm{L}\lambdaB-2\muB^T\bm{L}\lambdaB+\sum\limits_i\bm{\mu^{(i)}}\,^T\bm{L^{(i)}}\bm{\mu^{(i)}} \\
&=(\lambdaB-\muB)^T\bm{L}(\lambdaB-\muB)+\sum\limits_i\bm{\mu^{(i)}}\,^T\bm{L^{(i)}}\bm{\mu^{(i)}}-\muB^T\bm{L}\muB\,,
\end{split}
\end{equation}	
where,
\begin{align}
\bm{L}&\equiv\sum\bm{L^{(i)}}\,,\label{eq-joint-Fisher}\\
\muB&\equiv \bm{L}^{-1}\sum\bm{L^{(i)}}\bm{\mu^{(i)}}\,,\label{eq-joint-mean}
\end{align}
 and we have used the fact that $\big(\bm{L^{(i)}}\big)^T=\bm{L^{(i)}}$. Since the last two terms do not depend on $\lambdaB$, we can combine them with $\prod\mathcal{L}^{(i)}_{max}$, and Eq.\,\eqref{combine-likelihood-def} becomes
\begin{equation}\label{eq-combine-likelihood-result}
\mathcal{L}=\mathcal{L}_{max}\exp\Big[-\frac{1}{2}(\lambdaB-\muB)^T\bm{L}(\lambdaB-\muB)\Big]\,,
\end{equation}
where,
\begin{equation}\label{eq-joint-max}
\mathcal{L}_{max}=(\prod_i\mathcal{L}^{(i)}_{max})\exp\Big[-\frac{1}{2}\big(\sum\limits_i\bm{\mu^{(i)}}\,^T\bm{L^{(i)}}\bm{\mu^{(i)}}-\muB^T\bm{L}\muB\big)\Big]\,.
\end{equation}
So the likelihood of a joint analysis is also Gaussian [Eq.\,\eqref{eq-combine-likelihood-result}], with its mean, Fisher matrix and maximum value given by Eqs.\,\eqref{eq-joint-Fisher} - \eqref{eq-joint-max}. The above result is also applicable to the case where one of the likelihoods is replaced by a Gaussian prior, where one of the likelihoods is replaced by a prior with $\mathcal{L}^{(i)}_{max}=\mathcal{P}_{max}=\frac{|\bm{P}|^{1/2}}{(2\pi)^{N/2}}$ and set $\bm{L^{(i)}}=P$.

Note that we have defined IOI as
\begin{equation}
\frac{1}{N}\sum\limits_i\Delta\chi^2_{(i)}(\muB)\xrightarrow{\rm Gaussian}{\rm IOI}\,.
\end{equation}
The maximum value of the joint likelihood $\mathcal{L}$ is
\begin{equation}\label{eq-jointlike-max-general}
\begin{split}
\mathcal{L}_{max}&=\mathcal{L}(\muB)\\
&=\left(\prod\mathcal{L}^{(i)}_{max}\right)\exp\Big(-\frac{1}{2}\sum\limits_i\Delta\chi^2_{(i)}(\muB)\Big)\\
&\xrightarrow{\rm Gaussian} \left(\prod\mathcal{L}^{(i)}_{max}\right)\exp\big(-\tfrac{N}{2}{\rm{IOI}}\big)\,,
\end{split}
\end{equation}
By comparing Eq.\,\eqref{eq-joint-max} and Eq.\,\eqref{eq-jointlike-max-general} we have
\begin{equation}\label{eq-IOI-multiple-deduction}
{\rm IOI}\equiv\frac{1}{N}\Big(\sum\limits_i\bm{\mu^{(i)}}\,^T\bm{L^{(i)}}\bm{\mu^{(i)}}-\muB^T\bm{L}\muB\Big)\,,
\end{equation}
for multiple experiments.

It is important to emphasize that, $\frac{1}{N}\sum\Delta\chi^2_{(i)}$ is only a motivation for our definition of IOI. It is true that IOI and $\frac{1}{N}\sum\Delta\chi^2_{(i)}$ are the same for Gaussian likelihoods, but for non-Gaussian cases, we define IOI as a moment-based quantity [Eq.\,\eqref{eq-IOI-multiple-deduction}], with $\bm{L^{(i)}}$ given by the inverse of the covariance matrix and $\bm{\mu^{(i)}}$ given by the mean of the $i$th likelihood.

\subsection{Special cases: Two likelihoods}\label{appendix-sub-combine-two-likes}
More often we will compare two experiments to see their inconsistency. For the case of two experiments ($i=1,2$), the right-hand side of Eq.\,\eqref{eq-IOI-multiple-deduction} becomes
\begin{widetext}
\begin{equation}\label{eq-max-like-and-IOI}
\begin{split}
&~~~\sum\limits_i\bm{\mu^{(i)}}\,^T\bm{L^{(i)}}\bm{\mu^{(i)}}-\muB^T\bm{L}\muB\\
&=\bm{\mu^{(1)}}\,^T\bm{L^{(1)}}\bm{\mu^{(1)}}+\bm{\mu^{(2)}}\,^T\bm{L^{(2)}}\bm{\mu^{(2)}}-(\bm{\mu^{(1)}}\,^T\bm{L^{(1)}} +\bm{\mu^{(2)}}\,^T\bm{L^{(2)}})\bm{L}^{-1}\bm{L}~\bm{L}^{-1}(\bm{L^{(1)}}\bm{\mu^{(1)}}+\bm{L^{(2)}}\bm{\mu^{(2)}})\\
&=\bm{\mu^{(1)}}\,^T\bm{L^{(1)}}\bm{\mu^{(1)}}+\bm{\mu^{(2)}}\,^T\bm{L^{(2)}}\bm{\mu^{(2)}} -(\bm{\mu^{(1)}}\,^T\bm{L}-\bm{\delta}\,^T\bm{L^{(2)}})\bm{L}^{-1}\bm{L}~\bm{L}^{-1}(\bm{L^{(1)}}\bm{\delta}+\bm{L}\bm{\mu^{(2)}})\,,\\
&=\bm{\mu^{(1)}}\,^T\bm{L^{(1)}}\bm{\mu^{(1)}}-\bm{\mu^{(1)}}\,^T\bm{L^{(1)}}\bm{\delta}+\bm{\mu^{(2)}}\,^T\bm{L^{(2)}}\bm{\mu^{(2)}} +\bm{\delta}^T\bm{L^{(2)}}\bm{\mu^{(2)}}-\bm{\mu^{(1)}}\,^T\bm{L}\bm{\mu^{(2)}} +\bm{\delta}^T\bm{L^{(2)}}\bm{L}^{-1}\bm{L^{(1)}}\deltaB\\
&=\bm{\mu^{(1)}}\,^T\bm{L^{(1)}}\bm{\mu^{(2)}}+\bm{\mu^{(1)}}\,^T\bm{L^{(2)}}\bm{\mu^{(2)}}-\bm{\mu^{(1)}}\,^T\bm{L}\bm{\mu^{(2)}}+\deltaG\,,\\
&=\deltaB^T\bm{G}\deltaB\,,
\end{split}
\end{equation}
\end{widetext}
where $\deltaB\equiv\bm{\mu^{(1)}}-\bm{\mu^{(2)}}$ and in the second last row we have used $\bm{L^{(2)}}\bm{L}^{-1}\bm{L^{(1)}}=\bm{L^{(1)}}\bm{L}^{-1}\bm{L^{(2)}} =\big[(\bm{L^{(1)}})^{-1}+(\bm{L^{(2)}})^{-1}\big]^{-1}\equiv \bm{G}$. So the above derivation [Eq.\,\eqref{eq-max-like-and-IOI}] shows that Eq.\,\eqref{eq-IOI-definition-general} for IOI of multiple experiments reduces to Eq.\,\eqref{eq-IOI-definition} for two experiments. We will use
\begin{equation}\label{eq-IOI-2D}
{\rm IOI}\equiv\tfrac{1}{2}\deltaB^T\bm{G}\deltaB
\end{equation}
as the definition of IOI for two experiments.

\subsection{Integrals of joint Gaussian distributions}\label{appendix-integral-likelihood}
The integral of a single Gaussian likelihood gives
\begin{equation}
\begin{split}
&\int d^N\lambda\mathcal{L}^{(i)}_{max}\exp\left(-\tfrac{1}{2}(\lambdaB-\muB)\bm{L^{(i)}}(\lambdaB-\muB)\right)\\
=&\frac{(2\pi)^{\tfrac{N}{2}}\mathcal{L}^{(i)}_{max}}{\sqrt{|\bm{L^{(i)}}|}}\,.
\end{split}
\end{equation}
With  the combined likelihood given by Eq.\,\eqref{eq-combine-likelihood-result}, we immediately obtain,
\begin{equation}\label{combine-likelihood-integral}
\begin{split}
\int d^N\lambda \mathcal{L}=&\frac{(2\pi)^{\tfrac{N}{2}}\mathcal{L}_{max}}{\sqrt{|\bm{L}|}}\\
=&\frac{(2\pi)^{\tfrac{N}{2}}\prod\mathcal{L}^{(i)}_{max}}{\sqrt{|\sum\bm{L^{(i)}}|}}\exp\big(-\tfrac{1}{2}{\rm IOI}\big)\,.
\end{split}
\end{equation}

A useful case is to integrate the product of two normalized posteriors (as is used in Sec.\,\ref{subsection-tension}), with $\mathscr{P}^{(i)}=\frac{|\bm{F^{(i)}}|^{1/2}}{(2\pi)^{N/2}}\exp\big[-\tfrac{1}{2} (\lambdaB-\bm{\bar{\mu}^{(i)}})^T\bm{F^{(i)}}(\lambdaB-\bm{\bar{\mu}^{(i)}})\big]$, where $\bm{F^{(i)}}=\bm{L^{(i)}}+\bm{P}$ and $\bm{\bar{\mu}^{(i)}}=(\bm{F^{(i)}})^{-1}(\bm{L^{(i)}}\bm{\mu^{(i)}}+\bm{P}\bm{\mu^{(p)}})$. By setting $\bm{L^{(i)}}=\bm{F^{(i)}}$ and $\muB^{i}=\bm{\bar{\mu}^{(i)}}$, we obtain from Eq.\,\eqref{combine-likelihood-integral} that
\begin{equation}
\begin{split}
&\int \mathscr{P}^{(1)}\mathscr{P}^{(2)} d^N\lambda\\
=&\sqrt{\frac{|\bm{F^{(1)}}\bm{F^{(2)}}|}{|\bm{F^{(1)}}+\bm{F^{(2)}}|}} \exp\big(-\deltaGhalfbar\big)\\
=&|\bm{\bar{G}}|^{1/2}\exp\big(-\deltaGhalfbar\big)\,,
\end{split}
\end{equation}
where, $\bm{\bar{G}}=((\bm{F^{(1)}})^{-1}+(\bm{F^{(2)}})^{-1})^{-1}$ and $\bar{\deltaB}=\bm{\bar{\mu}^{(1)}}-\bm{\bar{\mu}^{(2)}}$. Then in the weak prior limit, we just have to take those bars off.

\section{Marginalization never increases IOI}\label{appendix-marginalization}
\subsection{Proof}\label{appendix-sub-the-proof-marg}
We will adopt the following notations. An element of the matrix $\bm{C}$ is denoted as lowercase $C_{ij}$, while $\bm{\widetilde{C_{ij}}}$ stands for a matrix obtained by taking out the $i$th row and the $j$th column of $\bm{C}$ and multiplied by $(-1)^{i+j}$. [Then $\bm{\widetilde{C_{ij,mn}}}=\bm{\widetilde{C_{mn,ij}}}$ is a matrix obtained by taking out the $i$th \& $m$th rows and $j$th \& $n$th columns of $\bm{C}$ and multiplied by $(-1)^{i+j+m+n}$.] $|\CB|$ is the determinant of $\bm{C}$. Then $|\bm{\widetilde{C_{ij}}}|$ is the ($i$,$j$) cofactor $\bm{C}$. Let $\bm{\widetilde{\delta_i}}$ be the vector results from deleting the $i$th component from $\deltaB$.

Suppose we are to marginalize $\lambda_1$. The full IOI is $\deltaGhalf$, and the IOI after marginalizing over $\lambda_1$ is IOI$_{mar,1}=\tfrac{1}{2}(\deltaG)_{mar,1}$. The quantity $(\deltaG)_{mar,1}$ is calculated as follows: (1) delete the first element of $\deltaB$ and get $\bm{\widetilde{\delta_1}}$; (2) delete the first row and the first column in $\CB=\CBone+\CBtwo$ and let it be $\bm{\widetilde{C_{11}}}$; and (3) then $(\deltaG)_{mar,1}=\bm{\widetilde{\delta_1}}\big(\bm{\widetilde{C_{11}}}\big)^{-1}\bm{\widetilde{\delta_1}}$. What we are going to show is IOI $\geq$ IOI$_{mar,1}$, that is, $\deltaG\geq(\deltaG)_{mar,1}$.

We first make a linear parameter transformation that keeps $\lambda_1$ unchanged. More explicitly, we let $\lambdaB\rightarrow \lambdaB'=\bm{M}\lambdaB$ with a transformation matrix of the form
\begin{equation}\label{eq-M-form}
  \bm{M}=\begin{pmatrix}
    1 &0 &0 & \cdots\\
    0 &m_{22} &m_{23}  &\cdots\\
    0 &m_{32} &m_{33}  &\cdots\\
    \vdots &\vdots &\vdots &\ddots
  \end{pmatrix}
  \xrightarrow{\rm symbolic}
  \bm{M}=\begin{pmatrix}
    1 & \bm{0}\\
   \bm{0} & \bm{\widetilde{M}}
   \end{pmatrix}\,,
\end{equation}
where $\bm{\widetilde{M}}\equiv\bm{\widetilde{M_{11}}}$. Then the new deviation vector $\deltaB'$ reads
\begin{equation}\label{eq-Delta-prime}
\deltaB=\begin{pmatrix}
  \delta_1\\
  \bm{\widetilde{\delta_1}}
\end{pmatrix}~\rightarrow~
\deltaB'=\begin{pmatrix}
  \delta_1\\
  \bm{\widetilde{\delta_1}}'
\end{pmatrix}
=\begin{pmatrix}
  \delta_1\\
  \bm{\widetilde{M}}\bm{\widetilde{\delta_1}}
\end{pmatrix}\,.
\end{equation}
Similarly, the new matrix $\bm{C}'$ reads,
\begin{equation}\label{eq-C-prime}
\bm{C}=\begin{pmatrix}
  C_{11} & \bm{V_{1}}^T\\
  \bm{V_{1}} &\bm{\widetilde{C_{11}}}
\end{pmatrix}
~\rightarrow~
\bm{C'}=\begin{pmatrix}
  C_{11} &\bm{V_{1}}^T\bm{\widetilde{M}}^T\\
  \bm{\widetilde{M}}\bm{V_{1}} &\bm{\widetilde{M}}^T\bm{\widetilde{C_{11}}}\bm{\widetilde{M}}
\end{pmatrix}\,,
\end{equation}
where $\big(\bm{V_1}\big)_i=C_{1i}$ for $i\neq1$. So under $\bm{M}$, $\delta_1$ and $C_{11}$ stay the same, $\bm{\widetilde{\delta_1}}$ and $\bm{V_1}$ transform as vectors, and $\bm{\widetilde{C_{11}}}$ transforms as a rank-2 tensor. To simply the proof, we further let $\CB'$ take the following form,
\begin{equation}\label{eq-rotate}
\begin{split}
\bm{C}&=\begin{pmatrix}
C_{11} & C_{12} & C_{13} &\cdots \\
C_{21} & C_{22} & C_{23} &\cdots \\
C_{31} & C_{32} & C_{33} &\cdots \\
\vdots & \vdots & \vdots & \ddots
\end{pmatrix}~\\
&\qquad\rightarrow~
\CB'=\begin{pmatrix}
C_{11} & C'_{12} & C'_{13}  &\cdots \\
C'_{12} & C'_{22} & 0 & \bm{0} \\
C'_{13} & 0 & C'_{33} & \bm{0} \\
\vdots & \bm{0} & \bm{0} & \ddots
\end{pmatrix}\,.
\end{split}
\end{equation}
We have used the fact that $C_{ij}=C_{ji}$. In other words, we transform the parameter space alone the direction of $\lambda_1$, so that the other (new) parameters are the eigenmodes of $\bm{\widetilde{{C_{11}}}}$ in the projected ($N$-1)-dimensional parameter subspace orthogonal to $\lambda_1$.  For the proof, we will work in the new parameter space and drop the primes. In this new parameter space, we have
\begin{equation}\label{DGD-N-expan}
\begin{split}
\deltaB^T\bm{G}\deltaB&=\frac{1}{|\bm{C}|}\Big(\sum\limits_{i,j}|\bm{\widetilde{C_{ij}}}|\delta_i\delta_j\Big)\\
&=\frac{1}{|\bm{C}|}\Big(\sum\limits_{i\neq1}|\bm{\widetilde{C_{ii}}}|\delta_i^2+\sum\limits_{i\neq1}\sum\limits_{j\neq1,i}|\bm{\widetilde{C_{ij}}}|\delta_i\delta_j\\
&\qquad\qquad~~ +2\sum\limits_{i\neq1}|\bm{\widetilde{C_{1i}}}|\delta_i\delta_1+|\bm{\widetilde{C_{11}}}|\delta_1^2\Big)\,.
\end{split}
\end{equation}
With our new matrix $C$ of the form \eqref{eq-rotate}, we can calculate those determinants explicitly, (let $\prod\limits_{i\neq1}C_{ii}= A$, and $i,j\neq1$)
\begin{equation*}
\begin{split}
&|\bm{\widetilde{C_{11}}}|=A\,,\\
&|\bm{\widetilde{C_{1i}}}|=-\frac{C_{1i}}{C_{ii}}A\,,\\
&|\bm{C}|=\big(C_{11}-\sum\limits_{i\neq1}\frac{C_{1i}}{C_{ii}}\big)A\,,\\
&|\bm{\widetilde{C_{ii}}}|=\big(C_{11}-\sum\limits_{j\neq1,i}\frac{C_{1j}}{C_{jj}}\big)\frac{A}{C_{ii}}\,,\\
&|\bm{\widetilde{C_{ij}}}|_{j\neq i}=-\frac{C_{1i}C_{1j}}{C_{ii}C_{jj}}A\,.
\end{split}
\end{equation*}
After we substitute the above determinants into Eq.\,\eqref{DGD-N-expan} and cancel out $A$, $\deltaB^T\bm{G}\deltaB$ reads
\begin{equation}\label{eq-DeltaGDelta-extend}
\begin{split}
\deltaB^T\bm{G}\deltaB&=\frac{1}{C_{11}-\sum\limits_{i\neq1}\tfrac{C_{1i}^2}{C_{ii}}} \Big(\sum\limits_{i\neq1}\frac{C_{11}}{C_{ii}}\delta_i^2-\sum\limits_{i\neq1}\sum\limits_{j\neq1,i} \frac{C_{1j}^2}{C_{ii}C_{jj}}\delta_i^2\\
&\qquad-\sum\limits_{i\neq}\sum\limits_{j\neq1,i}\frac{C_{1i}C_{1j}}{C_{ii}C_{jj}}\delta_i\delta_j-2\sum\limits_{i\neq1}\frac{C_{1i}}{C_{ii}}\delta_i\delta_1 +\delta_1^2\Big)\,.
\end{split}
\end{equation}
In our new parameter space, the IOI after marginalizing over $\lambda_1$ is rather easy to calculate and it is $2{\rm{IOI}}_{mar,1}=(\deltaB^T\bm{G}\deltaB)_{mar,1}=\sum\limits_{i\neq1}\frac{\delta_i^2}{C_{ii}}$. Taking the difference between $\deltaB^T\bm{G}\deltaB$ and $\big(\deltaB^T\bm{G}\deltaB\big)_{mar,1}$, considering only the numerator of it, and rearranging terms, we have
\begin{equation}\label{eq-numerator}
\begin{split}
&\rm{The~numerator~of}~\Big(\deltaB^T\bm{G}\deltaB-\big(\deltaB^T\bm{G}\deltaB\big)_{mar,1}\Big)\\
=&\sum\limits_{i\neq1}\frac{C_{11}}{C_{ii}}\delta_i^2-\sum\limits_{i\neq1}\frac{C_{11}}{C_{ii}}\delta_i^2\\
&\qquad+\sum\limits_{i\neq1}\sum\limits_{j\neq1} \frac{C_{1j}^2}{C_{ii}C_{jj}}\delta_i^2 -\sum\limits_{i\neq1}\sum\limits_{j\neq1,i} \frac{C_{1j}^2}{C_{ii}C_{jj}}\delta_i^2\\
&\qquad-\sum\limits_{i\neq}\sum\limits_{j\neq1,i}\frac{C_{1i}C_{1j}}{C_{ii}C_{jj}}\delta_i\delta_j -2\sum\limits_{i\neq1}\frac{C_{1i}}{C_{ii}}\delta_i\delta_1 +\delta_1^2\\
=&\sum\limits_{i\neq1}\frac{C_{1i}^2}{C_{ii}}\delta_i^2-\sum\limits_{i\neq}\sum\limits_{j\neq1,i}\frac{C_{1i}C_{1j}}{C_{ii}C_{jj}}\delta_i\delta_j -2\sum\limits_{i\neq1}\frac{C_{1i}}{C_{ii}}\delta_i\delta_1 +\delta_1^2\\
=&\sum\limits_{i,j\neq1}\frac{C_{1i}C_{1j}}{C_{ii}C_{jj}}\delta_i\delta_j-2\sum\limits_{i\neq1}\frac{C_{1i}}{C_{ii}}\delta_i\delta_1+\delta_1^2\\
=&\Big(\sum\limits_{i\neq1}\frac{C_{1i}}{C_{ii}}\delta_i-\delta_1\Big)^2
\geq0\,,
\end{split}
\end{equation}
So $\deltaB^T\bm{G}\deltaB\geq\big(\deltaB^T\bm{G}\deltaB\big)_{mar,1}$ and marginalization never increases IOI.

The condition for IOI $=$ IOI$_{mar,1}$ (called equality condition from now on) is $\sum\limits_{i\neq1}\frac{C_{1i}}{C_{ii}}\delta_i=\delta_1$. Since we have transformed the parameter space along the $\lambda_1$ axis, this equality condition is not generic. To find the generic equality condition, we first note that the right-hand side of the equality condition $\delta_1$ is invariant under the transformation. Therefore we can generalize the left-hand side to a quantity that is: (1) invariant under the transformation specified by Eq.\,\eqref{eq-M-form}, and (2) reduces to $\sum\limits_{i\neq1}\frac{C_{1i}}{C_{ii}}\delta_i$ in the case when $\CB$ takes the form of $\CB'$ in Eq.\,\eqref{eq-C-prime}. An easy guess of that quantity is $\sum\big(\bm{\widetilde{C_{11}}}^{-1}\big)_{ij}C_{1i}\delta_{j}$. And if we are marginalizing over the $k$th parameter, the generic equality condition becomes,
\begin{equation}\label{eq-IOI-marg-equal-condition-app}
\sum\limits_{i,j\neq k}\big(\bm{\widetilde{C_{kk}}}^{-1}\big)_{ij}C_{ki}\delta_{j}=\delta_k\,.  ~~{\rm{(Equality~Condition)}}
\end{equation}
We have tested and verified that if the above equality condition is satisfied, we indeed have IOI $=$ IOI$_{mar,k}$. An intuitive subset of the equality condition \eqref{eq-IOI-marg-equal-condition-app} is that $C_{ki}=\delta_j=0$, which represents the two conditions mentioned in Sec.\,\ref{subsection-marginalization-and-IOI}: if the two experiments do not have conflict on a particular parameter, and if that parameter does not have correlation between the other parameters, experimental inconsistency will be preserved after marginalizing that parameter.

\begin{figure*}[t]
\includegraphics[width=0.32\textwidth]{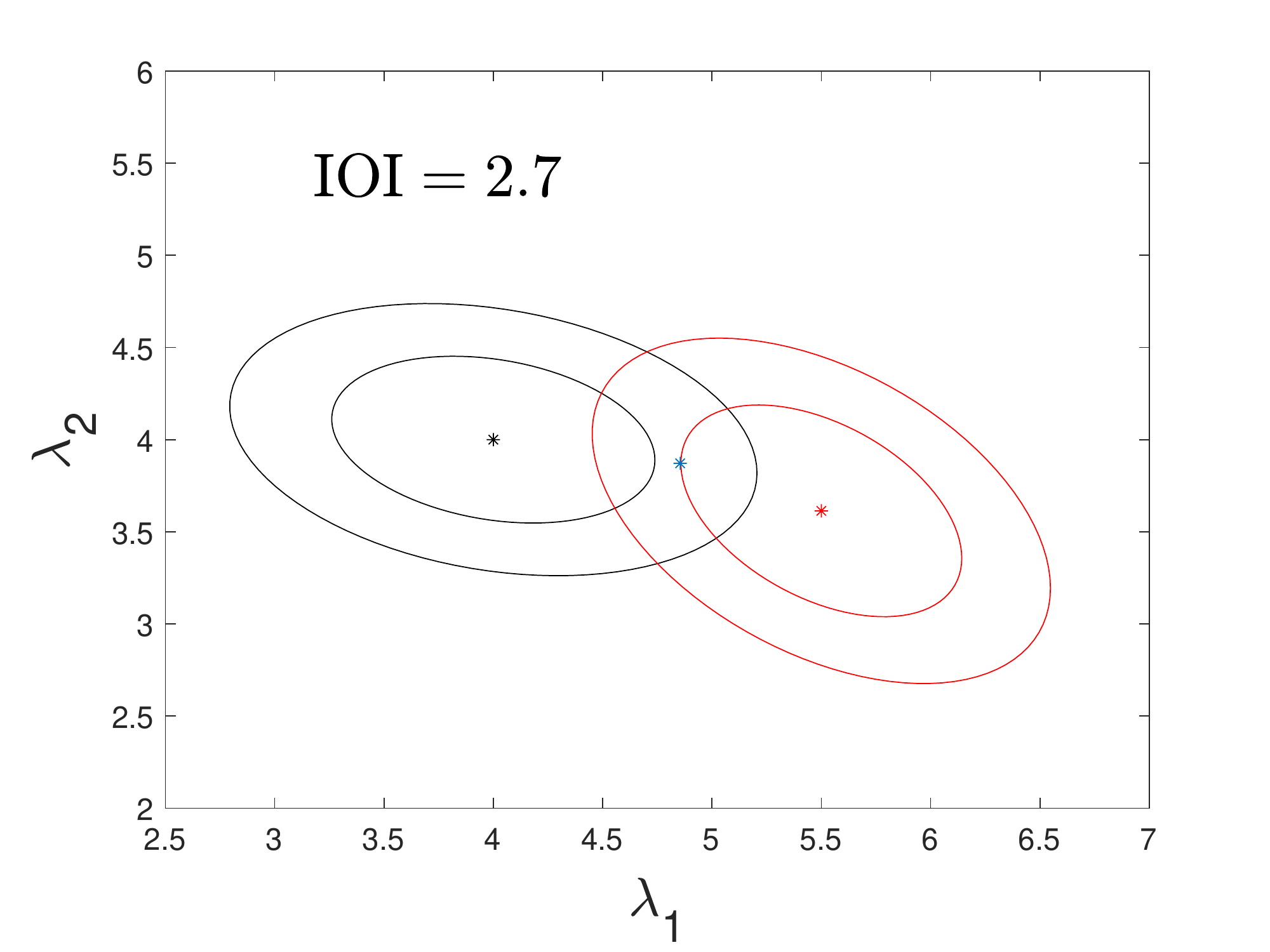}
\includegraphics[width=0.32\textwidth]{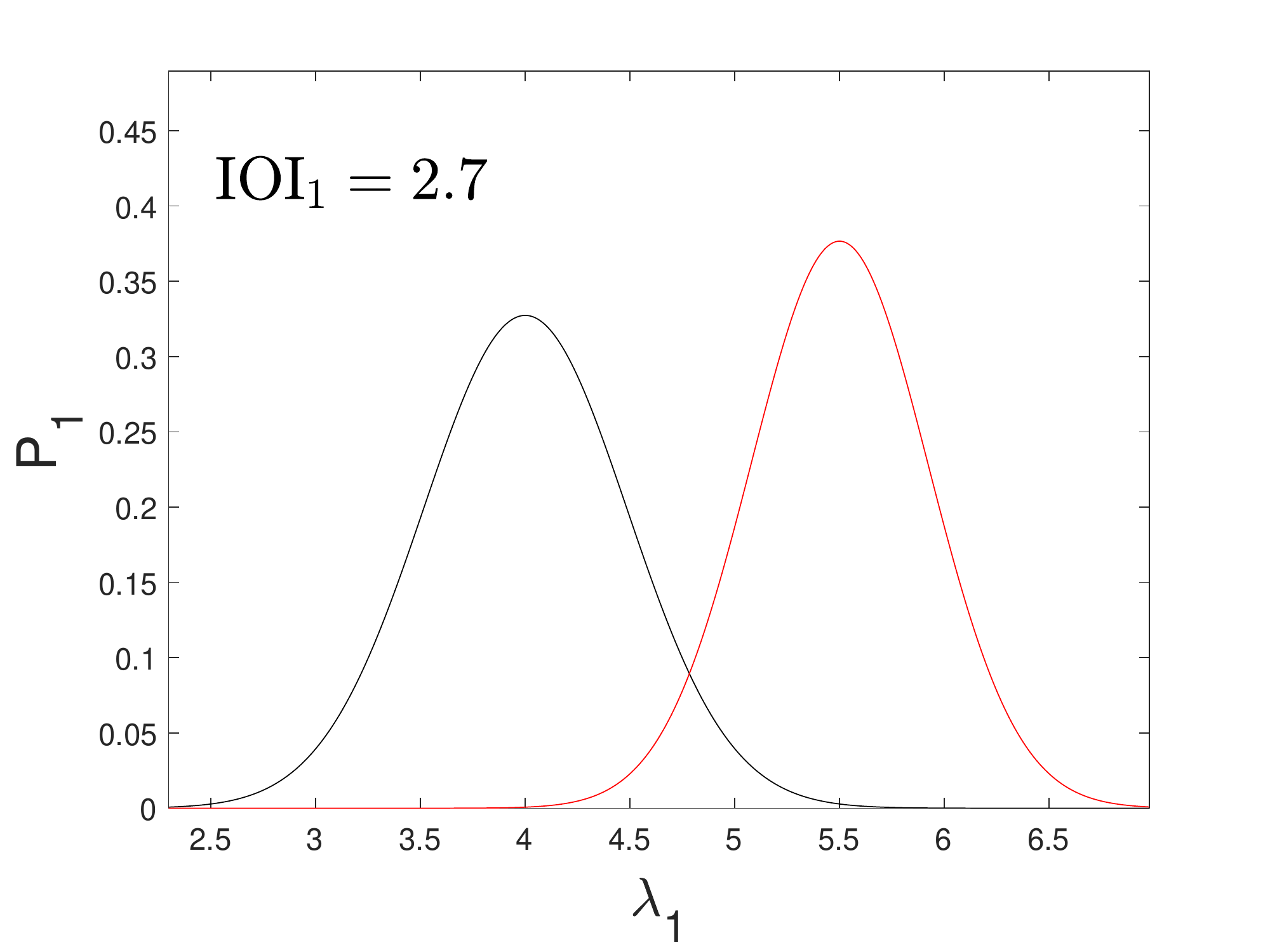}
\includegraphics[width=0.32\textwidth]{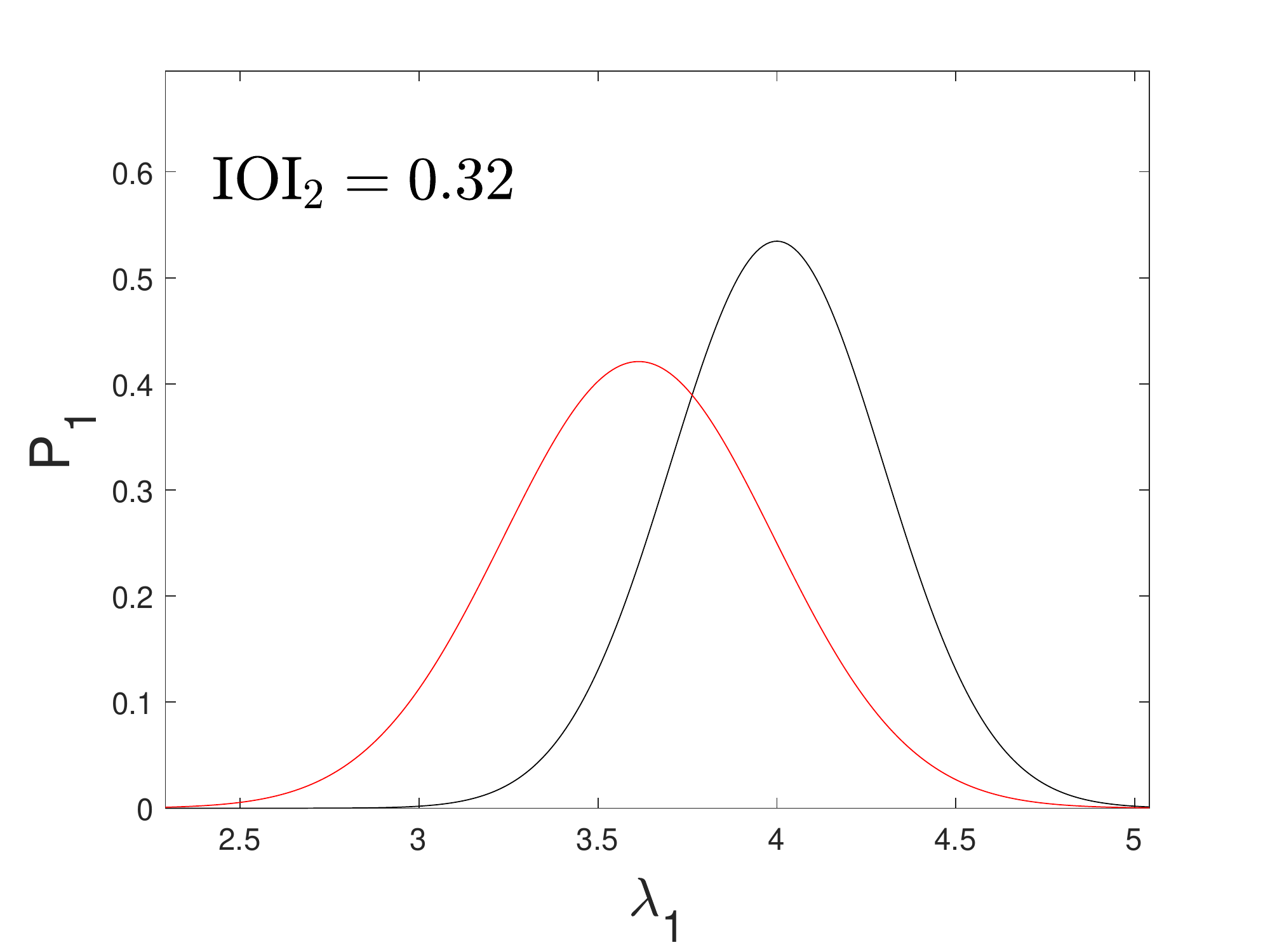}
\caption[Special case for IOI equality condition]{\label{fig-equal-condition-general}An example of situation which satisfies the equality condition \eqref{eq-IOI-marg-equal-condition-app}. Left: A two-parameter model constrained by two experiments shown in black and red. Middle: The constraint of $\lambda_1$ marginalized over $\lambda_2$. Right: The constraint of $\lambda_2$ marginalized over $\lambda_1$. In this particular example, the condition $\tfrac{C_{12}}{C_{11}}\delta_1=\delta_2$. If we marginalize over $\lambda_2$, IOI remains the same, i.e., IOI$=$IOI$_1$ even though IOI$_2\neq0$.}
\end{figure*}

\subsection{The analytic expression for the drop of IOI}\label{appendix-sub-IOI-drop}
In Sec.\,\ref{subsection-supplemental-measure}, we have defined the drop of IOI after marginalizing over $\lambda_k$ ($\Delta$IOI$_k$) as a supplemental measure of the one-parameter inconsistency in a multiparameter model. From the proof in Sec.\,\ref{appendix-sub-the-proof-marg}, especially from Eq.\,\eqref{eq-numerator} and Eq.\,\eqref{eq-DeltaGDelta-extend}, we can obtain $\Delta$IOI$_k$ analytically and it is,
\begin{equation}\label{eq-DeltaIOI-special}
  \Delta{\rm IOI}_k=\frac{\Big(\delta_k-\sum\limits_{i\neq k}\frac{C_{ki}}{C_{ii}}\delta_i\Big)^2}{C_{kk}-\sum\limits_{i\neq k}C_{ki}^2/C_{ii}}\,.
\end{equation}
The above expression is not generically right, since it is not in an invariant form under a linear transformation along the $\lambda_k$ axis. To generalize the result [Eq.\,\eqref{eq-DeltaIOI-special}], we only need to find an invariant quantity that reduces to Eq.\,\eqref{eq-DeltaIOI-special} in the case when $\CB$ takes the form of $\CB'$ in Eq\,\eqref{eq-C-prime}. We find that the general expression of $\Delta$IOI$_k$ is
\begin{equation}\label{eq-DeltaIOI-general}
\begin{split}
\Delta{\rm IOI}_k&=\frac{\Big[\delta_k-\sum\limits_{i,j\neq k}\Big(\mathbf{\widetilde{C_{kk}}}^{-1}\Big)_{ij}C_{ki}\delta_j\Big]^2} {C_{kk}-\sum\limits_{i,j\neq k}\big(\mathbf{\widetilde{C_{kk}}}^{-1}\big)_{ij}C_{ki}C_{kj}}\\
&\xrightarrow{\rm{Symbolically}}~ \frac{\Big(\delta_k-\mathbf{V_k}^T\big(\mathbf{\widetilde{C_{kk}}}^{-1}\big)\mathbf{\widetilde{\delta_k}}\Big)^2}{C_{kk}-\mathbf{V_k}^T\big(\mathbf{\widetilde{C_{kk}}}^{-1}\big)\mathbf{V_k}}\,,
\end{split}
\end{equation}
where $(\bm{V_k})_{i}=C_{ki}$ for $i\neq k$. We have tested and verified that Eq.\,\eqref{eq-DeltaIOI-general} is the same as the definition $\Delta$IOI$_k$ $\equiv$ IOI $-$ IOI$_{marg,k}$.

\begin{table*}[!htbp]
\caption[Situations and effects of marginalization over a parameter]{\label{table-marginalization-situations}The four situations and effects of marginalizing over a parameter $\lambda_i$. Only when both $\Delta$IOI$_i$ and IOI$_i$ are small can we safely marginalize over $\lambda_i$ without losing any information about the inconsistency of the model.}
\begin{ruledtabular}
\begin{tabular}{lp{0.6\textwidth}}
  Cases & Effect \\
  \hline
  Both $\Delta$IOI$_i$ and IOI$_i$ are small & Safe to marginalize over $\lambda_i$ to analyze the inconsistency for the whole model\\
  IOI$_i$ is small, but $\Delta$IOI$_i$ is large &  Marginalization will hide the inconsistency due the correlation (specified by $\bm{C}=\bm{C^{(1)}}+\bm{C^{(2)}}$) between $\lambda_i$ and some other parameters\\
  $\Delta$IOI$_i$ is small, but IOI$_i$ is large & Marginalization over $\lambda_i$ will get a similar IOI as the whole model, but ignore the shape of the original probability distribution\\
  Both IOI$_i$ and $\Delta$IOI$_i$ are large  & Marginalization over $\lambda_i$ will not be good for the analysis of the inconsistency\\
\end{tabular}
\end{ruledtabular}
\end{table*}

\subsection{A discussion on different situations in parameter marginalization}\label{appendix-sub-discuss-equality-condition}
In general, the equality equation can be satisfied without $C_{ki}=0$ or $\delta_j=0$. An example is shown by the three panels in Fig.\,\ref{fig-equal-condition-general}. It is a coincidence there for IOI to be the same after marginalizing over $\lambda_2$. The reason is as follows. If $\delta_2=0$, we will have IOI$>$IOI$_1$ because marginalizing over $\lambda_2$ hides some inconsistency due to the correlation between $\lambda_1$ and $\lambda_2$ specified by $\bm{G}$. In Fig.\,\ref{fig-equal-condition-general}, $\delta_2\neq0$, and the red ellipse is shifted a little downward compared to the case $\delta_2=0$. As a result, the joint mean is located at a point which is more easily ``supported'' by the two experiments,  and consequently IOI becomes smaller. Meanwhile, IOI$_1$ does not depend on the vertical shift of both distributions. For a particular amount of vertical shift, IOI becomes the same as IOI$_1$.

If the drop of IOI is small after parameter marginalization, we may be able to marginalize over the corresponding parameter to analyze the inconsistency for the whole model. If so, we can analyze the model in a low parameter dimension. And hopefully it can be reduced to two or three dimensions so that we can graphically show the full inconsistency. That is one of the motivations for us a we propose a supplemental measure of the inconsistency for one parameter ($\Delta$IOI$_i$) in Sec.\,\ref{subsection-supplemental-measure}. However, even when the drop is small, we still need to pay attention to whether it is a special case such as the one shown in Fig.\,\ref{fig-equal-condition-general}. In such a special case, marginalization over $\lambda_2$ ignores the original probability distributions, although we get the same IOI. Fortunately, it is easy to check whether it is a special case. For example, if the drop of IOI after marginalizing over $\lambda_2$ (i.e., $\Delta$IOI$_2$) is small, we can check whether the IOI after marginalizing over all other parameters but $\lambda_2$ (i.e., IOI$_2$) is also small. If both $\Delta$IOI$_2$ and IOI$_2$ are small, we can safely marginalize over $\lambda_2$ and analyze the inconsistency for the whole model. If not, marginalizing over $\lambda_2$ will either hide some inconsistencies or ignore the original shape of the probability distribution. Table \ref{table-marginalization-situations} summarizes the four situations regarding different combinations of $\Delta$IOI$_i$ and IOI$_i$ and the corresponding effects of marginalizing over the $\lambda_i$. In the right panel of Fig.\,\ref{fig-equal-condition-general}, IOI$_2$ is not $0$ so the example shown in Fig.\,\ref{fig-equal-condition-general} is a special case. Combining IOI$_i$ and $\Delta$IOI$_i$, we can get a lot of information about the full distribution.

\bibliography{IOI}{}

\end{document}